\documentclass[acmsmall,screen]{acmart}

\AtBeginDocument{%
  \providecommand\BibTeX{{%
    \normalfont B\kern-0.5em{\scshape i\kern-0.25em b}\kern-0.8em\TeX}}}

\setcopyright{rightsretained}
\setcctype{by-nc}
\acmYear{2025}
\acmJournal{CSUR}
\acmDOI{10.1145/3711683}
\acmVolume{57} 
\acmNumber{7} 
\acmMonth{2} 

\usepackage{graphicx}
\usepackage{array,multirow}
\usepackage{threeparttable}
\usepackage{enumitem}
\usepackage{makecell}
\usepackage{pifont}

\begin{document}

\title{Approximate Computing Survey, Part II: Application-Specific \& Architectural Approximation Techniques and Applications}

\author{Vasileios Leon}
  \affiliation{%
  \institution{National Technical University of Athens}
  \department{School of Electrical and Computer Engineering}
  \country{Greece}}
\author{Muhammad Abdullah Hanif}
  \affiliation{%
  \institution{New York University Abu Dhabi}
  \department{Division of Engineering}
  \country{United Arab Emirates}}
\author{Giorgos Armeniakos}
  \affiliation{%
  \institution{National Technical University of Athens}
  \department{School of Electrical and Computer Engineering}
  \country{Greece}}
\author{Xun Jiao}
  \affiliation{%
  \institution{Villanova University}
  \department{Department of Electrical and Computer Engineering}
  \country{United States}}
\author{Muhammad Shafique}
  \affiliation{%
  \institution{New York University Abu Dhabi}
  \department{Division of Engineering}
  \country{United Arab Emirates}}
\author{Kiamal Pekmestzi}
  \affiliation{%
  \institution{National Technical University of Athens}
  \department{School of Electrical and Computer Engineering}
  \country{Greece}}
\author{Dimitrios Soudris}
  \affiliation{%
  \institution{National Technical University of Athens}
  \department{School of Electrical and Computer Engineering}
  \country{Greece}}

\renewcommand{\shortauthors}{Vasileios Leon et al.}
\renewcommand{\shorttitle}{Approximate Computing Survey, Part II: Application-Specific \& Architectural Techniques and Applications}

\begin{abstract}
The challenging deployment of compute-intensive applications from domains such as Artificial Intelligence (AI) and Digital Signal Processing (DSP), forces the community of computing systems to explore new design approaches. \emph{Approximate Computing} appears as an emerging solution, allowing to tune the quality of results in the design of a system in order to improve the energy efficiency and/or performance. This radical paradigm shift has attracted interest from both academia and industry, resulting in significant research on approximation techniques and methodologies at different design layers (from system down to integrated circuits). Motivated by the wide appeal of Approximate Computing over the last 10 years, we conduct a two-part survey to cover key aspects (e.g., terminology and applications) and review the state-of-the art approximation techniques from all layers of the traditional computing stack. Part II of the survey classifies and presents the technical details of application-specific and architectural approximation techniques, which both target the design of resource-efficient processors/accelerators and systems. Moreover, it reports a quantitative analysis of the techniques and a detailed analysis of the application spectrum of Approximate Computing, and finally, it discusses open challenges and future directions. 
\end{abstract}



\received{18 July 2023}
\received[Revised]{16 October 2024}
\received[Accepted]{24 December 2024}
\received[Published]{20 February 2025}

\maketitle

\newpage
\section{Introduction}

The recent technological advancements in processing, storage, communication, and sensing have transformed the landscape of computing systems. With the emergence of Internet of Things (IoT), a huge amount of data is generated, imposing technical challenges in the resource-constrained embedded systems that are placed at the edge of the network. The ever-growing number of global IoT connections, which increased by 18\% in 2022 and is expected to be 29.7 billion in 2027 \cite{iotanal}, only deteriorates this situation. At the same time, the already stressed cloud data centers are drawing worldwide concerns due to their enormous electricity demands. Indicatively, it is estimated that the energy consumption of data centers in Europe was increased by 43\% from 2010 to 2020 \cite{avgerinou}, while according to \cite{datacenters}, data centers may generate up to 8\% of the global carbon emissions in 2030. 

The above issues become even more critical when considering the massive growth of demanding applications from domains such as Digital Signal Processing (DSP), Artificial Intelligence (AI), and Machine Learning (ML). More specifically, the emergence of compute- and memory-intensive DSP and AI/ML applications marks a new era for computing, where traditional design approaches and conventional processors cannot meet the performance requirements \cite{heterog, georgis}. Therefore, it is obvious that the proliferation of data and workloads, the increase in the complexity of applications,
and the urgent need for power efficiency and fast processing,
all together,
force the industry of computing systems to examine alternative design solutions and computing schemes. 

\begin{figure}[!t]
    \centering
    \includegraphics{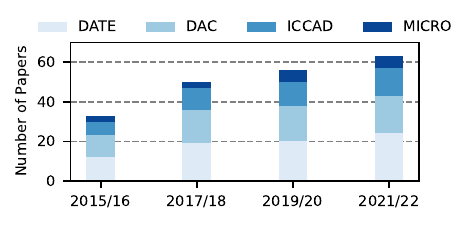}
    \vspace{-14pt}
    \caption{Number of publications that apply any type of approximation in 4 major design automation conferences.}
    \label{fig:stats_conf}
\end{figure}

In this transition, \textbf{\emph{Approximate Computing (AxC)}} is already considered a novel design paradigm that improves the energy efficiency and performance of computing systems~\cite{swagath:2020}.
During the past decade, AxC has been established as a vigorous approach to leverage the inherent ability of numerous applications to produce results of \emph{acceptable} quality, despite some \emph{inaccuracies} in the computations~\cite{csur:arme}.
The motivation behind AxC is further analyzed in Part I of the survey \cite{mysurvey_pt1}. 
Driven by its potential to address the energy and performance demands in a favorable manner, 
AxC is increasingly gaining significant attention over the last years.
A representative example of this trend is depicted in Figure~\ref{fig:stats_conf}.  
This figure illustrates the growing number of works that apply any kind of approximation (e.g., software, hardware, architectural), published in four major design automation conferences.
Furthermore, AxC has attracted significant interest from the industry.  
Companies such as Google \cite{google}, IBM \cite{ibm}, and Samsung \cite{samsung} exploit the AI/ML error resilience and design accelerators for compressed networks with reduced precision and less computations, improving the energy efficiency and performance in exchange for negligible or zero accuracy loss.

Considering the need to go beyond single optimizations or efficient processor design,
which solely do not suffice~\cite{edge:survey}, prior research has focused on approximation techniques at different design abstraction layers~\cite{swagath:2020}, involving algorithmic- or compiler-level approximations (\emph{software})~\cite{2018_Akhlaghi_ISCA, 2005_Alvarez_IEEEtc}, circuit-level approximations (\emph{hardware})
~\cite{2016_Jiang_IEEEtc, 2013_Chen_IEEEtvlsi} and systematic design of approximate processors (\emph{architectural})~\cite{ersa, naccel:micro}.
Despite their share objective of achieving energy/area gains and latency reduction, these fundamental techniques are orthogonal and can complement each other.
As combining AxC techniques is expected to be highly beneficial (several works report energy gains over 90\%~\cite{edge:survey}), a further adoption and understanding of cross-layer development (i.e., hardware/software co-design), including tools \& frameworks tailored to specific applications, is highly required to maximize these benefits. In this context, 
motivated by the attractive outcomes, novel methods, and future potential of AxC, 
we conduct a two-part survey that covers all its key aspects.

\section{Approximate Computing Survey}\label{sec:surv}

\textbf{Scope and Contribution:}
As also discussed in Part I \cite{mysurvey_pt1}, the proposed survey examines the entire AxC computing stack,
reviewing software, hardware, and architectural approximation techniques,
contrary to other AxC surveys
that focus on a single area
(either design layer, technique or application).
Table~\ref{tb_srv} 
compares surveys studying multiple areas of AxC, 
like our work. 
As shown,
the proposed two-part survey includes all the areas examined in the previous works,
while it reviews all of them in depth
(e.g., AI/ML has not received significant attention)
and it is up-to-date. 

Regarding the contribution of the current survey, it provides the reader with a complete view on AxC, while it allows them to emphasize on specific areas (e.g., architectural techniques or application domains),
as they are all presented in detail. 
Furthermore, 
this survey can be considered as a tutorial on the state-of-the-art approximation techniques, as it not only classifies them, but it reports in brief their technical details. 
Overall, 
the key contributions of the proposed two-part survey are: 
1) the introduction of the AxC paradigm by discussing its background and application domains, 
2) the review of a plethora of state-of-the-art approximation techniques, their coarse- and fine-grained classification,
and the presentation of their technical details, 
3) the study of application domains where approximations are tolerated and the analysis of representative techniques and results per domain,  
and 4) the discussion of open challenges in the design of approximate systems, as they emerge from the comprehensive literature review.  

\begin{table*}[!t]
\renewcommand{\arraystretch}{1.1}
\setlength{\tabcolsep}{3.4pt}
\centering
\footnotesize
\caption{Qualitative comparison of Approximate Computing surveys on the entire computing stack.}
\vspace{-8pt}
\begin{threeparttable}
\begin{tabular}{c@{\hspace{3.7pt}}cccccccccccccccc}
\hline
\multicolumn{2}{c|}{\rotatebox{45}{\textbf{\hspace{6pt}AxC Survey}}} &
  \multicolumn{1}{c|}{\rotatebox{80}{\textbf{Year Coverage}}} &
  \multicolumn{1}{c|}{\rotatebox{80}{\textbf{Pages \#}}} &
  \multicolumn{1}{c|}{\rotatebox{80}{\textbf{References \#}}} &
  \multicolumn{1}{c|}{\rotatebox{80}{\textbf{SW Tech.}}} &
  \multicolumn{1}{c|}{\rotatebox{80}{\textbf{HW Tech.}}} &
  \multicolumn{1}{c|}{\rotatebox{80}{\textbf{Arch. Approx.}}} &
  \multicolumn{1}{c|}{\rotatebox{80}{\textbf{AI/ML}}} &
  \multicolumn{1}{c|}{\rotatebox{80}{\textbf{Memories}}} &
  \multicolumn{1}{c|}{\rotatebox{80}{\textbf{\begin{tabular}[c]{@{}c@{}}Frameworks\\[-3pt] \& Tools\end{tabular}}}} &
  \multicolumn{1}{c|}{\rotatebox{80}{\textbf{Metrics}}} &
  \multicolumn{1}{c|}{\rotatebox{80}{\textbf{Benchmarks}}} &
  \multicolumn{1}{c|}{\rotatebox{80}{\textbf{\begin{tabular}[c]{@{}c@{}}CPU/FPGA/\\[-3pt] GPU/ASIC\end{tabular}}}} &
  \multicolumn{1}{c|}{\rotatebox{80}{\textbf{Terminology}}} 
  &
  \multicolumn{1}{c|}{\rotatebox{80}{\textbf{Challenges}}} 
  &
  \multicolumn{1}{c|}{\rotatebox{80}{\textbf{Quant. Analysis\tnote{2}\phantom{b}}}}
  \\ \hline \hline

\multicolumn{2}{c}{\cite{2013_Han_ETS}}            & 2013 & 6  & 65  & \ding{51}  & \ding{51}  & $\approx$ & $\approx$ & \ding{55}  & \ding{55}  & \ding{51} & \ding{55} & $\approx$ & \ding{55}  & \ding{55} & $\approx$ \\ \hline
  
\multicolumn{2}{c}{\cite{2016_Mittal_ACMsrv}}     & 2015 & 33\tnote{1}  & 84  & \ding{51}  & $\approx$ & \ding{51}  & $\approx$ & \ding{51}  & \ding{51}  & \ding{51} & \ding{51} & \ding{51}  & \ding{55}  & \ding{51} & \ding{55} \\ \hline

\multicolumn{2}{c}{\cite{2016_Xu_IEEEdt}}         & 2015 & 15 & 59  & \ding{51}  & \ding{51}  & \ding{51}  & \ding{55}  & $\approx$ & \ding{55}  & \ding{55} & \ding{55} & $\approx$ & \ding{55}  & \ding{55} & \ding{55} \\ \hline

\multicolumn{2}{c}{\cite{2015_Venkataramani_DAC}} & 2015 & 6  & 54  & \ding{51}  & \ding{51}  & \ding{51}  & \ding{55}  & \ding{55}  & \ding{51}  & \ding{55} & \ding{55} & $\approx$ & \ding{55}  & \ding{55} & \ding{55} \\ \hline

\multicolumn{2}{c}{\cite{2016_Shafique_DAC}}      & 2016 & 6  & 47  & \ding{51}  & \ding{51}  & \ding{51}  & \ding{55}  & \ding{55}  & $\approx$ & \ding{55} & \ding{55} & $\approx$ & \ding{55}  & \ding{55} & $\approx$ \\ \hline

\multicolumn{2}{c}{\cite{2018_Moreau_IEEEesl}}    & 2017 & 4  & 40  & \ding{51}  & \ding{51}  & \ding{55}  & $\approx$ & $\approx$ & \ding{55}  & \ding{55} & \ding{55} & $\approx$ & \ding{55}  & \ding{55} & \ding{55} \\ \hline

\multicolumn{2}{c}{\cite{2018_Moreno_LATS}}       & 2017 & 6  & 72  & \ding{51}  & \ding{51}  & \ding{51}  & $\approx$ & $\approx$ & $\approx$ & \ding{55} & \ding{51} & $\approx$ & \ding{55}  & \ding{51} & \ding{55} \\ \hline 

\multicolumn{2}{c}{\cite{2020_Stanley_ACMsrv}}   & 2020 & 39\tnote{1}  & 235 & $\approx$ & \ding{51}  & \ding{51}  & \ding{55}  & \ding{51}  & $\approx$ & \ding{55} & \ding{55} & $\approx$ & \ding{55}  & \ding{51} & \ding{55} \\ \hline \hline

\multirow{2}{*}{This work} & Pt. 1       & 2023 & 36\tnote{1}  & 222 & \ding{51}  & \ding{51}  &  &   &   & \ding{51}  &  &  & \ding{51} & \ding{51}  &  & \ding{51}\\  

& Pt. 2       & 2023 & 36\tnote{1}  & 301 & \ding{51}  & \ding{51}  & \ding{51} & \ding{51}  & \ding{51}  &   & \ding{51} & \ding{51}  & \ding{51} &   & \ding{51} & \ding{51} \\ \hline \hline

\end{tabular}
\begin{tablenotes}
\scriptsize 
    \item[1] Single-column pages. 
    \item[2] Quantitative analysis involving count of works, frequencies and numerical assessments.
\end{tablenotes}
\end{threeparttable}
\label{tb_srv}
\end{table*}

\begin{figure}[t]
 \vspace{-4pt}
    \centering
    \includegraphics[width=1\textwidth]{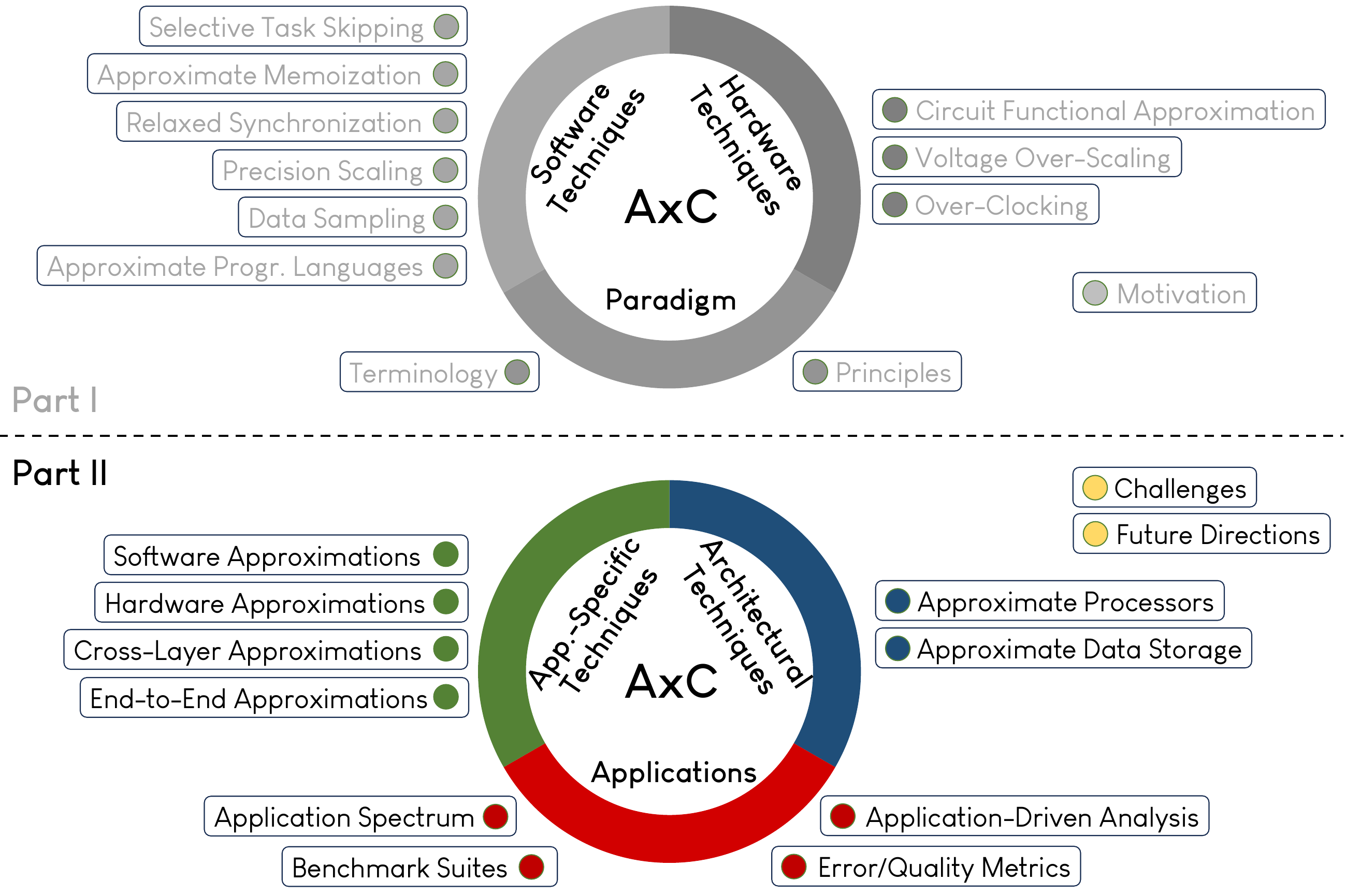}
    \caption{Organization of the proposed two-part survey on Approximate Computing.}
    \label{fg_struct}
\end{figure}

\noindent\textbf{Organization:}
Figure \ref{fg_struct} presents the content of each survey part:  

\begin{description}[leftmargin=35pt]
\item[\hspace{3.8pt}Part I:] It is presented in \cite{mysurvey_pt1}, and it introduces the AxC paradigm
(terminology and principles)
and reviews software \& hardware approximation techniques. 
\item[Part II:] It is presented in the current paper, and it reviews application-specific \& architectural approximation techniques and introduces the AxC applications (domains, quality metrics, benchmarks). 
\end{description}

The remainder of the article (Part II of the survey)
is organized as follows. 
Section~\ref{sec:app_spec}
reports software-, hardware-
and cross-layer application-specific approximation techniques. 
Section~\ref{sec:arch_lvl}
focuses on the architecture level and classifies approximate processors and storage. 
Section~\ref{sec:comp} performs a quantitative analysis of the reviewed works. 
Section~\ref{sec:apps} studies the applications of AxC,
while Section~\ref{sec:challenges}
discusses challenges and future directions. 
Finally, Section~\ref{sec:conc} draws the conclusions. 

\section{Application-Specific Approximation Techniques}\label{sec:app_spec}

\begin{figure}[!t]
    \centering
    \includegraphics[width=0.85\textwidth]{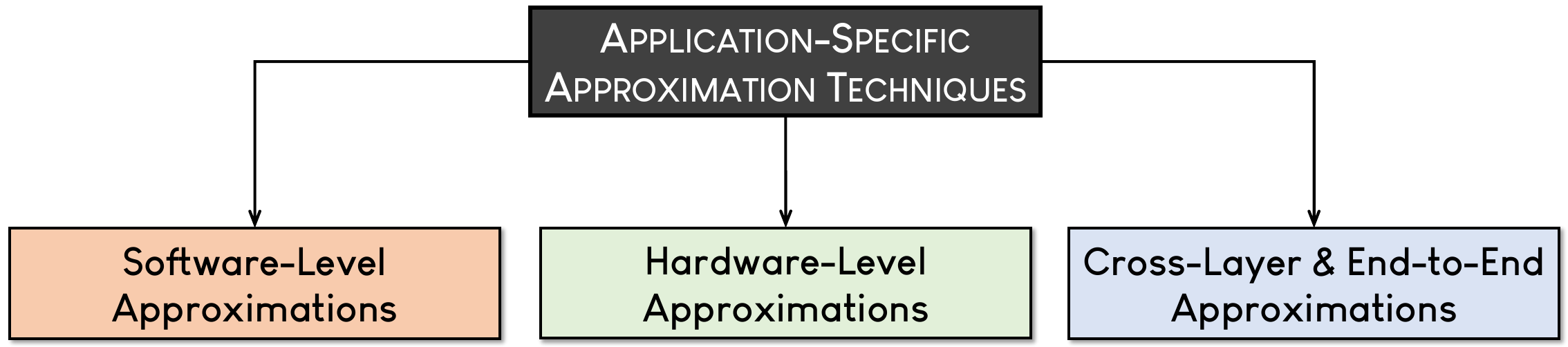}
    \caption{Classification of application-specific approximation techniques for DL and AI applications in three main classes:
    \emph{Software-Level}, \emph{Hardware-Level}, and \emph{Cross-Layer \& End-to-End}.}
    \label{fg_app_specific}
\end{figure}

This section reviews application-specific approximation techniques. It mainly discusses techniques that are proposed for improving the performance and efficiency of Deep Learning (DL) and AI applications. The techniques presented in the literature can be broadly categorized into software-level, hardware-level and cross-layer approximation techniques (see Figure~\ref{fg_app_specific} and Table~\ref{tb_appspec}). 

\begin{table*}[!t]
\renewcommand{\arraystretch}{1.05}
\centering
\footnotesize
\setlength{\tabcolsep}{3.5pt}
\caption{Classification of application-specific approximation techniques.}
\vspace{-8pt}
\label{tb_appspec}
\begin{tabular}{l l} 
\hline
\begin{tabular}[c]{@{}l@{}}\textbf{Application-Specific}\\\textbf{Approximation Class}\end{tabular} & \makecell[c]{\textbf{Technique/Approach}}                                   \\ 
\hline\hline
\multirow{4}{*}{\begin{tabular}[c]{@{}l@{}}Software-Level \\Approximation\end{tabular}}              & Computation Skipping / Pruning~\cite{han2015learning, han2015deep, yu2017scalpel, li2016pruning, yvinec2021red, yvinec2022red++, he2017channel, hu2016network, tan2020dropnet, sui2021chip, he2022filter, lebedev2016fast, wen2016learning, wang2020neural, chen2021only}               \\
                                                                                                     & Precision Scaling~\cite{rastegari2016xnor, hubara2016binarized, zhou2016dorefa, jain2019biscaled, han2015deep, wang2019haq}                                            \\
                                                                                                     & Input-Adaptive Approximation~\cite{ganapathy2020dyvedeep, panda2017energy1, panda2017energy2, tann2016runtime, venkataramani2015sapphire, venkataramani2015scalable, lou2021dynamic, plastiras2018efficient}         
                                                                                                     \\
                                                                                                     & Input Scaling~\cite{tan2019efficientnet, ghosh2020approximate}            
                                                                                                     
                                                                                                     \\ 
\hline
\multirow{4}{*}{\begin{tabular}[c]{@{}l@{}}Hardware-Level \\Approximation\end{tabular}}              & Functional Approx. \& Approx. Datapaths~\cite{venkataramani2014axnn, mrazek2016design, mrazek2019alwann, mrazek2020libraries, sonnino2023daism, spantidi2021positive, saadat2018minimally, kim2018efficient, hanif2019cann, leon_lascas, lentaris_icecs} 
\\
                                                                                                     & Approximation-based Novel Data Representation Formats~\cite{jain2019biscaled, jain2018compensated, hanif2022conlocnn, guo2022ant}

\\
                                                                                                     & Voltage Scaling \cite{2018_Zhang_DAC, shin2019sensitivity, ji2020error, 2019_Pandey_DAC, kim2018matic, koppula2019eden}                                          \\
                                                                                                     & Data Approximation \cite{koppula2019eden, kim2018matic}                                          \\ 
\hline
\multirow{3}{*}{\begin{tabular}[c]{@{}l@{}}Cross-Layer \& \\ End-to-End \\Approximation\end{tabular}}      & Approx. at Various Layers of Computing Stack~\cite{arm2023codesign, arm2023crossapprox, raha2017towards, raha2018approximating, ghosh2020approximate, hashemi2018approximate, gong2019ara, hanif2022cross, fan2019axdnn, swagath:2020, raha_axis_ext}       \\
                                                                                                     & Full-System Perspective~\cite{raha2017towards, raha2018approximating, ghosh2020approximate, hashemi2018approximate, raha_axis_ext, raha_2024}                                       \\
                                                                                                     & Approx. of Multiple Subsystems in a Synergistic Manner~\cite{raha2017towards, raha2018approximating, ghosh2020approximate, hashemi2018approximate, raha_axis_ext, raha_2024}   \\
\hline\hline
\end{tabular}
\end{table*}

\subsection{Software-Level Approximations}

Various types of software-level approximation techniques have been proposed in the literature to reduce the computational complexity and memory footprint of AI and DL workloads. 
These techniques can be broadly divided into four groups: 
(i) \emph{computation skipping / pruning}~\cite{han2015learning, han2015deep, yu2017scalpel, li2016pruning, yvinec2021red, yvinec2022red++, he2017channel, hu2016network, tan2020dropnet, sui2021chip, he2022filter, lebedev2016fast, wen2016learning, wang2020neural, chen2021only}, (ii) \emph{precision scaling}~\cite{rastegari2016xnor, hubara2016binarized, zhou2016dorefa, jain2019biscaled, han2015deep, wang2019haq}, (iii) \emph{input-adaptive approximation}~\cite{ganapathy2020dyvedeep, panda2017energy1, panda2017energy2, tann2016runtime, venkataramani2015sapphire, venkataramani2015scalable, lou2021dynamic, kyrkou2018dronet, plastiras2018efficient}, and (iv) \emph{input scaling}~\cite{tan2019efficientnet, ghosh2020approximate}. 

\subsubsection{Pruning}
Computation skipping is one of the most effective software-level techniques for approximating DNNs. Various techniques have been proposed under the umbrella of DNN pruning that exploits computation skipping for optimizing the DNN inference process. 
In general, DNN pruning refers to the process of removing non-essential (or less important) weights from a DNN to reduce its computational complexity, memory footprint as well as energy consumption during the inference process. 
These techniques are generally divided into two main categories: (1) unstructured pruning and (2) structured pruning. 

Unstructured pruning (a.k.a. fine-grained pruning) refers to techniques that analyze the importance/saliency of each individual weight and remove it if its importance is below a predefined threshold. 
These techniques are usually designed to be iterative, where in each iteration only a small percentage of less significant weights are removed and then the network is retrained for a few epochs to regain close to the baseline accuracy. 
For example, Han et al.~\cite{han2015learning} propose a three-step method to prune DNNs. 
First, they train a DNN to learn the importance of connections. 
Then, they remove non-essential (less important) connections from the DNN. 
Finally, they fine-tune the remaining weights to regain the lost accuracy. 
The authors further propose that learning the right connections is an iterative process. 
Therefore, the process of pruning followed by fine tuning should be repeated multiple times (with a small pruning ratio) to achieve higher compression ratios. 
Deep~Compression~\cite{han2015deep} combines a similar method with weight sharing and Huffman coding to achieve ultra-high compression ratios.  

Although unstructured pruning significantly reduces the number of parameters and the memory footprint of DNNs, it does not guarantee energy or latency benefits in all cases. 
This is mainly because; 
(1) the weights of a pruned network are usually stored in a compressed format, 
such as the Compressed Sparse Row (CSR) format, 
to achieve high compression ratios and, therefore, have to be uncompressed before corresponding computations, which results in significant overheads, 
and (2) the underlying hardware (in most use cases) is not designed to take advantage of unstructured sparsity in DNNs, which results in under utilization of the hardware. 
Therefore, specialized hardware accelerators, such as EIE~\cite{han2016eie} and SCNN~\cite{parashar2017scnn}, are required to efficiently process data using unstructured sparse DNNs. 
Similarly, Yu et al.~\cite{yu2017scalpel} highlight that it is essential to align the pruning method to the underlying hardware architecture to achieve efficiency gains. 
A mismatch between the type of sparsity in the network and the sparsity the hardware can support efficiently usually leads to significant overheads. 
Based on this observation, the authors propose two novel pruning methods, i.e., SIMD-aware weight pruning and node pruning. 
SIMD-aware weight pruning maintains weights in aligned fixed-size groups to fully utilize the available SIMD units in the underlying hardware, while node pruning removes less significant nodes from the network without affecting the dense format of weight matrices. 

Various other techniques have also been proposed in the literature to achieve structured sparsity in DNNs. These methods generally include weight-dependent pruning techniques~\cite{li2016pruning, yvinec2021red, yvinec2022red++}, activation-based techniques~\cite{he2017channel, hu2016network, tan2020dropnet, sui2021chip, he2022filter}, and regularization-based techniques~\cite{lebedev2016fast, wen2016learning, wang2020neural, chen2021only}. The weight-dependent techniques evaluate the importance of filters solely based on the weights of the filters, while activation-based techniques use activation maps to identify critical filters in the network. Similarly, regularization-based techniques encourage sparsity during training by adding regularization terms in the loss function or by exploiting batch normalization parameters. 

\subsubsection{Precision Scaling}
It refers to the techniques of mapping values from a high-precision range to a lower-precision range with a lesser number of quantization levels. 
The number of quantization levels in a range defines the number of bits required to represent each data value. 
Hence, by using a lower-precision format, such as 4-bit or 8-bit fixed-point or floating-point format instead of the standard 32-bit floating-point format, the memory footprint as well as the computational requirements of DNNs can be significantly reduced. 

Various techniques have been proposed to achieve precision scaling of DNNs without affecting their application-level accuracy. 
The most commonly used technique is 8-bit range linear post-training quantization, where floating-point weights and activations are converted to 8-bit fixed-point format. 
The range linear quantization is further divided into two types: (1) asymmetric quantization~\cite{jacob2018quantization} and (2) symmetric quantization~\cite{krishnamoorthi2018quantizing}. 
In asymmetric quantization, the minimum and maximum observed values in the input range are mapped to the minimum and maximum values in the output (integer/fixed-point) range, respectively. 
However, in symmetric quantization, the maximum absolute observed value in the input range is used to define both minimum and maximum values of the input range. 
Significant efforts have been made to push the limits of DNN precision scaling to ultra-low precision levels. 
Works like XNOR-Net~\cite{rastegari2016xnor}, Binarized Neural Network (BNN)~\cite{hubara2016binarized} and DoReFa-Net~\cite{zhou2016dorefa} explore the potential of aggressive quantization, up to 1-bit precision. 
However, as aggressive quantization leads to significant accuracy loss, these techniques usually employ quantization-aware training to achieve reasonable accuracy, as in such cases training also acts as an error-healing process. 

Non-linear (or non-uniform) quantization techniques have also been explored in the literature~\cite{jain2019biscaled, han2015deep}. These techniques are inspired by the non-uniform probability distribution of DNN data structures and propose to distribute the quantization levels accordingly. 
Apart from the above-mentioned techniques, mixed-precision techniques are also commonly used, where different precision formats can be used for different layers and filters/channels of a DNN~\cite{wang2019haq}. It is noted that mixed-precision quantization is only effective when the underlying hardware is capable of supporting different precision formats. 

\subsubsection{Input-Adaptive Approximation}

Besides pruning and precision scaling, input-adaptive approximations are also commonly used to approximate DNN workloads. In general, input-adaptive approximations refer to the techniques that dynamically adapt the level of approximation based on the characteristics of the input data. 
Various methods have been proposed in the literature to realize input-adaptive approximations. These methods mainly include (1) early-exit classifiers, (2) dynamic network architectures, and (3) selective attention mechanisms. Works such as~\cite{panda2017energy1, venkataramani2015scalable} that fall under early-exit classifiers argue that most of the inputs can be classified with very low effort. Towards this, the authors in~\cite{panda2017energy1} propose Conditional Deep Learning (CDL), where the outputs of convolutional layers are used to estimate the difficulty of input samples (with the help of linear classifiers) and conditionally activate the deeper layers of the DNN. Similarly, \cite{panda2017energy2} propose a hierarchical classification framework by exploiting semantic decomposition. Towards dynamic network architectures, \cite{ganapathy2020dyvedeep} presents an approach that adapts the structure of a DNN based on the difficulty of the input sample. Along similar lines, \cite{tann2016runtime} presents a method for building runtime configurable DNNs and a score margin-based mechanism for dynamically adjusting the DNN capability. 
Unlike the above techniques that involve network structure or size modification, selective attention mechanisms focus on reducing computational requirements by processing only the important/relevant regions of the input while ignoring all less relevant parts. A prominent work in this direction includes selective tile processing~\cite{plastiras2018efficient}, where a low cost attention mechanism is used to identify the tiles that require processing. 

\subsubsection{Input Scaling} 
Given that neighboring pixels in images are not always statistically independent and contain a significant amount of redundant information, input scaling can be exploited to reduce the computational requirements of DNN inference. 
Towards this, 
Tan et al. in~\cite{tan2019efficientnet} propose a compound scaling method that uniformly scales network width, depth, and input resolution to design a complete class of EfficientNets. 
Moreover, techniques such as  AxIS~\cite{ghosh2020approximate} employ input scaling as an approximation knob in their cross-layer approximations to achieve ultra-high efficiency gains. 

\subsection{Hardware-Level Approximations}\label{subsec:shw_lvl}

Venkataramani et al. in~\cite{venkataramani2014axnn} highlight that DNNs are mostly used for error-resilient applications, and hence, hardware approximations can be used to effectively trade quality for efficiency of the DNN inference process. 
Towards this, they propose a method to transform a given Neural Network (NN) to an Approximate Neural Network (AxNN)~\cite{venkataramani2014axnn}. 
The method mainly employs backpropagation to estimate the importance of each neuron and then selectively approximates less significant neurons through precision scaling of weights and activations to achieve better energy-accuracy trade-offs. The method also employs approximation-aware retraining to regain a significant portion of the accuracy lost due to approximations. 

As a single DNN inference can take from some million to billions of MAC operations, the use of approximate MAC units in DNN accelerators (instead of accurate ones) can have a significant impact on the overall energy consumption of DNN inference systems. In general, approximate MAC units refer to the simplified variants of the accurate counterparts that offer approximately the same functionality but at a lower energy or performance cost. 
Various significant efforts have been made toward designing approximate MAC units. 
Most of these works focus on approximating the multiplication part in the MAC operations, as multipliers are in general more complex and resource hungry than adders. 
Marazek et al. in~\cite{mrazek2016design} employ the Cartesian Genetic Programming (CGP) approach to approximate accurate multiplier designs. The methodology employs approximation-aware retraining of the given NN in the methodology to reach the best possible approximate multiplier design for the given application. 
However, approximation-aware retraining is not possible in all scenarios, e.g., due to limited computational resources or due to the lack of availability of training data. 
Therefore, to address this limitation, ALWANN~\cite{mrazek2019alwann} proposes a layer-wise approximation technique to select an appropriate approximate multiplier from a given library (e.g., EvoApprox~\cite{2017_Mrazek_DATE}) for each individual layer of the given DNN without involving any training. 
The authors also present a computationally inexpensive method to fine-tune the DNN weights after the approximate module selection process to reduce the accuracy loss due to approximations. 
In the same context, the Max-DNN framework \cite{leon_lascas} relies on ALWANN's functionalities to provide fine-grained  approximation of the multiplications at different DNN layers.  
In~\cite{mrazek2020libraries}, the authors present a methodology to generate libraries of more specific approximate circuits and show that such techniques can produce better results than techniques such as  ALWANN~\cite{mrazek2019alwann}. 
Other prominent approximate multiplier designs for DNN acceleration include Digital Approximate In-SRAM Multiplier (DAISM)~\cite{sonnino2023daism}, positive/negative approximate multipliers~\cite{spantidi2021positive} and approximate log-multipliers~\cite{saadat2018minimally, kim2018efficient}. 
Moreover, towards mitigating the impact of approximation errors, CANN~\cite{hanif2019cann} presents the concept of \textit{curable} approximations for DNN accelerators, where approximation errors can be internally compensated with some additional functionality. 
To facilitate the design and exploration of approximate modules for efficient DNN inference, fast simulation techniques and frameworks are also being developed and studied~\cite{gong2023approxtrain, de2020proxsim, taheri2023deepaxe}. 
Finally, the use of approximate multipliers provides resource saving and increased throughput in FPGA-based DNN accelerators \cite{lentaris_icecs}, 
while the same benefits are also delivered in classic DSP kernels \cite{leon_qam}. 

Another set of techniques that includes works such as BiScaled-DNNs~\cite{jain2019biscaled}, CoNLoCNN~\cite{hanif2022conlocnn}, Compensated-DNNs~\cite{jain2018compensated} and ANT~\cite{guo2022ant}, targets novel data representation formats to significantly reduce the hardware complexity and energy requirements of DNN inference systems. BiScaled-DNNs~\cite{jain2019biscaled} argues that DNN data structures are long-tailed and, therefore, they can be effectively quantized using two different scale factors (namely, scale-fine and scale-wide). Scale-fine enables the system to precisely capture small numbers using more fractional bits, while scale-wide allows the system to cover the entire range of large numbers but at a coarser granularity. CoNLoCNN~\cite{hanif2022conlocnn} employs a modified version of power-of-two quantization to replace multiplication with shift operations. Compensated-DNNs~\cite{jain2018compensated} propose the concept of dynamic compensation to mitigate the impact of quantization errors caused by aggressive quantization. 
On the other hand, ANT~\cite{guo2022ant} argues that most of the existing quantization techniques use fixed-point or floating-point data representation formats, which offer limited benefits, as both are effective for specific data distributions, and thus, require a reasonably large number of bits to maintain the DNN accuracy. 
Therefore, \cite{guo2022ant} proposes a novel adaptive numerical data type that combines the benefits of float and integer data formats and, thereby, it allows the system to adapt based on the importance of each value in a tensor. 

Apart from functional approximation of arithmetic modules in DNN accelerators, voltage-scaling can also be employed to improve the energy efficiency of DNNs. ThunderVolt~\cite{2018_Zhang_DAC} presents TE-Drop, a timing error recovery technique for DNN accelerators. It employs Razor flip-flops to detect timing errors in the MAC units and then avoids re-execution penalty by dropping the MAC operation subsequent to the erroneous MAC. 
\cite{shin2019sensitivity} presents a similar approach but employs Razor flip-flops in between the multiplier and the adder of each MAC unit. 
As dropping MAC operations can result in significant performance degradation, the authors in~\cite{ji2020error} present a Compensation MAC (CMAC) unit that compensates for the dropped multiplication using an additional adder in the MAC subsequent to the erroneous MAC. 
Along similar lines, GreenTPU~\cite{2019_Pandey_DAC} presents a Timing Error Control Unit (TECU) to keep track of the timing error causing input patterns and boost the operating voltage of the subsequent MACs in the processing array. 
Apart from exploiting voltage scaling for improving energy efficiency of computational units, works such as MATIC~\cite{kim2018matic} explore the potential of aggressive voltage scaling of on-chip weight memories  
to improve the energy efficiency of DNN accelerators. 
Similarly, EDEN~\cite{koppula2019eden} presents a framework for using approximate DRAM with reduced voltage and reduced latency to improve the energy and performance efficiency of DNN inference systems. 
These techniques mainly employ memory adaptive training to achieve ultra-high efficiency gains and without any significant loss in the DNN accuracy. 

\subsection{Cross-Layer and End-to-End Approximations}
The adoption of approximations in real-world systems has been mostly  limited to either the software or hardware level and to only one subsystem (usually the computational one). To uncover the true potential of Approximate Computing, a full-system approach is essential where approximations are applied at multiple abstraction levels and to multiple subsystems
(e.g., the memory, sensor, and communication subsystems) 
together with the computational subsystem. Various studies have shown the potential of systematic joint approximation of subsystems to achieve ultra-high efficiency gains. 
Earlier works highlight the potential of combining software- and hardware-level approximations. For example, in~\cite{panda2016cross}, the authors highlight different algorithmic and circuit-level approximation techniques for significantly reducing the energy consumption of DNN inference systems. 
For algorithmic approximations, they mainly discuss scalable effort computing techniques~\cite{venkataramani2015scalable, panda2017energy1}, in which less complex inputs are processed using smaller networks, and more complex inputs are passed through a larger number of processing layers to generate an accurate output. For circuit/hardware-level approximations, they mainly highlight voltage scaling of on-chip SRAM and functional approximation of multipliers to reduce the energy consumption of DNNs.  

In~\cite{raha2017towards}, Raha et al. present a full-system perspective for approximating smart camera systems designed for computer vision and image processing applications. They present a systematic methodology for judiciously approximating sensor, memory, and computation subsystems. For sensor approximation, they consider pixel subsampling (or downscaling image resolution), as it not only reduces the energy consumption of the sensor, but it also reduces the entire system's energy consumption by decreasing the input size (or data) and the number of computations required for processing the input. For approximating memory, they focus on increasing the refresh interval of DRAM, as sub-optimal refresh intervals result in a significant reduction in refresh energy consumption while still offering space to store critical data in non-erroneous DRAM pages. For computation approximation, they consider loop perforation and early loop termination techniques. To systematically tune all the subsystem approximation knobs, they first find the optimal image resolution for the given quality constraint and then using gradient descent over the individual quality vs. energy curves find the optimal operating points for DRAM and computation skipping. In~\cite{raha2018approximating}, the methodology is extended to cover the communication subsystem as well. The work shows that for compute-intensive systems the methodology offers 3.5$\times$--5.5$\times$ energy benefits, while for communication-intensive systems, e.g., where the input is sent to a server for processing, the methodology offers 1.8$\times$--3.7$\times$ energy benefits at the cost of less than 1\% application-level quality loss. In~\cite{ghosh2020approximate}, Ghosh et al. design a similar methodology for DNN-based inference system and demonstrate that their methodology can lead to 1.6$\times$--1.7$\times$ energy benefits for the on-device inference scenario and 1.4$\times$--3.4$\times$ energy benefits for the on-cloud inference scenario at the cost of less than 1\% accuracy loss.
The extension of this work is presented in \cite{raha_axis_ext},
where various approximation techniques are applied:
(i) compute approximation via a quality-driven model compression framework,  
(ii) memory approximation by reducing the DRAM refresh rate and allocating the weights and data in quality bins,  
(iii) sensor approximation using image subsampling in the camera, 
and (iv) communication approximation using lossy network data compression. 
Finally, an end-to-end approach for approximating an iris scanning system is presented in \cite{hashemi2018approximate}. The authors employ approximations both at the software and hardware levels to build an accurate system using approximate pipeline. They identify eight different approximation knobs and then employ a reinforcement learning-based methodology for design space exploration on the approximate pipeline to achieve optimal trade-off between accuracy and runtime/energy consumption. Their method can lead to 48$\times$ improvement in runtime while maintaining 100\% accuracy.

The concept of distributed approximate systems is presented in \cite{raha_2024},
where a multi-device approach is adopted using synergistic approximations of edge computing systems to enable energy-efficient collaborative inference of distributed DNNs. 
This work performs significance-aware approximation schemes to prune the nodes and reduce the huge design space, 
while it also applies a 
heuristic to dynamically determine the approximation knobs for all the 
subsystems of all nodes under 
a given quality specification.

Apart from the above methods, various other techniques have also been proposed that combine DNN compression/approximation algorithms with hardware-level approximations to achieve significant efficiency gains. 
Gong et al.~\cite{gong2019ara} present a framework that combines dynamic layered CNN structure, kernel shrinking and layer-by-layer quantization at the algorithm level and uses an approximate computing-based reconfigurable architecture at the hardware level. 
Similarly, \cite{hanif2022cross} presents a cross-layer framework that combines network pruning and quantization with hardware-level functional approximations of arithmetic units. 
Along similar lines, \cite{fan2019axdnn} integrates activation pruning and voltage scaling with hardware-level functional approximation. To overcome the accuracy loss incurred due to hardware approximations, the authors propose to incorporate approximations in the training process, which results in an increase of about 5.32\% in accuracy. The authors report around 52.5\% improvement in the energy efficiency of the system on average. 

As the spectrum of AI-based products and services is expanding at a rapid pace, there is an increased demand for computing systems that can efficiently handle various types of DNN workloads. Towards this, \cite{swagath:2020} presents an AI system design, RaPID, that integrates different approximation techniques developed at different layers of the computing stack. 
At the algorithm level, it employs precision scaling and gradient compression to reduce the compute, memory, and data-transfer costs. 
At the hardware level, it employs a reconfigurable approximate AI core that supports different levels of approximations. 
Finally, it incorporates an approximation-aware compiler to map approximate algorithms on approximate hardware. 
The system is designed to support various types of DNNs and both DNN inference and training workloads.

Of particular relevance, the need for further hardware efficiency across multiple abstraction levels also exists in resource-limited scenarios, suited to low-power ML, such as flexible and printed applications~\cite{Mubarik:MICRO:2020:printedml,arm2022crosslayer}.
Printed classifiers are significantly less costly than conventional CMOS technologies, paving the way for high circuit customizations, also known as Bespoke implementations~\cite{isca17bespoke}. Recent examples introduced Approximate Computing in printed ML classifiers tailored to Bespoke architectures~\cite{arm2023codesign, arm2023crossapprox}. 
\cite{arm2023crossapprox}~proposes an automated framework that applies a model-to-circuit cross approximation and generates close-to-accuracy-optimal approximate ML circuits, allowing the deployment of complex classifiers on battery-powered devices ($<30$mW). 
On the other hand, through a hardware-friendly retraining (at algorithmic level), \cite{arm2023codesign} replaces coefficients with more area-efficient ones, while an approximate neuron (at hardware-level) boosts the area/power efficiency by discarding the least significant information among the least significant summands. 

\section{Architectural Approximation Techniques}\label{sec:arch_lvl}

Architectures that offer the flexibility to willingly trade-off among resource usage, storage and correctness in processor designs have the potential to achieve both higher performance and improved energy efficiency compared to error-free baselines.
This section elaborates how these gains can be accomplished through the adoption of \emph{approximate processors} and \emph{approximate data storage}, where certain computational and/or storage tasks can be loaded to specialized/approximate computing units, resulting in faster processing at the expense of reduced accuracy~\cite{accel:isca, axaccel:isca, naccel:micro, gpu:accel, snnap}.
Such approximations are typically implemented at a higher level of abstraction and involve reorganizing the system's functional blocks or optimizing the use of memory, rather than making modifications to the system's physical components (like in hardware approximation techniques). 

\subsection{Approximate Processors}\label{sec:approxproc}

Early research in the design of approximate processors includes a complete system and toolchain, proposed in~\cite{accel:isca}.
The design spans from circuits to a compiler, which integrates an area- and energy-efficient analog neural accelerator tailored to approximate general-purpose code.
This solution includes a compiler workflow that configures both the topology and weights of the neural network, while a customized training algorithm reduces errors derived from analog range limitations.
In a similar approach~\cite{naccel:micro}, digital Neural Processing Units (NPUs) are tightly integrated with the processor pipeline, resulting in low-power approximations for specific sections of general-purpose code.
Finally, approximate neural network processing along with high data-level parallelism has been explored for GPUs~\cite{gpu:accel} and FPGAs~\cite{snnap}, as well.

Relax~\cite{relax} is an architectural framework that comprises three key components - ISA extension, hardware support and software support.
The ISA extension allows the registration of a fault handler for a code region, enabling software applications to incorporate try/catch behavior.
The hardware support for Relax simplifies hardware design and enhances energy efficiency by relaxing reliability constraints and providing fault detection mechanisms, 
while the software support offers a language-level recovery construct in C/C++, with relax blocks marking regions susceptible to hardware faults.
Similarly, ERSA~\cite{ersa} is a system architecture that targets applications with inherent error resilience and ensures high degrees of resilience at low cost. 
It achieves high error resilience to high-order bit errors and control flow errors using asymmetric reliability (and the concept of ``configurable reliability'' for general-purpose applications) in many-core architectures, error-resilient algorithms, and intelligent software optimizations.

Prior research studies on the design of approximate processors also include the voltage over-scaling paradigm as a potential energy-saving method with adjustable approximation level~\cite{nvdla:pedram}.
The authors in~\cite{axprogramm:asplos} propose an ISA extension provided with approximate operations and storage,
as well as a microarchitecture design, called Truffle, which effectively supports these extensions.
This work utilizes dual-voltage operation to save power and allows the programmers to specify which parts of a program can be computed approximately at a lower energy cost. 
Similar manipulations in operational (timing) errors to decrease power consumption of processor architectures from voltage/reliability trade-offs, have been studied and can be found in~\cite{vos:proc1, vos:proc2}.

Among previous architectures, RISC-V has been established as an increasingly growing competitor and alternative of CPUs with funded projects and support from star companies (e.g., Intel, Microsoft, ST Microelectronics~\cite{riscv}).
The availability of open-source RISC-V tools and libraries provides developers with greater flexibility in designing and optimizing approximate computing systems.
More specifically, the authors in~\cite{riscv:bw} extend prior works in variable
bit-width approximate arithmetic operators with configurable data memory units as well, while~\cite{riscv:simil} leverages the inherent ``value similarity'' characteristic of ML workloads and by using lightweight micro-architectural ISA extensions skips entire instruction sequences. 
Accordingly, it substitutes the results with previously executed computations, improving both performance and energy efficiency.
Furthermore, \cite{riscv:control} utilizes multiple hardware-level approximations and minimizes quality degradation of several applications by adopting a hardware controller, which adjusts the degree of approximation by selecting various approximate units.
Lastly, RISC-V processors targeting DNN applications have also been explored in~\cite{riscv:mixed, riscv:lightweight, riscv:arme}.
In~\cite{riscv:mixed}, the authors exploit the efficacy of SIMD instructions and propose mixed-precision 
Quantized Neural Network (QNN) 
execution that enables functional units down to 2 bits, opposed to commercial cores (e.g., ARM Cortex M55) that cannot natively support smaller than 8-bit SIMD instructions.
On the other hand, both \cite{riscv:lightweight} and~\cite{riscv:arme} extend the base RISC-V ISA to enable low bit-width and mixed-precision operations tailored to DNN tasks, but this time by modifying the core's processing unit that can be integrated in a full RISC-V core with minimal overheads.

\subsection{Approximate Data Storage}

This survey considers the two main types 
of memories: 
the non-volatile and the volatile memories.
The works on approximate non-volatile data storage 
target 
Phase-Change Memories (PCMs) \cite{2014_Sampson_ACMtc, 2012_Fang_ATS, 2018_Teimoori_DATE}, 
flash memories \cite{2011_Salajegheh_FAST, 2013_Tseng_DAC, 2012_Liu_FAST},
and 
Spin Transfer Torque (SST) RAMs \cite{2015_Sampaio_CASES, 2015_Ranjan_DAC, 2018_Zeinali_IEEEtcasii, 2018_Teimoori_DATE}.
Regarding the volatile memories,
the review includes 
static RAMs (SRAMs) \cite{2009_kumar_SQED, 2011_chang_TVID, 2016_Frustaci_TVLSI, 2022_Reviriego_TEPC},
dynamic RAMs (DRAMs) \cite{2022_Reviriego_TEPC, 2011_Liu_ASPLOS, 2015_Jung_MEMSYS, raha_dram},
and caches \cite{2015_Joshua_MICRO, 2016_Joshua_MICRO, Li_DAC_2019}.
It is noted that the literature also includes approximation techniques that are applied and evaluated
in different types of memory \cite{2020_Ranjan_TVLSI}. 

\subsubsection{Approximate Non-Volatile Memories}

Sampson et al. \cite{2014_Sampson_ACMtc}
introduce two mechanisms to approximate PCMs,
which can be used 
for persistent storage (e.g., for databases or filesystems) 
or as main memory. 
The first mechanism
reduces the number of programming pulses 
for writing the multi-level cells 
(i.e., cells storing multiple bits of information),
while the second one
stores approximate data in blocks with exhausted error correction resources.
SoftPCM  \cite{2012_Fang_ATS} integrates a method to reduce the write traffic
towards improved energy efficiency and lifetime.
More specifically,
a circuit compares 
the new data with stored data,
and the writing operation is not performed
if the difference is below a threshold.  

The authors of \cite{2011_Salajegheh_FAST} provide approximate storage in embedded flash memories
by lowering the voltage below its nominal value. 
To deal with unpredictable behaviors and large error rates,
they employ three software-based methods, 
i.e., ``in-place writes'', ``multiple-place writes'', and ``RS-Berger codes''.
The first method repeatedly writes the data to the same memory address,
while the second one writes the data in more than one addresses.
The last method is an error detection \& correction code that recovers the originally stored data.
In the same context, 
Tseng et al. \cite{2013_Tseng_DAC} 
characterize various multi-level cell flash memories 
when the supply voltage drops during read, program and erase operations. 
Based on the characterization results,
they propose a dynamic scaling mechanism 
that adjusts the voltage supply of flash
according to the executed operation.
Retention relaxation \cite{2012_Liu_FAST}
is another interesting technique applied in 
multi-level cell flash memories.
In particular, 
the write operation for data with short lifetime
(e.g., data from proxy and MapReduce workloads) 
is accelerated based on modeling 
the relationship between raw bit error rates and retention time.

The literature also includes works on approximate SST RAMs,
which are non-volatile alternative of SRAMs
for on-chip scratchpad or cache memory. 
Sampaio et al. \cite{2015_Sampaio_CASES}
propose an approximate STT-RAM cache architecture that 
reduces the reliability overhead
under a given error bound.
The application's critical data are protected using a latency-aware double error correction module,
while approximation-aware read and write policies provide approximate storage. 
Moreover,
the architecture is equipped with a unit for monitoring the quality of the application's output.
To provide an approximate scratchpad memory, 
Ranjan et al. \cite{2015_Ranjan_DAC}
design a quality configurable SST-RAM,
in which read and write operations
are executed at varying accuracy levels. 
Approximate reads are performed 
by either 
lowering the current used to sense the stored value
or lowering the sensing duration and increasing the current.    
Similarly,
approximate writes are performed by lowering 
the current and/or sensing duration. 
Furthermore, 
the authors of 
\cite{2018_Zeinali_IEEEtcasii}
propose a 
scheme for
progressively scaling the cells of
STT-RAM arrays
based on a scaling factor.
In this approach, the most significant bits
of the data words
are implemented to provide lower bit error rates
from the least-significant ones.
In \cite{2018_Teimoori_DATE}, 
the authors propose 
AdAM,
which provides 
adaptive approximation management across the entire non-volatile memory hierarchy.
In more detail, 
they model a system with 
STT-RAM scratchpad and PCM main memory 
and apply various 
approximation knobs (e.g., read/write pulse magnitude/duration) 
to exchange data accuracy for improved 
STT-RAM access delay and PCM lifetime.

\subsubsection{Approximate Volatile Memories}

Regarding SRAM memories,
significant research efforts have focused on lowering the supply voltage. 
In this context, 
Kumar et al. \cite{2009_kumar_SQED} apply system-level design techniques to reduce the SRAM leakage power.
Their framework is constrained by a data-reliability factor,
while they use a statistical/probabilistic setup to model soft errors and process variations.
In \cite{2011_chang_TVID},
the authors 
propose a hybrid memory array for video processors.
Their policy is to store
the high-order bits of luminance pixels,
i.e., those that are more sensitive to human vision, 
in robust 8T SRAM bit-cells,
while the low-order bits are stored in 6T bit-cells.
This approach allows to lower the supply voltage without significantly affecting the high-order bits.
Frustaci et al. \cite{2016_Frustaci_TVLSI}
evaluate a wide range of approximation techniques 
(bit dropping, selective write assist, selective error correcting code, voltage scaling)
for dynamic management of the energy--quality trade-off in SRAMs.
Their analysis is based on measurements on a 28-nm SRAM testchip,
and it also includes combinations of the examined techniques and different array/word sizes.

The authors of \cite{2022_Reviriego_TEPC}
use approximate SRAM and DRAM memories
to implement sketches
for similarity estimation. 
In more detail,
based on theoretical analysis and simulation,
they examine the error rates
in unprotected and parity-protected memories
and tune their voltage supply. 
Moreover, 
the work in \cite{2011_Liu_ASPLOS}
applies lower refresh rates in the  
DRAM portions storing the non-critical data. 
Similarly, 
Jung et al. \cite{2015_Jung_MEMSYS}
define reliable and unreliable memory regions,
however, 
in their strategy they completely disable 
the DRAM's auto-refresh feature. 
The authors in \cite{raha_dram} propose a quality-configurable approximate DRAM system and evaluate it along with an FPGA. 
Their design leverages characterized memory errors to partition the DRAM into a number of quality bins based on the frequency, location, and nature of bit errors in DRAM pages,   
while also taking into account the variable retention time of the DRAM cells. 
The data placement in the bins is performed with respect to data criticality, with the refresh rate serving as the quality control knob.  

Besides SRAMs and DRAMs, 
the literature also includes research works on approximate caches.
In \cite{2015_Joshua_MICRO}, 
the authors propose the Doppelganger cache
based on the approximate similarity among the cache values.  
In particular, 
they 
associate the tags of multiple similar blocks with a single data array entry,
and hence, 
they reduce the cache memory footprint. 
In the same context,
the authors of \cite{2016_Joshua_MICRO} 
propose the Bunker cache
based on the spatio-value similarity, 
where approximately similar data 
exhibit spatial regularity in memory.
Their approach is to 
map similar data to the same cache location 
with respect only to their memory address. 
Finally, 
ASCache \cite{Li_DAC_2019} is an approximate SSD cache
that allows 
bit errors
within a tolerable threshold, 
while avoiding 
unnecessary cache misses and guaranteeing end-to-end data integrity. 

\section{Comparative Quantitative Analysis of Approximation Techniques}\label{sec:comp}
This section reports notable numerical results and outcomes for the application-specific and architectural approximation techniques.

\subsection{Application-Specific Approximation Techniques}

Table~\ref{table:as_quant_sw} and Table~\ref{table:as_quant_hw} present a quantitative analysis of some of the most prominent software-level and hardware-level techniques covered in Section~\ref{sec:app_spec}. This comparison considers only the methods evaluated on the ImageNet (or a relatively complex) dataset. From each sub-category at the software and hardware level (in Table~\ref{tb_appspec}), one of the most prominent or state-of-the-art techniques is selected. The goal is to highlight the efficiency and accuracy, providing a comparison of the strengths and trade-offs across sub-categories and levels. For instance, the tables show that structured pruning outperforms in model size reduction, while conditional deep learning and adaptive numerical data types provide superior energy efficiency. 

It is important to note that each technique has unique advantages, and often, methods from different sub-categories and levels can be combined to achieve higher overall benefits with better accuracy--efficiency trade-offs. In this context, \cite{ghosh2020approximate} demonstrates that applying various approximations across different modules of a deep learning inference system (based on ResNet-101 for ImageNet classification) can achieve approximately 70\% energy reduction with only a 0.77\% absolute accuracy loss. 

\begin{table*}[!t]
\renewcommand{\arraystretch}{1.05}
\centering
\footnotesize
\setlength{\tabcolsep}{1.0pt}
\caption{Quantitative analysis of software-level approximation techniques for deep learning workloads.}
\vspace{-8pt}
\label{table:as_quant_sw}
\resizebox{1\linewidth}{!}{
\begin{tabular}{cccccc} 
\hline
\begin{tabular}[c]{@{}l@{}}\textbf{Approximation Class}\end{tabular} & \makecell[c]{\textbf{Workload}} & \textbf{Metrics/Improvements} & \textbf{Accuracy Gain/Drop} & \begin{tabular}[c]{@{}c@{}}\textbf{Dependency on} \\ \textbf{Other Layers}\end{tabular} & \textbf{Ref.}                    \\ 
\hline\hline
\begin{tabular}[c]{@{}c@{}}Structured Pruning\end{tabular} & \begin{tabular}[c]{@{}c@{}}ResNet50 on \\ ImageNet dataset\end{tabular} & \begin{tabular}[c]{@{}c@{}}63.5\% reduction in FLOPs, \\ 64.5\% reduction in no. of \\ parameters\end{tabular} & \begin{tabular}[c]{@{}c@{}}1\% drop in top-1 \\ and 0.4\% in top-5 \\ accuracy\end{tabular} & No & \cite{chen2021only} \\ 

\hline
\begin{tabular}[c]{@{}c@{}}Unstructured Pruning\end{tabular} & \begin{tabular}[c]{@{}c@{}}VGG16 on \\ ImageNet dataset\end{tabular} & \begin{tabular}[c]{@{}c@{}}79\% reduction in FLOPs, \\ 92.5\% reduction in no. of \\ parameters, 97.95\% reduction in \\ model size (with compression) \end{tabular} & \begin{tabular}[c]{@{}c@{}}0.33\% increase in top-1 \\ and 0.41\% in top-5 \\ accuracy\end{tabular} & \begin{tabular}[c]{@{}c@{}}Improves inference \\ latency and energy \\ efficiency when \\ hardware-aware\end{tabular} & \cite{han2015deep} \\
\hline
\begin{tabular}[c]{@{}c@{}}Binarization\end{tabular} & \begin{tabular}[c]{@{}c@{}}ResNet-18 on \\ ImageNet dataset\end{tabular} & \begin{tabular}[c]{@{}c@{}}32$\times$ model size reduction, \\ 58$\times$ speed up in convolution \\ (excluding memory allocation \\ and access operations) \end{tabular} & \begin{tabular}[c]{@{}c@{}}18.4\% drop in top-1 \\ and 16\% in top-5 \\ accuracy\end{tabular} & \begin{tabular}[c]{@{}c@{}}Existing hardware \\ supports required \\ binary operations\end{tabular} & \cite{rastegari2016xnor} \\
\hline
\begin{tabular}[c]{@{}c@{}}Mixed Precision\end{tabular} & \begin{tabular}[c]{@{}c@{}}MobileNet-V1/V2 on \\ ImageNet dataset\end{tabular} & \begin{tabular}[c]{@{}c@{}}1.4$\times$ to 1.95$\times$ \\ latency reduction \end{tabular} & \begin{tabular}[c]{@{}c@{}}Less than 1\% \\ accuracy loss\end{tabular} & \begin{tabular}[c]{@{}c@{}}Yes, requires \\ specialized hardware\end{tabular} & \cite{wang2019haq} \\
\hline
\begin{tabular}[c]{@{}c@{}}Conditional Deep \\ Learning Networks\end{tabular} & \begin{tabular}[c]{@{}c@{}}ResNet-50 on \\ Tiny ImageNet\end{tabular} & \begin{tabular}[c]{@{}c@{}}6.8$\times$ reduction in no. of operations, \\ 4.96$\times$ reduction in energy \end{tabular} & \begin{tabular}[c]{@{}c@{}}1.12\% accuracy \\ improvement\end{tabular} & \begin{tabular}[c]{@{}c@{}}No\end{tabular} & \cite{panda2017energy1} \\
\hline
\hline
\end{tabular}}
\end{table*}

\begin{table*}[!t]
\renewcommand{\arraystretch}{1.05}
\centering
\footnotesize
\setlength{\tabcolsep}{2.1pt}
\caption{Quantitative analysis of hardware-level approximation techniques for deep learning workloads.}
\vspace{-8pt}
\label{table:as_quant_hw}
\begin{tabular}{cccccc} 
\hline
\begin{tabular}[c]{@{}l@{}}\textbf{Approximation Class}\end{tabular} & \makecell[c]{\textbf{Workload}} & \textbf{Metrics/Improvements} & \textbf{Accuracy Gain/Drop} & \begin{tabular}[c]{@{}c@{}}\textbf{Dependency on} \\ \textbf{Other Layers}\end{tabular} & \textbf{Ref.}                    \\ 
\hline\hline
\begin{tabular}[c]{@{}c@{}}Low-Precision DNNs \\ with Error \\ Compensation\end{tabular} & \begin{tabular}[c]{@{}c@{}}ResNet50 on \\ ImageNet dataset\end{tabular} & \begin{tabular}[c]{@{}c@{}} $\approx$10\% energy reduction \\ compared to 8-bit \\fixed-point\end{tabular} & \begin{tabular}[c]{@{}c@{}}0.16\% increase in \\ accuracy compared to \\ 8-bit FxP\end{tabular} & No & \cite{jain2018compensated} \\ 
\hline
\begin{tabular}[c]{@{}c@{}}Adaptive Numerical \\ Data Type\end{tabular} & \begin{tabular}[c]{@{}c@{}}ResNet on \\ ImageNet dataset\end{tabular} & \begin{tabular}[c]{@{}c@{}}On average 2.8$\times$ speedup \\ and 2.5$\times$ energy efficiency \\ improvement over other \\ quantization accelerators\end{tabular} & \begin{tabular}[c]{@{}c@{}}less than 1\% \\ drop in accuracy\end{tabular} & \begin{tabular}[c]{@{}c@{}}No runtime \\ dependencies\end{tabular} & \cite{guo2022ant} \\
\hline
\begin{tabular}[c]{@{}c@{}}Voltage Scaling in \\ DNN Accelerators\end{tabular} & \begin{tabular}[c]{@{}c@{}}ResNet-18 on \\ ImageNet dataset\end{tabular} & \begin{tabular}[c]{@{}c@{}}47\% power \\ saving \end{tabular} & \begin{tabular}[c]{@{}c@{}}Less than 2\% \\ accuracy loss\end{tabular} & \begin{tabular}[c]{@{}c@{}}No\end{tabular} & \cite{ji2020error} \\
\hline
\begin{tabular}[c]{@{}c@{}}Inference using \\ Approximate DRAM\end{tabular} & \begin{tabular}[c]{@{}c@{}}VGG16 on \\ ImageNet dataset\end{tabular} & \begin{tabular}[c]{@{}c@{}} 29\% (on average) DRAM \\ energy reduction\end{tabular} & \begin{tabular}[c]{@{}c@{}}less than 1\% \\ accuracy drop\end{tabular} & \begin{tabular}[c]{@{}c@{}}No runtime \\ dependencies\end{tabular} & \cite{koppula2019eden} \\
\hline
\hline
\end{tabular}
\end{table*}

\subsection{Architectural Approximation Techniques}

\begin{table}[]
\renewcommand{\arraystretch}{1.05}
\setlength{\tabcolsep}{6.2pt}
\centering
\footnotesize
\caption{Quantitative analysis of remarkable approximation techniques for processors.}
\vspace{-8pt}
\begin{tabular}{cccccc}
\hline
\textbf{Platform} &
  \textbf{Key Technique} &
  \textbf{Precision} &
  \textbf{Resource Gains} &
  \textbf{Quality Loss} &
  \textbf{Ref.} \\ \hline \hline
GPU &
  \begin{tabular}[c]{@{}c@{}}scalable units with \\ quality control\end{tabular} &
  32-bit &
  2.1$\times$ energy reduction & 2.5\% quality loss &
  \cite{gpu:accel} \\ \hline
FPGA &
  \begin{tabular}[c]{@{}c@{}}configurable network's\\ topology and weights\end{tabular} &
  16-bit &
  2.8$\times$--28$\times$ energy savings & <10\% quality loss &
  \cite{snnap} \\ \hline
FPGA &
    \begin{tabular}[c]{@{}c@{}}RISC-V with posit \\ processing unit\end{tabular} &
  8-/16-bit &
  10$\times$ speedup & <1\% accuracy loss &
  \cite{riscv:lightweight} \\ \hline
ASIC 7nm &
  \begin{tabular}[c]{@{}c@{}}mixed-precision \&\\ multi-pumped units\end{tabular} &
  2-/4-/8-bit &
  415--1470 GOPs/W & <1\% accuracy loss &
  \cite{riscv:arme} \\ \hline
ASIC 22nm &
  \begin{tabular}[c]{@{}c@{}}mixed-precision QNN \\ with SIMD\end{tabular} &
  2-/4-/8-bit &
  200--600 GOPs/W & <1\% accuracy loss &
  \cite{riscv:mixed} \\ \hline\hline
\end{tabular}
\label{tab:approxproc}
\end{table}

Table~\ref{tab:approxproc} summarizes the key state-of-the-art works in the domain of approximate processors. 
In GPUs, approximate processing emphasizes scalability across multiple cores, integrating quality control mechanisms that reduce accuracy loss to just 2.5\% while achieving over 2$\times$ energy efficiency~\cite{gpu:accel}. 
For FPGAs, the reconfigurability enables substantial gains, delivering up to 28$\times$ speedup with medium-precision operations. 
While both FPGA and RISC-V cores offer significant advantages~\cite{snnap}, ASIC-based designs demonstrate the most potential. 
Mixed-precision cores, capable of operating with as low as 2-bit precision, achieve performance levels between 600~\cite{riscv:mixed} and 1470~\cite{riscv:arme} GOPs/W, with minimal accuracy loss.
Especially when prioritizing efficiency over accuracy (e.g., allowing up to 5\% degradation), RISC-V processors can achieve peak performance of up to 1.9 TOPs/W.
In general, ASICs provide the optimal balance between energy efficiency and quality preservation, particularly in mixed-precision applications.
Table \ref{tb_mem_comp} reports the corresponding analysis for approximate data storage with respect to the memory type.
Overall, the state-of-the-art techniques mainly involve voltage scaling, approximate memory cells, and storage bins of different quality. Significant resource gains are achieved (e.g., up to 73\% in DRAM power \cite{raha_dram}), however, some techniques insert overheads (e.g., \cite{2013_Tseng_DAC} in performance). 

\begin{table*}[!t]
\renewcommand{\arraystretch}{1.05}
\centering
\footnotesize
\caption{Quantitative analysis of remarkable memory approximation techniques.}
\vspace{-8pt}
\begin{tabular}{>{\centering\arraybackslash}m{1.2cm} >{\centering\arraybackslash}m{3.3cm} >{\centering\arraybackslash}m{3.3cm} >
{\centering\arraybackslash}m{3.3cm} >{\centering\arraybackslash}m{0.6cm} }
\hline  
\makecell[c]{\textbf{Memory}} & 
\makecell[c]{\textbf{Main Techniques}} & 
\makecell[c]{\textbf{Resource Gains}} &
\makecell[c]{\textbf{Loss / Overhead}} &
\makecell[c]{\textbf{Ref.}}
\\
\hline
\hline
PCM & approx. multi-level cells, failed memory cells & 1.7$\times$ speedup, +23\% array lifetime & <10\% quality loss & \cite{2014_Sampson_ACMtc} \\ \hline
Flash & power fading, dynamic voltage scaling & 45\% energy & performance drop: \hspace{25pt} avg=0.1\%, 
worst=2\%--19\% & \cite{2013_Tseng_DAC} \\ \hline
SST RAM & progressive cell scaling & 37\% power, 25\% area & BER=10$^{-6}$ & \cite{2018_Zeinali_IEEEtcasii} \\ \hline
SRAM & bit dropping, voltage scaling, sel. write assist, sel. ECC & >2$\times$ energy & 1\% area & \cite{2016_Frustaci_TVLSI} \\ \hline
DRAM & quality bins construction & up to 73\% power & $\sim$0\% quality loss  & \cite{raha_dram} \\ \hline
Cache & spatio-value similarity & up to 1.2$\times$ speedup, \hspace{25pt} up to 1.4$\times$ energy & small quality loss & \cite{2016_Joshua_MICRO} \\ 
\hline\hline
\end{tabular}
\label{tb_mem_comp}
\end{table*}

\section{Application Domains of Approximate Computing}\label{sec:apps}

This section introduces the error-tolerant application domains 
that 
have been systematically examined for Approximate Computing.
Table \ref{tb_app} classifies per domain the works presented in Part I \cite{mysurvey_pt1} and Part II of the survey, while approximation opportunities and representative techniques for each domain are discussed. Furthermore, the case studies of each domain are analyzed, and widely used benchmark suites and error / Quality-of-Service (QoS) metrics are reported.   

\subsection{The Application Spectrum of Approximate Computing}

\begin{table*}[!t]
\renewcommand{\arraystretch}{1.05}
\centering
\footnotesize
\setlength{\tabcolsep}{5pt}
\caption{Classification of state-of-the-art Approximate Computing works per
application domain.}
\vspace{-8pt}
\begin{tabular}{c m{10cm}}
\hline  
\textbf{Application Domain} & \makecell[c]{\textbf{\textbf{References}}}\\
\hline
\hline
Image Processing &  \cite{2020_Adams_IEEEtc, 2017_Akbari_IEEEtvlsi, 2018_Akbari_IEEEtcasii, 2005_Alvarez_IEEEtc, 2015_Boston_OOPSLA, 2020_Castro_ICCAD, 2015_Chen_GLSVLSI, 2016_Chen_IEEEtc, 2018_Chen_IEEEtmscs, 2020_Ebrahimi_IEEEtcasii, 2018_Esposito_IEEEtcasi, 2020_Frustaci_IEEEtcasii, 2013_Gupta_IEEEtcad, 2015_Hashemi_ICCAD, 2016_Hashemi_DAC, 2009_Hoffmann_MIT, 2019_Imani_Date, 2019_Jiang_IEEEtc, 2020_Jiao_IEEEtcad, 2012_Kahng_DAC, 2010_Kurdahi_IEEEtvlsi, 2017_Lee_DATE, 2021_Leon_ACMtecs, 2018_Leon_IEEEtvlsi, 2018_Leon_IEEEmicro, 2018_Li_ICS, 2011_Liu_ASPLOS, 2018_Liu_ARITH, 2017_Liu_IEEEtc, 2018_Liu_IEEEtcasi, 2019_Liu_ISCA, 2014_Miguel_MICRO, 2014_Misailovic_OOPSLA, 2010_Misailovic_ICSE, 2014_Mishra_WACAS, 2015_Momeni_IEEEtc, 2015_Park_FSE, 2014_Park_GIT, 2018_Pashaeifar_IEEEtvlsi, 2021_Pilipovic_IEEEtcasi, 2015_Raha_DATE, 2013_Rahimi_IEEEtcasii, 2014_Ranjan_DATE, 2020_Saadat_DATE, 2019_Saadat_DAC, 2019_Sabetzadeh_IEEEtcasi, 2014_Samadi_ASPLOS, 2013_Samadi_MICRO, 2015_Sampson_UOW, 2017_Schlachter_IEEEtvlsi, 2013_Shi_FCCM, 2014_Shi_DAC, 2011_Sidiroglou_FCE, 2020_Strollo_IEEEtcasi, 2015_Tan_ASP-DAC, 2018_Tziantzioulis_IEEEmicro, 2019_Vahdat_IEEEtvlsi, 2017_Vahdat_DATE, 2015_Vassiliadis_PPoPP, 2016_Vassiliadis_CGO, 2019_Venkatachalam_IEEEtc, 2020_Waris_IEEEtcasii, 2015_Yazdanbakhsh_DATE, 2013_Ye_ICCAD, 2017_Zendegani_IEEEtvlsi, 2016_Zendegani_DATE, 2019_Zervakis_IEEEtcasii, 2018_Ullah_DAC, 2019_Chen_IEEEtc, 2015_Vasicek_IEEEtec, 2019_Fernando_ACMpapl, 2019_Joshi_ICSE, 2017_Chen_IEEEtscl, 2017_Mitra_CGO, 2012_Fang_ATS, 2015_Sampaio_CASES, 2011_chang_TVID, naccel:micro, gpu:accel, snnap, relax, axprogramm:asplos, riscv:bw, riscv:control, 2015_Joshua_MICRO, 2016_Joshua_MICRO, Li_DAC_2019, 2021_Waris_IEEEtcasii, 2022_Zhu_IEEEtcasii, tevot, devot, date21, qlut_2017, raha_dram} \\
\hline
Signal Processing & \cite{2005_Alvarez_IEEEtc, 2010_Baek_PLDI, 2020_Baharvand_IEEEtetc, 2015_Boston_OOPSLA, 2013_Chen_IEEEtvlsi, 2014_Chiang_PPoPP, 2018_Esposito_IEEEtcasi, 2018_Guo_ISSTA, 2013_Gupta_IEEEtcad, 2018_Hashemi_DAC, 2016_Jiang_IEEEtc, 2020_Kang_CGO, 2019_Karakoy_ACMmacs, 2017_Lee_DATE, 2018_Leon_IEEEtvlsi, 2018_Leon_IEEEmicro, 2010_Linderman_CGO, 2017_Liu_ICCAD, 2010_Liu_IEEEtvlsi, 2019_Liu_ISCA, 2020_Manikanta_IEEEtvlsi, 2014_Nepal_DATE, 2015_Park_FSE, 2018_Pashaeifar_IEEEtvlsi, 2016_Ragavan_ISVLSI, 2013_Rahimi_IEEEtcasii, 2013_Ramasubramanian_DAC, 2014_Ranjan_DATE, 2016_Rubio_ICSE, 2013_Rubio_SC, 2014_Samadi_ASPLOS, 2011_Sampson_PLDI, 2013_Shi_FCCM, 2020_Stitt_ACMtecs, 2019_Venkatachalam_IEEEtc, 2012_Venkataramani_DAC, 2015_Yazdanbakhsh_DATE, 2015_Yazdanbakhsh_DATE, 2018_Yesil_IEEEmicro, 2018_Zervakis_IEEEtvlsi, 2019_Zervakis_IEEEtcasii, 2014_Sampson_ACMtc, 2013_Venkataramani_DATE, 2019_Nepal_IEEEtetc, naccel:micro, snnap, axprogramm:asplos, vos:proc2, leon_qam} \\
\hline
Computer Vision &  \cite{2015_Achour_OOPSLA, 2020_Baharvand_IEEEtetc, 2009_Hoffmann_MIT, 2019_Karakoy_ACMmacs, 2018_Li_ICS, 2014_Miguel_MICRO, 2010_Misailovic_ICSE, 2014_Mishra_WACAS, 2011_Mohapatra_DATE,  2014_Nepal_DATE, 2014_Schkufza_PLDI, 2011_Sidiroglou_FCE, 2007_Sorber_SenSys, 2015_Tan_ASP-DAC, 2017_Mitra_CGO, relax} \\
\hline
Computer Graphics &  \cite{2010_Baek_PLDI, 2020_Baharvand_IEEEtetc, 2015_Campanoni_CGO, 2015_Keramidas_WAPCO, 2011_Liu_ASPLOS, 2019_Liu_ISCA, 2013_Misailovic_ACMtecs, 2015_Park_FSE, 2014_Park_GIT, 2011_Sampson_PLDI, 2014_Sampson_ACMtc, relax, axprogramm:asplos} \\
\hline 
Big Data Analysis &  \cite{2013_Agarwal_EuroSys, 2016_Anderson_ICDE, 2015_Goiri_ASPLOS, 2019_Hu_MASCOTS, 2016_Kandula_MOD, 2016_Krishnan_WWW, 2012_Laptev_VLDB, 2019_Park_MOD, 2017_Quoc_ATC, 2017_Quoc_MIDDL, 2018_Wen_ICDCS, 2016_Zhang_VLDB, 2019_Joshi_ICSE, 2013_Tseng_DAC, 2012_Liu_FAST} \\
\hline
Databases &  \cite{2013_Agarwal_EuroSys, 2010_Baek_PLDI, 2010_Byna_GPGPU, 2016_Kandula_MOD, 2011_Sidiroglou_FCE, 2016_Zhang_VLDB, 2013_Tseng_DAC} \\
\hline
Data Mining &  \cite{2015_Achour_OOPSLA, 2011_Ansel_CGO, 2020_Baharvand_IEEEtetc, 2017_Brumar_IPDPS, 2010_Byna_GPGPU, 2009_Hoffmann_MIT, 2018_Kislal_ELSEclss, 2021_Leon_ACMtecs, 2018_Li_ICS, 2018_Liu_IEEEtcasi, 2019_Liu_ISCA, 2009_Meng_IPDPS, 2010_Mengt_IPDPS, 2011_Mohapatra_DATE, 2015_Raha_DATE, 2014_Ranjan_DATE, 2012_Renganarayana_RACES, 2013_Samadi_MICRO, 2015_Sampson_UOW, 2015_Shi_IEEEcal, 2011_Sidiroglou_FCE, 2010_Sreeram_IISWC, 2015_Tan_ASP-DAC, 2015_Tian_GLSVLSI, 2015_Yazdanbakhsh_DATE, 2010_Chakradhar_DAC, 2014_Chippa_IEEEtvlsi, 2015_Ranjan_DAC, 2022_Reviriego_TEPC, accel:isca, naccel:micro, snnap, relax, ersa, vos:proc1, 2015_Joshua_MICRO, 2016_Joshua_MICRO, 2020_Ranjan_TVLSI, qlut_2017, raha_dram} \\
\hline 
Machine Learning &  \cite{arm2023codesign, arm2023crossapprox, 2010_Baek_PLDI, 2014_Bornholt_ASPLOS, 2015_Hashemi_ICCAD, 2018_Kanduri_DAC, 2013_Kim_ICCAD, 2018_Kislal_ELSEclss, 2014_Mishra_WACAS, 2011_Mohapatra_DATE, 2014_Nepal_DATE, 2019_Park_MOD, 2015_Raha_DATE, 2014_Ranjan_DATE, 2014_Samadi_ASPLOS, 2013_Samadi_MICRO, 2016_Tolpin_IFL, 2015_Yazdanbakhsh_DATE, 2015_Zhang_DATE, 2019_Nepal_IEEEtetc, 2014_Chippa_IEEEtvlsi, 2015_Ranjan_DAC, accel:isca, naccel:micro, snnap, ersa, vos:proc1, riscv:simil, 2015_Joshua_MICRO, 2016_Joshua_MICRO, 2020_Ranjan_TVLSI, qlut_2017, raha_dram} \\
\hline 
Artificial Neural Networks & \cite{2018_Akhlaghi_ISCA, 2021_Ansari_IEEEtc, 2015_Campanoni_CGO, 2021_Leon_ACMtecs, 2017_Lin_ISCAS, 2019_Pandey_DAC, 2021_Pilipovic_IEEEtcasi, 2019_Saadat_DAC, 2019_Vahdat_IEEEtvlsi, 2020_Wang_IEEEtc, 2017_Wang_IEEEtvlsi, 2015_Yazdanbakhsh_DATE, 2018_Zhang_DAC, 2015_Zhang_DATE, 2019_Vasicek_DATE, 2015_Ranjan_DAC, ersa, nvdla:pedram, vos:proc1, riscv:lightweight, riscv:mixed, 2020_Ranjan_TVLSI, han2015learning, han2015deep, yu2017scalpel, li2016pruning, yvinec2021red, yvinec2022red++, he2017channel, hu2016network, tan2020dropnet, sui2021chip, he2022filter, lebedev2016fast, wen2016learning, wang2020neural, chen2021only, rastegari2016xnor, hubara2016binarized, zhou2016dorefa, jain2019biscaled, han2015deep, wang2019haq, ganapathy2020dyvedeep, panda2017energy1, panda2017energy2, tann2016runtime, venkataramani2015sapphire, venkataramani2015scalable, lou2021dynamic, plastiras2018efficient, tan2019efficientnet, ghosh2020approximate, venkataramani2014axnn, mrazek2016design, mrazek2019alwann, mrazek2020libraries, sonnino2023daism, spantidi2021positive, saadat2018minimally, kim2018efficient, hanif2019cann, jain2019biscaled, jain2018compensated, hanif2022conlocnn, guo2022ant, shin2019sensitivity, ji2020error, kim2018matic, koppula2019eden, koppula2019eden, kim2018matic, raha2017towards, raha2018approximating, ghosh2020approximate, hashemi2018approximate, gong2019ara, hanif2022cross, fan2019axdnn, swagath:2020, leon_lascas, lentaris_icecs, qlut_2017, raha_dram, raha_axis_ext, raha_2024, riscv:arme} \\
\hline
Scientific Computing & \cite{2011_Ansel_CGO, 2015_Boston_OOPSLA, 2017_Brumar_IPDPS, 2013_Carbin_OOPSLA, 2017_Chiang_POPL, 2014_Chiang_PPoPP, 2017_Darulova_ACMtoplas, 2018_Guo_ISSTA, 2019_Jiang_IEEEtc, 2020_Kang_CGO, 2019_Karakoy_ACMmacs, 2019_Laguna_HPC, 2018_Lam_SAGE, 2013_Lam_ICS, 2021_Leon_ACMtecs, 2018_Menon_SC, 2014_Misailovic_OOPSLA, 2013_Misailovic_ACMtecs, 2015_Park_FSE, 2014_Park_GIT, 2013_Rahimi_IEEEtcasii, 2006_Rinard_ICS, 2016_Rubio_ICSE, 2013_Rubio_SC, 2011_Sampson_PLDI, 2014_Schkufza_PLDI, 2018_Yesil_IEEEmicro, 2014_Sampson_ACMtc, 2019_Fernando_ACMpapl, 2019_Joshi_ICSE, gpu:accel, axprogramm:asplos, vos:proc1, 2015_Joshua_MICRO} \\
\hline
Financial Analysis &  \cite{2015_Achour_OOPSLA, 2010_Baek_PLDI, 2020_Baharvand_IEEEtetc, 2017_Brumar_IPDPS, 2009_Hoffmann_MIT, 2018_Li_ICS, 2010_Linderman_CGO, 2019_Liu_ISCA, 2014_Miguel_MICRO, 2014_Misailovic_OOPSLA, 2014_Samadi_ASPLOS, 2015_Sampson_UOW, 2015_Shi_IEEEcal, 2011_Sidiroglou_FCE, 2018_Tziantzioulis_IEEEmicro, 2016_Vassiliadis_CGO, 2018_Yesil_IEEEmicro, 2018_Zhang_IEEEcal, 2019_Fernando_ACMpapl, 2019_Joshi_ICSE, gpu:accel, snnap, 2015_Joshua_MICRO} \\
\hline
Physics Simulations & \cite{2015_Achour_OOPSLA, 2011_Ansel_CGO, 2020_Brunie_SC, 2015_Campanoni_CGO, 2018_Li_ICS, 2019_Liu_ISCA, 2018_Menon_SC, 2014_Miguel_MICRO, 2013_Misailovic_ACMtecs, 2012_Misailovic_RACES, 2006_Rinard_ICS, 2007_Rinard_OOPSLA, 2014_Samadi_ASPLOS, 2015_Sampson_UOW, 2016_Vassiliadis_CGO, 2018_Yesil_IEEEmicro, 2018_Zhang_IEEEcal, 2017_Mitra_CGO, relax, vos:proc1, 2015_Joshua_MICRO} \\
\hline\hline
\end{tabular}
\label{tb_app}
\end{table*}

\textbf{Image Processing:} This vital process is useful in various emerging applications from fields such as multimedia, bio-informatics and robotics, which often require processing of a tremendous amount of data with diverse standards and formats~\cite{2005_Alvarez_IEEEtc}. 
Fortunately, 
image processing workloads exhibit tolerance against errors, as most systems can accept minor perturbations or distortion within a quality degradation limit. Thus, Approximate Computing has received increasing attention here to mitigate the high computation cost and enhance the efficiency of the deployed systems by reasonable trade-offs based on constraints of the specific application.
A noticeable amount of Approximate Computing schemes and methodologies has focused on the filters inside the image processing function, 
which are applied to the input image to perform specific processing tasks, such as edge detection or morphological processing~\cite{2018_Esposito_IEEEtcasi}. 
Most of them require intensive matrix computations and convolutions. 
Therefore, several works propose novel approximate functional units to replace the exact ones~\cite{2018_Akbari_IEEEtcasii}. 

\textbf{Signal Processing:} In edge computing and IoT systems, various signal processing methodologies and algorithms are applied to process raw signal from sensors, which are often associated with noise~\cite{2014_Sampson_ACMtc}. Recently, signal processing algorithms are migrating from cloud-based platforms with high computation capabilities,  such as servers and desktops, 
to the edge and low power-devices (i.e., embedded systems) that are often resource-constrained and cost-sensitive. 
Real-time requirements are common in signal processing systems, thus Approximate Computing schemes are proposed to suffice critical performance requirements~\cite{2020_Baharvand_IEEEtetc}. 

\textbf{Computer Vision:} 
Computer vision is critical in many emerging applications such as augmented/virtual reality, autonomous driving and robotics. These applications often require intensive computation efforts on a large volume of data. 
One core component of a computer vision system is the computation kernel known as meta-functions, 
which are dedicated to execute one specific algorithm. As those meta-functions are typically inherently error resilient, voltage scaling techniques are applied to obtain attractive quality--energy trade-offs~\cite{2011_Mohapatra_DATE}. Another application, ray tracing, which is also a common computer vision task,
requires intensive arithmetic with floating-point numbers, 
whose precision can be formulated as a stochastic optimization problem for acceleration~\cite{2014_Schkufza_PLDI}. 

\textbf{Computer Graphics:} 
Modern digital photography and filming as well as 3D modelling, rendering and simulation depend on extensive support from computer graphics algorithms.
These workloads demand a significant amount of computations due to the visual contents that are increasingly complex.
Although computationally intensive, computer graphics show opportunities for Approximate Computing by having a strong data locality,
which indicates that there are high redundancies that can be harvested to mitigate the computation costs~\cite{2015_Keramidas_WAPCO}. 
By using cache to memorize outcomes of instructions, computer graphics applications such as OpenGL games can have a reduced number of instructions and witness acceleration~\cite{2015_Keramidas_WAPCO}. 

\textbf{Big Data Analysis:}
The data query is the first and foremost task in the big data analysis domain. Targeting improved performance, an approximate query engine BlinkDB is proposed in \cite{2013_Agarwal_EuroSys}. BlinkDB uses a set of stratified samples for the purpose of optimization and deploys a dynamic sampling strategy to select an appropriately sized sample, which permits users to trade accuracy for response time. 
Other works are mainly based on big data frameworks such as Hadoop and Spark.
The authors in \cite{2015_Goiri_ASPLOS} propose ApproxHadoop, which is a framework to run approximation-enabled MapReduce programs. ApproxHadoop leverages statistical theories to optimize the MapReduce programs when approximating with data sampling and task dropping. 

\textbf{Databases:}
Considering the Approximate Computing in database access, \cite{2016_Kandula_MOD} introduces a novel data query system that injects samplers in the run-time and without requiring former existing samples. 
This system allows multiple join inputs and approximates the answer according to complex outputs.
In the meantime, inspired by  parallel computing, \cite{2011_Sidiroglou_FCE} utilizes the nature of approximation to improve the performance and decrease the error according to the accuracy demands.

\textbf{Data Mining:}
During the data mining process, redundant computations, e.g., calling functions with the same parameters or using iterative input and output data, increase the execution time and lead to low performance. 
\cite{2017_Brumar_IPDPS} targets this redundant computing and proposes an approximate approach named ATM. 
ATM employs both dynamic redundancy and approximation for a task-level computing that utilizes the previous task for the future. 
In practice, ATM also introduces an adaptive algorithm in the run-time that dynamically balances the trade-off between accuracy and performance. 
Moreover, \cite{2020_Baharvand_IEEEtetc} uses the core-level redundancy, 
which is a combination of accurate and approximate versions of a task on different cores to increase the reliability. 

\textbf{Machine Learning:}
With the development of cloud computing, an increasing number of machine learning systems have become dominant in many sophisticated and crucial software architectures. 
The machine learning models can be either supervised or unsupervised, learn features from the training data and deploy inference on unseen query data. 
Nowadays, most machine learning models are data-driven, meaning that the key to their success is the input data. However, in the cloud computing environment, the large scale data makes the machine learning process tardiness and energy-consuming. To reduce resource consumption, recent works propose approaches in different domains. The work in \cite{2010_Baek_PLDI} presents an energy-efficient Approximate Computing framework that dynamically makes approximation decisions based on a predefined accuracy level. 
Moreover, \cite{2018_Kanduri_DAC} employs a run-time resource management approach based on functionally approximate kernels that exploit accuracy--performance trade-offs within error resilient applications. This approach makes performance-aware decisions on power management under varied situations. 

\textbf{Artificial Neural Networks:}
Considering the number of arithmetic operations in Artificial Neural Networks (ANNs), the conventional architectures perform over millions of computations on each input data, leading to a high rate of timing and energy consumption. 
Therefore, as explained in Section \ref{sec:app_spec},
approximation techniques
at different design layers
aim to provide sufficient performance and power efficiency. 
The approximation of ANNs imposes several challenges.
One of the them is to avoid retraining for reducing the accuracy loss \cite{mrazek2019alwann},
while the approximation localization \cite{leon_lascas}
is also a key factor towards improved resources gains. 
Finally, cross-layer and end-to-end approaches~\cite{panda2016cross}
further increase the resource savings by exploiting the error resilience of each layer/sub-system. 
 
\textbf{Scientific Computing:}
Due to the rapid development of the computing industry in recent decades, scientific computing has become much more powerful and precise. Natural science research has been greatly enhanced by the availability of new materials and methodologies as well as the development of new hardware.  
Though the specific tasks in scientific computing may vary, the underlying mathematical functions are similar 
(e.g., linear equations, nonlinear equations, eigenvalue problems, FFT analysis). 
To meet their computing demands,  
the community aims to develop more efficient, robust, and fast frameworks for scientific computations.  
In this context, 
many studies have examined the trade-off between precision and performance. One of the ideas is to get a balance between them by tuning the floating-point precision into mixed precision. The implementation in ~\cite{2013_Lam_ICS} uses binary instrumentation and modification to build mixed-precision configurations, whereas   \cite{2020_Brunie_SC} uses dynamic program information and temporal locality to customize floating-point precision for scientific computing. 

\textbf{Financial Analysis:}
Financial analysis is another area in which approximate computations can be useful. The internet and the digitalization of the financial industry have led to increased importance of computation in financial analysis, especially when analyzing large amounts of data and incorporating machine learning. With the advent of these technologies, quantitative finance models continue to develop. The use of approximate computations increases the efficiency of data analysis, as well as enables real-time processes. For example, \cite{2018_Tziantzioulis_IEEEmicro} demonstrates a compiler-directed output-based approximate memoization function that
provides speedup in two different financial analysis applications to pricing a portfolio of stock options and swaptions. 

\textbf{Physics Simulations:}
Apart from being a part of scientific computing, simulation of physical phenomena involves some more specific areas, such as hydrodynamics simulation, thermodynamics
simulation, and optical simulation.  Many of these tasks are very compute-intensive. Approximate Computing can be applied to the physics simulation to accelerate this process. \cite{2017_Mitra_CGO} presents a novel system, called OPPROX, for the application’s execution-phase-aware approximation. OPPROX is demonstrated to be an efficient method for reducing dozens of workloads in a simulation of the Sedov blast wave problem. 

\subsection{Application-Driven Analysis of Case Studies}

\begin{table*}[]
\renewcommand{\arraystretch}{1.035}
\footnotesize
\caption{Remarkable improvements among all examined works with respect to the application domain and the different layers of the computing stack.}
\vspace{-8pt}
\centering
\setlength{\tabcolsep}{1.2pt}
\begin{tabular}{cccclc}
\hline
\textbf{Application Domain} & \textbf{Stack} & \textbf{\#Works} & \textbf{Main Techniques} & \multicolumn{1}{c}{\textbf{Remarkable Gains}} & \textbf{Ref.} \\ \hline \hline
\multirow{4}{*}{Image Processing} & SW & 24 & Approx. Progr. Language & \begin{tabular}[c]{@{}l@{}}34\% energy gains \& \\ 113$\times$ less programming effort\end{tabular} & \cite{2014_Park_GIT} \\ \cline{5-6}
 & HW & 26 & Multiplier Approx. & \begin{tabular}[c]{@{}l@{}}95\% energy gains \& 85\% area \\ reduction for 1\% accuracy loss\end{tabular} & \cite{2019_Vahdat_IEEEtvlsi} \\ \cline{5-6}
 & Arch & 26 & Approx. Data Storage & 1.7$\times$ speedup for \textless{}10\%  quality loss & \cite{2014_Sampson_ACMtc} \\ \hline \hline
\multirow{4}{*}{Signal Processing} & SW & 13 & Precision Scaling & 1.33$\times$ speedup for no accuracy loss & \cite{2020_Kang_CGO} \\ \cline{5-6} 
 & HW & 11 & Approx. Synthesis & 60\% power gains for \textless 2\% avg. error & \cite{2013_Venkataramani_DATE} \\ \cline{5-6} 
 & Arch & 4 & Approx. Processor & \begin{tabular}[c]{@{}l@{}}2.3$\times$ speedup \& 3$\times$ energy gains\\  for \textless 10\% accuracy loss\end{tabular} & \cite{naccel:micro} \\ \hline \hline
\multirow{3}{*}{Computer Vision} & SW & 12 & Loop Perforation & 35\% energy gains \& 40\% area reduction & \cite{2020_Baharvand_IEEEtetc} \\ \cline{5-6} 
 & HW & 3 & Approx. Synthesis & 23\% power gains \& 19\% area savings & \cite{2014_Nepal_DATE} \\ \cline{5-6}
 & Arch & 1 & Approx. Processors & 40\% energy gains for negligible quality loss & \cite{relax} \\ \hline \hline
\multirow{3}{*}{Computer Graphics} & SW & 10 & Approx. Progr. Language & 2$\times$--17$\times$ energy gains & \cite{2015_Park_FSE} \\ \cline{5-6} 
 & HW & 0 & -- & -- & -- \\ \cline{5-6} 
 & Arch & 3 & Approx. Processor & up to 43\% energy gains & \cite{axprogramm:asplos} \\ \hline \hline
\multirow{4}{*}{Big Data Analysis} & SW & 13 & Data Sampling & up to 8$\times$ speedup \& 38\% wait time reduction & \cite{2016_Anderson_ICDE} \\ \cline{5-6} 
 & HW & 0 & -- & -- & -- \\ \cline{5-6} 
 & Arch & 2 & Approx. Data Storage & \begin{tabular}[c]{@{}l@{}}1.8$\times$--5.7$\times$ speedup for \textless{}1\% \\ memory overhead\end{tabular} & \cite{2012_Liu_FAST} \\ \hline \hline
\multirow{4}{*}{Databases} & SW & 5 & Data Sampling & up to 20$\times$ speedup for precise execution & \cite{2016_Zhang_VLDB} \\ \cline{5-6} 
 & HW & 0 & -- & -- &  \\ \cline{5-6} 
 & Arch & 2 & Approx. Data Storage & \begin{tabular}[c]{@{}l@{}}22\%--45\% energy gains for \\ 0.1\% performance loss\end{tabular} & \cite{2013_Tseng_DAC} \\ \hline \hline
\multirow{5}{*}{Data Mining} & SW & 19 & Loop Perforation & 6.4$\times$/7.1$\times$ for 5\%/10\% error budget & \cite{2018_Li_ICS} \\ \cline{5-6} 
 & HW & 6 & Multiplier Approx. & \begin{tabular}[c]{@{}l@{}}up to 86\% power, 46\% delay \& 63\% area\\  gains for NMED \textless{}0.0081 
 \end{tabular} & \cite{2018_Liu_IEEEtcasi} \\ \cline{5-6} 
 & Arch & 15 & Approx. Processor & \begin{tabular}[c]{@{}l@{}}1.2$\times$ speedup, 1.4$\times$ energy gains \\ for 6\% quality loss\end{tabular} & \cite{naccel:micro} \\ \hline \hline
\multirow{3}{*}{Machine Learning} & SW & 12 & Computation Skipping & 2.18$\times$ energy gains for \textless{}0.5\% accuracy loss & \cite{2015_Raha_DATE} \\ \cline{5-6} 
 & HW & 11 & Voltage Over-Scaling & 22\% power reduction for no accuracy loss & \cite{arm2023crossapprox} \\ \cline{5-6} 
 & Arch & 11 & Approx. Processor & 24\% energy reduction for no accuracy loss & \cite{riscv:bw} \\ \hline \hline
\multirow{4}{*}{Artificial Neural Networks} & SW & 41 & Computation Skipping & \begin{tabular}[c]{@{}l@{}}2$\times$ speedup, 1.9$\times$ energy gains \\ for 3\% accuracy loss\end{tabular} & \cite{2018_Akhlaghi_ISCA} \\ \cline{5-6} 
 & HW & 42 & Multiplier Approx. & 2.8$\times$ power gains for 5\% accuracy loss & \cite{2021_Leon_ACMtecs} \\ \cline{5-6} 
 & Arch & 11 & Approx. Data Storage & \begin{tabular}[c]{@{}l@{}}up to 2$\times$ energy gains for \\ negligible accuracy loss\end{tabular} & \cite{2020_Ranjan_TVLSI} \\ \hline \hline
\multirow{5}{*}{Scientific Computing} & SW & 26 & Approx. Progr. Language & \begin{tabular}[c]{@{}l@{}}up to 1.25$\times$ energy gains for \\ negligible quality loss\end{tabular} & \cite{2014_Misailovic_OOPSLA} \\ \cline{5-6} 
 & HW & 2 & Divider Approx. & \begin{tabular}[c]{@{}l@{}}3.9$\times$ speedup, 4.8$\times$ power gains \\ for \textless{}5\% NMED\end{tabular} & \cite{2019_Jiang_IEEEtc} \\ \cline{5-6} 
 & Arch & 6 & Approx. Data Storage & \begin{tabular}[c]{@{}l@{}}up to 2.55$\times$ energy gains, 1.55$\times$ cache \\ area reduction for \textless{}10\% quality error\end{tabular} & \cite{2015_Joshua_MICRO} \\ \hline \hline
\multirow{3}{*}{Financial Analysis} & SW & 19 & Approx. Memoization & \begin{tabular}[c]{@{}l@{}}16$\times$ speedup for 5\% memory overhead \&\\  \textless{}3\% accuracy loss\end{tabular} & \cite{2017_Brumar_IPDPS} \\ \cline{5-6} 
 & HW & 0 & -- & -- &  \\ \cline{5-6} 
 & Arch & 4 & Approx. Processor & 2.1$\times$ energy gains for 4\% quality loss & \cite{2006_Rinard_ICS} \\ \hline \hline
\multirow{3}{*}{Physics Simulations} & SW & 16 & Approx. Memoization & 3$\times$ speedup for 10\% quality loss & \cite{2014_Samadi_ASPLOS} \\ \cline{5-6} 
 & HW & 0 & -- &  -- &  \\ \cline{5-6} 
 & Arch & 5 & Approx. Processor & 1.25$\times$ energy gains for negligible quality loss & \cite{relax} \\ \hline\hline
\end{tabular}
\label{tb_analysis}
\end{table*}

Table~\ref{tb_analysis} presents the respective analysis of the aforementioned application domains, including also some statistical results about the dominant sub-categories for each layer of the computing stack.
Regarding the distribution of the main techniques in the various research areas, Table~\ref{tb_analysis} reveals a considerable heterogeneity that is contingent upon the particular use case.
As shown, works in image processing are almost evenly distributed through the entire computing stack (i.e., 24 software, 26 hardware, and 26 architectural works). 
However, within the specific domain, notably, multiplier approximations dominate over the other hardware techniques, accounting for almost 80\% of the total works.
Similar observations arise for big data analysis, where software-oriented approximations this time, and specifically data sampling sub-category, constitute the vast majority and the most prominent solutions.
On the other hand, it is noteworthy that in five application domains 
(computer graphics, big data analysis, databases, financial analysis and physics simulations),  there are no works focusing on hardware-level approximations.
This might be explained by the fact that research and development efforts in these domains are traditionally revolved around software algorithms, optimizations and data management techniques, rather than the lower level of the design abstraction hierarchy.
The latter highlights, thus, the space for further exploration and development of hardware-oriented approximate designs.

Table~\ref{tb_analysis} also reports some significant experimental results, most of them concerning speedup, area and power/energy gains over conventional designs, with their respective overheads and/or quality losses.
Out of all the techniques reported, it is observed that software-based approximations realize the highest speedups compared to hardware and architectural approximations.
For example, the memoization approach of~\cite{2017_Brumar_IPDPS} achieves 16$\times$ speedup for only 5\% memory overhead and 3\% accuracy loss, while~\cite{2016_Zhang_VLDB} manages up to 20$\times$ speedup for accurate execution.
Finally, the table underscores the potential of Approximate Computing in machine learning applications, at each layer of the computing stack, that can attain remarkable energy reduction (22\%--54\%), while maintaining unaffected the level of quality with zero accuracy loss.

\subsection{Benchmark Suites and Error Metrics}

Various open source benchmark suites for Approximate Computing have been proposed and released. 
Table \ref{tb_benchm} reports some of the most well-established suites,
which have been employed in the state-of-the-art works examined in this survey. 
Parallel programming benchmarks including PARSEC (Princeton Application Repository for Shared-Memory Computers)~\cite{2008_Bienia_PACT} and SPLASH-2 (Stanford
Parallel Applications for Shared Memory)~\cite{1995_Woo_ISCA} 
are cross-domain benchmark suites with different emerging applications for evaluating and studying parallel chip-multiprocessors. 
There are also benchmarks focusing on multimedia and communication applications such as MediaBench~\cite{1997_Chunho_MICRO}. 
SciMark is a benchmark suite primarily focusing on Java language applications of scientific computing, including kernels such as FFT, sparse matrix multiplication, Monte Carlo simulation and successive over-relaxation~\cite{2004_Pozo_SCI}.
With the development of multi-platform parallel/acceleration libraries and frameworks such as OpenMP, OpenCL, and CUDA, more recent benchmarks focus on them.  
Rodinia is a benchmark set for heterogenous computing with diverse applications targeting at platforms from multi-core CPU systems to GPUs~\cite{2009_Che_IISWC}. Regarding hardware, AxBench provides Verilog and C benchmark implementations~\cite{2017_Yazdanbakhsh_IEEEdt}.

The main error/QoS metrics that are used in Approximate Computing
are presented in Table \ref{tb_metrics}. 
At the bottom level, different error rates are used, such as 
Error Rate (ER),
which calculates the number of erroneous results over the total number of results,
and correspondingly, 
Bit Error Ratio (BER),
which calculates the number of erroneous bits over the total bits.
Another family of error metrics is the Error Distance (ED), including Hamming Distance (HD), Mean (Relative) ED (MED/MRED), Normalized MED (NMED), and other numerical-based metrics. 
Another metric that is employed in the evaluation of approximate designs
is the Probability of (Relative) ED higher than $X$ (PRED/PRED), where $X$ is a numerical value, 
Finally, 
Mean Squared Error (MSE) and Root MSE (RMSE) are also common metrics to evaluate the output quality.

There are also application-specific error metrics to describe the efficiency of the approximations at a higher level. For example, in telecommunications, Packet Error Ratio (PER) is usually considered to evaluate the communication quality. In ML applications, pattern recognition and information retrieval, classification accuracy is usually considered, as well as other metrics such as precision (relevant instances of retrieved instances) and recall (relevant instances of total instances). 

\begin{table*}[!t]
\renewcommand{\arraystretch}{1.05}
\footnotesize 
\centering
\setlength{\tabcolsep}{7.2pt}
\caption{Open-source benchmark suites used in state-of-the-art works of Approximate Computing.}
\vspace{-8pt}
\begin{tabular}{l| c m{7.0cm}}
\hline  
\makecell[c]{\textbf{Benchmark Suite}}   & \makecell[c]{\textbf{Source Code}}                                           & \makecell[c]{\textbf{Application Domain}} \\
\hline
\hline
\emph{PARSEC} \cite{2008_Bienia_PACT} & C/C++ & Image Processing, Computer Vision, Data Mining, Financial Analysis, Physics Simulations \\
\emph{SPLASH-2} \cite{1995_Woo_ISCA} & C & Scientific Computing, Computer Graphics, Physics Simulations\\
\emph{MediaBench} \cite{1997_Chunho_MICRO} & C & Image Processing, Signal Processing \\
\emph{MiBench} \cite{2001_Guthaus_WWC} & C & Image Processing, Signal Processing \\
\emph{MiniBench} \cite{2006_Narayan_IISWC} & C/C++ & Data Mining \\
\emph{Rodinia} \cite{2009_Che_IISWC} & OpenMP, OpenCL, CUDA & Data Mining, Scientific Computing, Physics Simulation, Machine Learning  \\
\emph{PolyBench} \cite{2015_Pouchet_POLY} & C, CUDA & Scientific Computing \\ 
\emph{SciMark 2.0} \cite{2004_Pozo_SCI} & Java & Scientific Computing \\
\emph{AxBench} \cite{2017_Yazdanbakhsh_IEEEdt} & C++, CUDA, Verilog & Image Processing, Signal Processing, Computer Graphics, Data Mining, Financial Analysis \\
\hline \hline
\end{tabular}
\label{tb_benchm}
\end{table*}
\begin{table*}[!t]
\renewcommand{\arraystretch}{1.05}
\footnotesize 
\centering
\setlength{\tabcolsep}{5pt}
\caption{Error/QoS metrics of Approximate Computing.}
\vspace{-8pt}
\begin{tabular}{l| m{7.5cm} c}
\hline  
\makecell[c]{\textbf{Metric}}   & \makecell[c]{\textbf{Description}}                                           & \makecell[c]{\textbf{Application Domain}} \\
\hline
\hline
\emph{ER}                       & Error Rate -- Erroneous results per total results                  &  \multirow{9}{*}{\centering General Computing} \\
\emph{EP}                       & Error Probability -- Probability of error occurrence                         &  \\ 
\emph{MED/MRED}                 & Mean (Relative) Error Distance                 &  \\ 
\emph{NMED/NMRED}               & Normalized Mean (Relative) Error Distance     &  \\ 
\emph{Max ED/RED}               & Maximum (Relative) Error Distance             &  \\ 
\emph{PED/PRED}                 & Probability of (Relative) Error Distance $> X$                               &  \\ 
\emph{MSE}                      & Mean Squared Error                            &  \\ 
\emph{RMSE}                     & Root Mean Squared Error                            &  \\
\emph{HD}                       & Hamming Distance                            &  \\
\hline
\emph{BER}                      & Bit Error Ratio -- Bit errors per total received bits              & \multirow{2}{*}{
\begin{minipage}[c]{3.3cm}
\centering Digital Systems, Telecommunications
\end{minipage}} \\
\emph{PER}                      & Packet Error Ratio -- Incorrect packets per total received packets &  \\ \hline
\emph{PSNR}                     & Peak Signal-to-Noise Ratio -- Quality measurement between images             & 
\multirow{3}{*}{
\begin{minipage}[c]{3.3cm}
\centering Image Processing, Video Processing, Computer Vision
\end{minipage}} \\ 
\emph{SSIM}                     & Structural Similarity -- Quality measurement between images                  &  \\ 
\emph{MPD}                     & Mean Pixel Difference                  &  \\ \hline
\emph{Classif. Accuracy}  & Correct classifications per total classifications                  & 
\multirow{3}{*}{
\begin{minipage}[c]{3.3cm}
\centering Pattern Recognition, Information Retrieval, Machine Learning
\end{minipage}} \\
\emph{Precision}                & Relevant instances per total retrieved instances                   &  \\
\emph{Recall}                   & Relevant instances per total relevant instances                    &  \\
\hline \hline
\end{tabular}
\label{tb_metrics}
\end{table*}
\section{Challenges and Future Directions}\label{sec:challenges}

This section highlights several key challenges that need to be addressed to accelerate the progress in Approximate Computing and enable its use in real-world systems. 

\textbf{Cross-Layer and End-to-End System-Level Approximations:} Most of the research in Approximate Computing have focused on individual abstraction layers of the computing stack. However, the true potential of Approximate Computing lies in cross-layer and end-to-end system-level approximations, where multiple complementary approximation techniques are combined to achieve a better quality--efficiency trade-off. Therefore, a system-level view is necessary for trading quality for efficiency, as approximation can be deployed at different abstraction layers and in different components. Only the right set of approximations deployed at the right layers and in the right components can offer optimal returns. Some existing works in this direction include~\cite{agrawal2016approximate, ghosh2020approximate, hashemi2018approximate}. However, still a significant amount of research is required to develop sophisticated automated methodologies that can efficiently identify the appropriate set of approximations for all the components in a system to offer the best quality--efficiency trade-off while satisfying the user-defined constraints. One of the key limitations towards enabling such methodologies is the lack of tools and methods that can efficiently and accurately quantify the impact of approximations, specifically multiple types of approximations, on the application-level accuracy of the systems. 

\textbf{Hardware Models:} The field of Approximate Computing lacks validated hardware models that are essential for performing efficient-yet-accurate design space exploration to select appropriate level of approximation for each hardware module in the system. 
For example, the models used for voltage overscaling are based on assumptions about the distribution of errors in computing and memory structures. 
Similarly, most of the ReRAM-based in-memory computing works are based on assumptions about the precision with which values can be written to or read from the memory arrays. 
Moreover, different works use different models and assumptions, which makes it difficult to compare techniques based on the reported results.
Therefore, there is a dire need for building hardware models that have been validated by the hardware community through extensive simulations. 
The models need to be validated for a wide variety of scenarios to enable their reliable deployment in cross-layer approximation methodologies. 

\textbf{Benchmarks:} The Approximate Computing community lacks representative sets of applications as well as datasets from different application domains that can be used for benchmarking approximation techniques. 
Having a carefully selected set of applications and datasets is crucial, as the impact of approximations can vary drastically from application to application and even from data sample to data sample. 
Some efforts have been made towards defining benchmarks, e.g., AxBench~\cite{yazdanbakhsh2016axbench}; however, these efforts cover only a small set of applications and scenarios. 
Therefore, there is a need for a much broader set of applications and datasets from a wider spectrum of application domains. 
Moreover, as it is difficult to cover all the application domains and applications, specifically when new applications are emerging on a daily basis due to the rapid pace of technology growth, there is a need for standards and rules for defining evaluation strategies and selecting datasets for an unbiased and thorough evaluation of approximation techniques. 

\textbf{Evaluation Metrics:} Design space exploration methodologies are guided through quality and performance evaluation metrics to find the optimal configurations that offer the best quality--efficiency trade-offs and meet all the user-defined constraints. As a result, it is crucial to have reliable quality and performance evaluation metrics in place. Metrics that do not capture suitable characteristics can lead to sub-optimal configurations, eventually affecting the user experience. Therefore, it is essential to define the right set of quality evaluation metrics for each application. 

\textbf{Post-Fabrication Testing:} 
Process variations, which lead to permanent faults and variations in the hardware characteristics of transistors, are a major concern in nano-scale devices. All the fabricated devices are passed through post-fabrication testing to detect manufacturing-induced defects and variations. However, approximations make this already challenging process further challenging. Only a limited amount of work has been carried out towards testing of Approximate Computing devices. Therefore, there is a dire need for specialized approximation-aware testing methodologies that can enable efficient-yet-accurate testing of these devices and reduce the yield loss, wherever possible, through binning. 

\textbf{Reproducibility for Fair Comparison and Future Developments:}
In most cases, it is challenging to replicate the exact evaluation methodology and setup reported in research works, making it difficult (if not impossible) to reproduce the results. Therefore, it is essential to promote open-source contributions to ensure the reproducibility of the results. The open-source contributions will also facilitate further research and development in the area of Approximate Computing, resulting in rapid progress towards uncovering the true potential of Approximate Computing for improving the efficiency of computing systems. 

\section{Conclusion}\label{sec:conc}
This article presented Part II of a comprehensive survey on Approximate Computing,
reviewing the state-of-the-art application-specific and architectural approximation techniques. 
Regarding the application-specific techniques,
the survey focused on the emerging AI/ML applications
and 
discussed about software, hardware, cross-layer and end-to-end approximations.
Regarding the architectural techniques,
the survey focused on approximate processors \& accelerators and different memory types for approximate data storage. 
Moreover, it included an extensive analysis of the application spectrum of Approximate Computing,
i.e., a domain classification of all the works reviewed,
it presented representative techniques and remarkable results,
as well as it reported 
well-established error metrics and benchmark suites  
for evaluating the quality-of-service of approximate designs. 
Despite the notable progress demonstrated by Approximate Computing, there remains a pressing need for ongoing and significant innovation to fully realize the potential of approximations in the design of complex computing systems.
These open challenges were discussed along with future directions at the end of the survey. 
\section*{Acknowledgement}

This research is partially supported by ASPIRE, the technology program management pillar of Abu Dhabi’s Advanced Technology Research Council (ATRC), via the ASPIRE Awards for Research Excellence.

\vspace{30pt}

\bibliographystyle{ACM-Reference-Format}
\bibliography{REFs.bib}


\begin{thebibliography}{301}


\ifx \showCODEN    \undefined \def \showCODEN     #1{\unskip}     \fi
\ifx \showDOI      \undefined \def \showDOI       #1{#1}\fi
\ifx \showISBNx    \undefined \def \showISBNx     #1{\unskip}     \fi
\ifx \showISBNxiii \undefined \def \showISBNxiii  #1{\unskip}     \fi
\ifx \showISSN     \undefined \def \showISSN      #1{\unskip}     \fi
\ifx \showLCCN     \undefined \def \showLCCN      #1{\unskip}     \fi
\ifx \shownote     \undefined \def \shownote      #1{#1}          \fi
\ifx \showarticletitle \undefined \def \showarticletitle #1{#1}   \fi
\ifx \showURL      \undefined \def \showURL       {\relax}        \fi
\providecommand\bibfield[2]{#2}
\providecommand\bibinfo[2]{#2}
\providecommand\natexlab[1]{#1}
\providecommand\showeprint[2][]{arXiv:#2}

\bibitem[Achour and Rinard(2015)]%
        {2015_Achour_OOPSLA}
\bibfield{author}{\bibinfo{person}{Sara Achour} {and} \bibinfo{person}{Martin~C. Rinard}.} \bibinfo{year}{2015}\natexlab{}.
\newblock \showarticletitle{Approximate Computation with Outlier Detection in {T}opaz}. In \bibinfo{booktitle}{\emph{ACM SIGPLAN Int'l. Conference on Object-Oriented Programming, Systems, Languages, and Applications (OOPSLA)}}. \bibinfo{pages}{711–730}.
\newblock


\bibitem[Adams et~al\mbox{.}(2020)]%
        {2020_Adams_IEEEtc}
\bibfield{author}{\bibinfo{person}{Elizabeth Adams}, \bibinfo{person}{Suganthi Venkatachalam}, {and} \bibinfo{person}{Seok-Bum Ko}.} \bibinfo{year}{2020}\natexlab{}.
\newblock \showarticletitle{Approximate Restoring Dividers Using Inexact Cells and Estimation From Partial Remainders}.
\newblock \bibinfo{journal}{\emph{IEEE Trans. on Computers}} \bibinfo{volume}{69}, \bibinfo{number}{4} (\bibinfo{year}{2020}), \bibinfo{pages}{468--474}.
\newblock


\bibitem[Afzali-Kusha and Pedram(2023)]%
        {nvdla:pedram}
\bibfield{author}{\bibinfo{person}{Hassan Afzali-Kusha} {and} \bibinfo{person}{Massoud Pedram}.} \bibinfo{year}{2023}\natexlab{}.
\newblock \showarticletitle{X-NVDLA: Runtime Accuracy Configurable NVDLA Based on Applying Voltage Overscaling to Computing and Memory Units}.
\newblock \bibinfo{journal}{\emph{IEEE Trans. on Circuits and Systems I: Regular Papers}} \bibinfo{volume}{70}, \bibinfo{number}{5} (\bibinfo{year}{2023}), \bibinfo{pages}{1989--2002}.
\newblock


\bibitem[Agarwal et~al\mbox{.}(2013)]%
        {2013_Agarwal_EuroSys}
\bibfield{author}{\bibinfo{person}{Sameer Agarwal}, \bibinfo{person}{Barzan Mozafari}, \bibinfo{person}{Aurojit Panda}, \bibinfo{person}{Henry Milner}, \bibinfo{person}{Samuel Madden}, {and} \bibinfo{person}{Ion Stoica}.} \bibinfo{year}{2013}\natexlab{}.
\newblock \showarticletitle{Blink{DB}: Queries with Bounded Errors and Bounded Response Times on Very Large Data}. In \bibinfo{booktitle}{\emph{ACM SIGOPS European Conference on Computer Systems (EuroSys)}}. \bibinfo{pages}{29–42}.
\newblock


\bibitem[Agrawal et~al\mbox{.}(2016)]%
        {agrawal2016approximate}
\bibfield{author}{\bibinfo{person}{Ankur Agrawal} {et~al\mbox{.}}} \bibinfo{year}{2016}\natexlab{}.
\newblock \showarticletitle{Approximate Computing: Challenges and Opportunities}. In \bibinfo{booktitle}{\emph{IEEE Int'l. Conference on Rebooting Computing (ICRC)}}. \bibinfo{pages}{1--8}.
\newblock


\bibitem[Agrawal et~al\mbox{.}(2021)]%
        {ibm}
\bibfield{author}{\bibinfo{person}{Ankur Agrawal} {et~al\mbox{.}}} \bibinfo{year}{2021}\natexlab{}.
\newblock \showarticletitle{{A 7nm 4-Core AI Chip with 25.6TFLOPS Hybrid FP8 Training, 102.4TOPS INT4 Inference and Workload-Aware Throttling}}. In \bibinfo{booktitle}{\emph{IEEE Int'l. Solid- State Circuits Conference (ISSCC)}}, Vol.~\bibinfo{volume}{64}. \bibinfo{pages}{144--146}.
\newblock


\bibitem[Akbari et~al\mbox{.}(2017)]%
        {2017_Akbari_IEEEtvlsi}
\bibfield{author}{\bibinfo{person}{Omid Akbari}, \bibinfo{person}{Mehdi Kamal}, \bibinfo{person}{Ali Afzali-Kusha}, {and} \bibinfo{person}{Massoud Pedram}.} \bibinfo{year}{2017}\natexlab{}.
\newblock \showarticletitle{Dual-Quality 4:2 Compressors for Utilizing in Dynamic Accuracy Configurable Multipliers}.
\newblock \bibinfo{journal}{\emph{IEEE Trans. on Very Large Scale Integration (VLSI) Systems}} \bibinfo{volume}{25}, \bibinfo{number}{4} (\bibinfo{year}{2017}), \bibinfo{pages}{1352--1361}.
\newblock


\bibitem[Akbari et~al\mbox{.}(2018)]%
        {2018_Akbari_IEEEtcasii}
\bibfield{author}{\bibinfo{person}{Omid Akbari}, \bibinfo{person}{Mehdi Kamal}, \bibinfo{person}{Ali Afzali-Kusha}, {and} \bibinfo{person}{Massoud Pedram}.} \bibinfo{year}{2018}\natexlab{}.
\newblock \showarticletitle{{RAP-CLA}: A Reconfigurable Approximate Carry Look-Ahead Adder}.
\newblock \bibinfo{journal}{\emph{IEEE Trans. on Circuits and Systems II: Express Briefs}} \bibinfo{volume}{65}, \bibinfo{number}{8} (\bibinfo{year}{2018}), \bibinfo{pages}{1089--1093}.
\newblock


\bibitem[Akhlaghi et~al\mbox{.}(2018)]%
        {2018_Akhlaghi_ISCA}
\bibfield{author}{\bibinfo{person}{Vahideh Akhlaghi}, \bibinfo{person}{Amir Yazdanbakhsh}, \bibinfo{person}{Kambiz Samadi}, \bibinfo{person}{Rajesh~K. Gupta}, {and} \bibinfo{person}{Hadi Esmaeilzadeh}.} \bibinfo{year}{2018}\natexlab{}.
\newblock \showarticletitle{Sna{PEA}: Predictive Early Activation for Reducing Computation in Deep Convolutional Neural Networks}. In \bibinfo{booktitle}{\emph{ACM/IEEE Int'l. Symposium on Computer Architecture (ISCA)}}. \bibinfo{pages}{662–673}.
\newblock


\bibitem[Alvarez et~al\mbox{.}(2005)]%
        {2005_Alvarez_IEEEtc}
\bibfield{author}{\bibinfo{person}{Carlos Alvarez}, \bibinfo{person}{Jesus Corbal}, {and} \bibinfo{person}{Mateo Valero}.} \bibinfo{year}{2005}\natexlab{}.
\newblock \showarticletitle{Fuzzy Memoization for Floating-Point Multimedia Applications}.
\newblock \bibinfo{journal}{\emph{IEEE Trans. on Computers}} \bibinfo{volume}{54}, \bibinfo{number}{7} (\bibinfo{year}{2005}), \bibinfo{pages}{922--927}.
\newblock


\bibitem[Amant et~al\mbox{.}(2014)]%
        {accel:isca}
\bibfield{author}{\bibinfo{person}{Renée~St. Amant} {et~al\mbox{.}}} \bibinfo{year}{2014}\natexlab{}.
\newblock \showarticletitle{General-Purpose Code Acceleration with Limited-Precision Analog Computation}. In \bibinfo{booktitle}{\emph{ACM/IEEE Int'l. Symposium on Computer Architecture (ISCA)}}. \bibinfo{pages}{505--516}.
\newblock


\bibitem[Anderson and Cafarella(2016)]%
        {2016_Anderson_ICDE}
\bibfield{author}{\bibinfo{person}{Michael~R. Anderson} {and} \bibinfo{person}{Michael Cafarella}.} \bibinfo{year}{2016}\natexlab{}.
\newblock \showarticletitle{Input Selection for Fast Feature Engineering}. In \bibinfo{booktitle}{\emph{IEEE Int'l. Conference on Data Engineering (ICDE)}}. \bibinfo{pages}{577--588}.
\newblock


\bibitem[Ansari et~al\mbox{.}(2021)]%
        {2021_Ansari_IEEEtc}
\bibfield{author}{\bibinfo{person}{Mohammad~Saeed Ansari}, \bibinfo{person}{Bruce~F. Cockburn}, {and} \bibinfo{person}{Jie Han}.} \bibinfo{year}{2021}\natexlab{}.
\newblock \showarticletitle{An Improved Logarithmic Multiplier for Energy-Efficient Neural Computing}.
\newblock \bibinfo{journal}{\emph{IEEE Trans. on Computers}} \bibinfo{volume}{70}, \bibinfo{number}{4} (\bibinfo{year}{2021}), \bibinfo{pages}{614--625}.
\newblock


\bibitem[Ansel et~al\mbox{.}(2011)]%
        {2011_Ansel_CGO}
\bibfield{author}{\bibinfo{person}{Jason Ansel}, \bibinfo{person}{Yee~Lok Wong}, \bibinfo{person}{Cy Chan}, \bibinfo{person}{Marek Olszewski}, \bibinfo{person}{Alan Edelman}, {and} \bibinfo{person}{Saman Amarasinghe}.} \bibinfo{year}{2011}\natexlab{}.
\newblock \showarticletitle{Language and Compiler Support for Auto-Tuning Variable-Accuracy Algorithms}. In \bibinfo{booktitle}{\emph{IEEE/ACM Int'l. Symposium on Code Generation and Optimization (CGO)}}. \bibinfo{pages}{85--96}.
\newblock


\bibitem[Aponte-Moreno et~al\mbox{.}(2018)]%
        {2018_Moreno_LATS}
\bibfield{author}{\bibinfo{person}{Alexander Aponte-Moreno}, \bibinfo{person}{Alejandro Moncada}, \bibinfo{person}{Felipe Restrepo-Calle}, {and} \bibinfo{person}{Cesar Pedraza}.} \bibinfo{year}{2018}\natexlab{}.
\newblock \showarticletitle{A Review of Approximate Computing Techniques Towards Fault Mitigation in {HW/SW} Systems}. In \bibinfo{booktitle}{\emph{IEEE Latin-American Test Symposium (LATS)}}. \bibinfo{pages}{1--6}.
\newblock


\bibitem[Armeniakos et~al\mbox{.}(2024)]%
        {riscv:arme}
\bibfield{author}{\bibinfo{person}{Giorgos Armeniakos}, \bibinfo{person}{Alexis Maras}, \bibinfo{person}{Sotirios Xydis}, {and} \bibinfo{person}{Dimitrios Soudris}.} \bibinfo{year}{2024}\natexlab{}.
\newblock \showarticletitle{Mixed-precision Neural Networks on RISC-V Cores: ISA Extensions for Multi-Pumped Soft SIMD Operations}. In \bibinfo{booktitle}{\emph{Int'l. Conference on Computer-Aided Design (ICCAD)}}. \bibinfo{pages}{1--9}.
\newblock


\bibitem[Armeniakos et~al\mbox{.}(2022a)]%
        {csur:arme}
\bibfield{author}{\bibinfo{person}{Giorgos Armeniakos}, \bibinfo{person}{Georgios Zervakis}, \bibinfo{person}{Dimitrios Soudris}, {and} \bibinfo{person}{J\"{o}rg Henkel}.} \bibinfo{year}{2022}\natexlab{a}.
\newblock \showarticletitle{Hardware Approximate Techniques for Deep Neural Network Accelerators: A Survey}.
\newblock \bibinfo{journal}{\emph{Comput. Surveys}} \bibinfo{volume}{55}, \bibinfo{number}{4} (\bibinfo{year}{2022}), \bibinfo{pages}{1--36}.
\newblock


\bibitem[Armeniakos et~al\mbox{.}(2022b)]%
        {arm2022crosslayer}
\bibfield{author}{\bibinfo{person}{Giorgos Armeniakos}, \bibinfo{person}{Georgios Zervakis}, \bibinfo{person}{Dimitrios Soudris}, \bibinfo{person}{Mehdi~B. Tahoori}, {and} \bibinfo{person}{Jörg Henkel}.} \bibinfo{year}{2022}\natexlab{b}.
\newblock \showarticletitle{Cross-Layer Approximation For Printed Machine Learning Circuits}. In \bibinfo{booktitle}{\emph{Design, Automation \& Test in Europe (DATE)}}. \bibinfo{pages}{1--6}.
\newblock


\bibitem[Armeniakos et~al\mbox{.}(2023a)]%
        {arm2023codesign}
\bibfield{author}{\bibinfo{person}{Giorgos Armeniakos}, \bibinfo{person}{Georgios Zervakis}, \bibinfo{person}{Dimitrios Soudris}, \bibinfo{person}{Mehdi~B. Tahoori}, {and} \bibinfo{person}{Jörg Henkel}.} \bibinfo{year}{2023}\natexlab{a}.
\newblock \showarticletitle{Co-Design of Approximate Multilayer Perceptron for Ultra-Resource Constrained Printed Circuits}.
\newblock \bibinfo{journal}{\emph{IEEE Trans. on Computers}} \bibinfo{volume}{72}, \bibinfo{number}{9} (\bibinfo{year}{2023}), \bibinfo{pages}{2717--2725}.
\newblock


\bibitem[Armeniakos et~al\mbox{.}(2023b)]%
        {arm2023crossapprox}
\bibfield{author}{\bibinfo{person}{Giorgos Armeniakos}, \bibinfo{person}{Georgios Zervakis}, \bibinfo{person}{Dimitrios Soudris}, \bibinfo{person}{Mehdi~B. Tahoori}, {and} \bibinfo{person}{Jörg Henkel}.} \bibinfo{year}{2023}\natexlab{b}.
\newblock \showarticletitle{Model-to-Circuit Cross-Approximation For Printed Machine Learning Classifiers}.
\newblock \bibinfo{journal}{\emph{IEEE Trans. on Computer-Aided Design of Integrated Circuits and Systems}} \bibinfo{volume}{42}, \bibinfo{number}{11} (\bibinfo{year}{2023}), \bibinfo{pages}{3532--3544}.
\newblock


\bibitem[Baek and Chilimbi(2010)]%
        {2010_Baek_PLDI}
\bibfield{author}{\bibinfo{person}{Woongki Baek} {and} \bibinfo{person}{Trishul~M. Chilimbi}.} \bibinfo{year}{2010}\natexlab{}.
\newblock \showarticletitle{Green: A Framework for Supporting Energy-Conscious Programming Using Controlled Approximation}. In \bibinfo{booktitle}{\emph{ACM SIGPLAN Conference on Programming Language Design and Implementation (PLDI)}}. \bibinfo{pages}{198–209}.
\newblock


\bibitem[Baharvand and Miremadi(2020)]%
        {2020_Baharvand_IEEEtetc}
\bibfield{author}{\bibinfo{person}{Farshad Baharvand} {and} \bibinfo{person}{Seyed~Ghassem Miremadi}.} \bibinfo{year}{2020}\natexlab{}.
\newblock \showarticletitle{{LEXACT}: Low Energy N-Modular Redundancy Using Approximate Computing for Real-Time Multicore Processors}.
\newblock \bibinfo{journal}{\emph{IEEE Trans. on Emerging Topics in Computing}} \bibinfo{volume}{8}, \bibinfo{number}{2} (\bibinfo{year}{2020}), \bibinfo{pages}{431--441}.
\newblock


\bibitem[Bertoldi et~al\mbox{.}(2017)]%
        {avgerinou}
\bibfield{author}{\bibinfo{person}{Paolo Bertoldi}, \bibinfo{person}{Maria Avgerinou}, {and} \bibinfo{person}{Luca Castellazzi}.} \bibinfo{year}{2017}\natexlab{}.
\newblock \showarticletitle{{Trends in Data Centre Energy Consumption under the European Code of Conduct for Data Centre Energy Efficiency}}.
\newblock \bibinfo{journal}{\emph{European Comission -- Joint Research Centre (JRC)}} (\bibinfo{year}{2017}), \bibinfo{pages}{1--43}.
\newblock


\bibitem[Bienia et~al\mbox{.}(2008)]%
        {2008_Bienia_PACT}
\bibfield{author}{\bibinfo{person}{Christian Bienia}, \bibinfo{person}{Sanjeev Kumar}, \bibinfo{person}{Jaswinder~Pal Singh}, {and} \bibinfo{person}{Kai Li}.} \bibinfo{year}{2008}\natexlab{}.
\newblock \showarticletitle{The {PARSEC} Benchmark Suite: Characterization and Architectural Implications}. In \bibinfo{booktitle}{\emph{Int'l. Conf. on Parallel Architectures and Compilation Techniques}}. \bibinfo{pages}{72–81}.
\newblock


\bibitem[Bornholt et~al\mbox{.}(2014)]%
        {2014_Bornholt_ASPLOS}
\bibfield{author}{\bibinfo{person}{James Bornholt}, \bibinfo{person}{Todd Mytkowicz}, {and} \bibinfo{person}{Kathryn~S. McKinley}.} \bibinfo{year}{2014}\natexlab{}.
\newblock \showarticletitle{Uncertain\raisebox{0.8pt}{$\scriptstyle <$}{T}\raisebox{0.8pt}{$\scriptstyle >$}: A First-Order Type for Uncertain Data}. In \bibinfo{booktitle}{\emph{ACM Int'l. Conference on Architectural Support for Programming Languages and Operating Systems}}. \bibinfo{pages}{51–66}.
\newblock


\bibitem[Boston et~al\mbox{.}(2015)]%
        {2015_Boston_OOPSLA}
\bibfield{author}{\bibinfo{person}{Brett Boston}, \bibinfo{person}{Adrian Sampson}, \bibinfo{person}{Dan Grossman}, {and} \bibinfo{person}{Luis Ceze}.} \bibinfo{year}{2015}\natexlab{}.
\newblock \showarticletitle{Probability Type Inference for Flexible Approximate Programming}. In \bibinfo{booktitle}{\emph{ACM SIGPLAN Int'l. Conference on Object-Oriented Programming, Systems, Languages, and Applications (OOPSLA)}}. \bibinfo{pages}{470–487}.
\newblock


\bibitem[Brumar et~al\mbox{.}(2017)]%
        {2017_Brumar_IPDPS}
\bibfield{author}{\bibinfo{person}{Iulian Brumar}, \bibinfo{person}{Marc Casas}, \bibinfo{person}{Miquel Moreto}, \bibinfo{person}{Mateo Valero}, {and} \bibinfo{person}{Gurindar~S. Sohi}.} \bibinfo{year}{2017}\natexlab{}.
\newblock \showarticletitle{{ATM}: Approximate Task Memoization in the Runtime System}. In \bibinfo{booktitle}{\emph{IEEE Int'l. Parallel and Distributed Processing Symposium (IPDPS)}}. \bibinfo{pages}{1140--1150}.
\newblock


\bibitem[Brunie et~al\mbox{.}(2020)]%
        {2020_Brunie_SC}
\bibfield{author}{\bibinfo{person}{Hugo Brunie}, \bibinfo{person}{Costin Iancu}, \bibinfo{person}{Khaled~Z. Ibrahim}, \bibinfo{person}{Philip Brisk}, {and} \bibinfo{person}{Brandon Cook}.} \bibinfo{year}{2020}\natexlab{}.
\newblock \showarticletitle{Tuning Floating-Point Precision Using Dynamic Program Information and Temporal Locality}. In \bibinfo{booktitle}{\emph{ACM/IEEE SC, Int'l. Conference for High Performance Computing, Networking, Storage and Analysis}}. \bibinfo{pages}{1--14}.
\newblock


\bibitem[Byna et~al\mbox{.}(2010)]%
        {2010_Byna_GPGPU}
\bibfield{author}{\bibinfo{person}{Surendra Byna}, \bibinfo{person}{Jiayuan Meng}, \bibinfo{person}{Anand Raghunathan}, \bibinfo{person}{Srimat Chakradhar}, {and} \bibinfo{person}{Srihari Cadambi}.} \bibinfo{year}{2010}\natexlab{}.
\newblock \showarticletitle{Best-Effort Semantic Document Search on {GPU}s}. In \bibinfo{booktitle}{\emph{Workshop on General-Purpose Computation on Graphics Processing Units (GPGPU)}}. \bibinfo{pages}{86–93}.
\newblock


\bibitem[Campanoni et~al\mbox{.}(2015)]%
        {2015_Campanoni_CGO}
\bibfield{author}{\bibinfo{person}{Simone Campanoni}, \bibinfo{person}{Glenn Holloway}, \bibinfo{person}{Gu-Yeon Wei}, {and} \bibinfo{person}{David Brooks}.} \bibinfo{year}{2015}\natexlab{}.
\newblock \showarticletitle{{HELIX-UP}: Relaxing Program Semantics to Unleash Parallelization}. In \bibinfo{booktitle}{\emph{IEEE/ACM Int'l. Symposium on Code Generation and Optimization (CGO)}}. \bibinfo{pages}{235–245}.
\newblock


\bibitem[Cao et~al\mbox{.}(2022)]%
        {datacenters}
\bibfield{author}{\bibinfo{person}{Zhiwei Cao}, \bibinfo{person}{Xin Zhou}, \bibinfo{person}{Han Hu}, \bibinfo{person}{Zhi Wang}, {and} \bibinfo{person}{Yonggang Wen}.} \bibinfo{year}{2022}\natexlab{}.
\newblock \showarticletitle{{Toward a Systematic Survey for Carbon Neutral Data Centers}}.
\newblock \bibinfo{journal}{\emph{IEEE Communications Surveys \& Tutorials}} \bibinfo{volume}{24}, \bibinfo{number}{2} (\bibinfo{year}{2022}), \bibinfo{pages}{895--936}.
\newblock


\bibitem[Carbin et~al\mbox{.}(2013)]%
        {2013_Carbin_OOPSLA}
\bibfield{author}{\bibinfo{person}{Michael Carbin}, \bibinfo{person}{Sasa Misailovic}, {and} \bibinfo{person}{Martin~C. Rinard}.} \bibinfo{year}{2013}\natexlab{}.
\newblock \showarticletitle{Verifying Quantitative Reliability for Programs That Execute on Unreliable Hardware}. In \bibinfo{booktitle}{\emph{ACM SIGPLAN Int'l. Conference on Object-Oriented Programming, Systems, Languages, and Applications (OOPSLA)}}. \bibinfo{pages}{33–52}.
\newblock


\bibitem[Castro-Godínez et~al\mbox{.}(2020)]%
        {2020_Castro_ICCAD}
\bibfield{author}{\bibinfo{person}{Jorge Castro-Godínez}, \bibinfo{person}{Julián Mateus-Vargas}, \bibinfo{person}{Muhammad Shafique}, {and} \bibinfo{person}{Jörg Henkel}.} \bibinfo{year}{2020}\natexlab{}.
\newblock \showarticletitle{Ax{HLS}: Design Space Exploration and High-Level Synthesis of Approximate Accelerators using Approximate Functional Units and Analytical Models}. In \bibinfo{booktitle}{\emph{Int'l. Conference On Computer Aided Design (ICCAD)}}. \bibinfo{pages}{1--9}.
\newblock


\bibitem[Chakradhar and Raghunathan(2010)]%
        {2010_Chakradhar_DAC}
\bibfield{author}{\bibinfo{person}{Srimat~T. Chakradhar} {and} \bibinfo{person}{Anand Raghunathan}.} \bibinfo{year}{2010}\natexlab{}.
\newblock \showarticletitle{Best-Effort Computing: Re-thinking Parallel Software and Hardware}. In \bibinfo{booktitle}{\emph{Design Automation Conference (DAC)}}. \bibinfo{pages}{865--870}.
\newblock


\bibitem[Chang et~al\mbox{.}(2011)]%
        {2011_chang_TVID}
\bibfield{author}{\bibinfo{person}{Ik~Joon Chang}, \bibinfo{person}{Debabrata Mohapatra}, {and} \bibinfo{person}{Kaushik Roy}.} \bibinfo{year}{2011}\natexlab{}.
\newblock \showarticletitle{A Priority-Based {6T/8T} Hybrid {SRAM} Architecture for Aggressive Voltage Scaling in Video Applications}.
\newblock \bibinfo{journal}{\emph{IEEE Trans. on Circuits and Systems for Video Technology}} \bibinfo{volume}{21}, \bibinfo{number}{2} (\bibinfo{year}{2011}), \bibinfo{pages}{101--112}.
\newblock


\bibitem[Che et~al\mbox{.}(2009)]%
        {2009_Che_IISWC}
\bibfield{author}{\bibinfo{person}{Shuai Che} {et~al\mbox{.}}} \bibinfo{year}{2009}\natexlab{}.
\newblock \showarticletitle{Rodinia: A Benchmark Suite for Heterogeneous Computing}. In \bibinfo{booktitle}{\emph{IEEE Int'l. Symposium on Workload Characterization (IISWC)}}. \bibinfo{pages}{44--54}.
\newblock


\bibitem[Chen and Hu(2013)]%
        {2013_Chen_IEEEtvlsi}
\bibfield{author}{\bibinfo{person}{Jienan Chen} {and} \bibinfo{person}{Jianhao Hu}.} \bibinfo{year}{2013}\natexlab{}.
\newblock \showarticletitle{Energy-Efficient Digital Signal Processing via Voltage-Overscaling-Based Residue Number System}.
\newblock \bibinfo{journal}{\emph{IEEE Trans. on Very Large Scale Integration (VLSI) Systems}} \bibinfo{volume}{21}, \bibinfo{number}{7} (\bibinfo{year}{2013}), \bibinfo{pages}{1322--1332}.
\newblock


\bibitem[Chen et~al\mbox{.}(2019)]%
        {2019_Chen_IEEEtc}
\bibfield{author}{\bibinfo{person}{Ke Chen}, \bibinfo{person}{Linbin Chen}, \bibinfo{person}{Pedro Reviriego}, {and} \bibinfo{person}{Fabrizio Lombardi}.} \bibinfo{year}{2019}\natexlab{}.
\newblock \showarticletitle{Efficient Implementations of Reduced Precision Redundancy ({RPR}) Multiply and Accumulate ({MAC})}.
\newblock \bibinfo{journal}{\emph{IEEE Trans. on Computers}} \bibinfo{volume}{68}, \bibinfo{number}{5} (\bibinfo{year}{2019}), \bibinfo{pages}{784--790}.
\newblock


\bibitem[Chen et~al\mbox{.}(2015)]%
        {2015_Chen_GLSVLSI}
\bibfield{author}{\bibinfo{person}{Linbin Chen}, \bibinfo{person}{Jie Han}, \bibinfo{person}{Weiqiang Liu}, {and} \bibinfo{person}{Fabrizio Lombardi}.} \bibinfo{year}{2015}\natexlab{}.
\newblock \showarticletitle{Design of Approximate Unsigned Integer Non-Restoring Divider for Inexact Computing}. In \bibinfo{booktitle}{\emph{Great Lakes Symposium on VLSI (GLSVLSI)}}. \bibinfo{pages}{51--56}.
\newblock


\bibitem[Chen et~al\mbox{.}(2016)]%
        {2016_Chen_IEEEtc}
\bibfield{author}{\bibinfo{person}{Linbin Chen}, \bibinfo{person}{Jie Han}, \bibinfo{person}{Weiqiang Liu}, {and} \bibinfo{person}{Fabrizio Lombardi}.} \bibinfo{year}{2016}\natexlab{}.
\newblock \showarticletitle{On the Design of Approximate Restoring Dividers for Error-Tolerant Applications}.
\newblock \bibinfo{journal}{\emph{IEEE Trans. on Computers}} \bibinfo{volume}{65}, \bibinfo{number}{8} (\bibinfo{year}{2016}), \bibinfo{pages}{2522--2533}.
\newblock


\bibitem[Chen et~al\mbox{.}(2017)]%
        {2017_Chen_IEEEtscl}
\bibfield{author}{\bibinfo{person}{Linbin Chen}, \bibinfo{person}{Jie Han}, \bibinfo{person}{Weiqiang Liu}, {and} \bibinfo{person}{Fabrizio Lombardi}.} \bibinfo{year}{2017}\natexlab{}.
\newblock \showarticletitle{Algorithm and Design of a Fully Parallel Approximate Coordinate Rotation Digital Computer ({CORDIC})}.
\newblock \bibinfo{journal}{\emph{IEEE Trans. on Multi-Scale Computing Systems}} \bibinfo{volume}{3}, \bibinfo{number}{3} (\bibinfo{year}{2017}), \bibinfo{pages}{139--151}.
\newblock


\bibitem[Chen et~al\mbox{.}(2018)]%
        {2018_Chen_IEEEtmscs}
\bibfield{author}{\bibinfo{person}{Linbin Chen}, \bibinfo{person}{Jie Han}, \bibinfo{person}{Weiqiang Liu}, \bibinfo{person}{Paolo Montuschi}, {and} \bibinfo{person}{Fabrizio Lombardi}.} \bibinfo{year}{2018}\natexlab{}.
\newblock \showarticletitle{Design, Evaluation and Application of Approximate High-Radix Dividers}.
\newblock \bibinfo{journal}{\emph{IEEE Trans. on Multi-Scale Computing Systems}} \bibinfo{volume}{4}, \bibinfo{number}{3} (\bibinfo{year}{2018}), \bibinfo{pages}{299--312}.
\newblock


\bibitem[Chen et~al\mbox{.}(2021)]%
        {chen2021only}
\bibfield{author}{\bibinfo{person}{Tianyi Chen}, \bibinfo{person}{Bo Ji}, \bibinfo{person}{Tianyu Ding}, \bibinfo{person}{Biyi Fang}, \bibinfo{person}{Guanyi Wang}, \bibinfo{person}{Zhihui Zhu}, \bibinfo{person}{Luming Liang}, \bibinfo{person}{Yixin Shi}, \bibinfo{person}{Sheng Yi}, {and} \bibinfo{person}{Xiao Tu}.} \bibinfo{year}{2021}\natexlab{}.
\newblock \showarticletitle{Only Train Once: A One-Shot Neural Network Training And Pruning Framework}.
\newblock \bibinfo{journal}{\emph{Advances in Neural Information Processing Systems}}  \bibinfo{volume}{34} (\bibinfo{year}{2021}), \bibinfo{pages}{19637--19651}.
\newblock


\bibitem[Cherupalli et~al\mbox{.}(2017)]%
        {isca17bespoke}
\bibfield{author}{\bibinfo{person}{Hari Cherupalli}, \bibinfo{person}{Henry Duwe}, \bibinfo{person}{Weidong Ye}, \bibinfo{person}{Rakesh Kumar}, {and} \bibinfo{person}{John Sartori}.} \bibinfo{year}{2017}\natexlab{}.
\newblock \showarticletitle{Bespoke Processors for Applications with Ultra-Low Area and Power Constraints}. In \bibinfo{booktitle}{\emph{ACM/IEEE Int'l. Symposium on Computer Architecture (ISCA)}}. \bibinfo{pages}{41--54}.
\newblock


\bibitem[Chiang et~al\mbox{.}(2017)]%
        {2017_Chiang_POPL}
\bibfield{author}{\bibinfo{person}{Wei-Fan Chiang}, \bibinfo{person}{Mark Baranowski}, \bibinfo{person}{Ian Briggs}, \bibinfo{person}{Alexey Solovyev}, \bibinfo{person}{Ganesh Gopalakrishnan}, {and} \bibinfo{person}{Zvonimir Rakamari\'{c}}.} \bibinfo{year}{2017}\natexlab{}.
\newblock \showarticletitle{Rigorous Floating-Point Mixed-Precision Tuning}. In \bibinfo{booktitle}{\emph{ACM SIGPLAN Symposium on Principles of Programming Languages (POPL)}}. \bibinfo{pages}{300–315}.
\newblock


\bibitem[Chiang et~al\mbox{.}(2014)]%
        {2014_Chiang_PPoPP}
\bibfield{author}{\bibinfo{person}{Wei-Fan Chiang}, \bibinfo{person}{Ganesh Gopalakrishnan}, \bibinfo{person}{Zvonimir Rakamaric}, {and} \bibinfo{person}{Alexey Solovyev}.} \bibinfo{year}{2014}\natexlab{}.
\newblock \showarticletitle{Efficient Search for Inputs Causing High Floating-Point Errors}. In \bibinfo{booktitle}{\emph{ACM SIGPLAN Symposium on Principles and Practice of Parallel Programming (PPoPP)}}. \bibinfo{pages}{43–52}.
\newblock


\bibitem[Chippa et~al\mbox{.}(2014)]%
        {2014_Chippa_IEEEtvlsi}
\bibfield{author}{\bibinfo{person}{Vinay~Kumar Chippa}, \bibinfo{person}{Debabrata Mohapatra}, \bibinfo{person}{Kaushik Roy}, \bibinfo{person}{Srimat~T. Chakradhar}, {and} \bibinfo{person}{Anand Raghunathan}.} \bibinfo{year}{2014}\natexlab{}.
\newblock \showarticletitle{Scalable Effort Hardware Design}.
\newblock \bibinfo{journal}{\emph{IEEE Trans. on Very Large Scale Integration (VLSI) Systems}} \bibinfo{volume}{22}, \bibinfo{number}{9} (\bibinfo{year}{2014}), \bibinfo{pages}{2004--2016}.
\newblock


\bibitem[Cho et~al\mbox{.}(2012)]%
        {ersa}
\bibfield{author}{\bibinfo{person}{Hyungmin Cho}, \bibinfo{person}{Larkhoon Leem}, {and} \bibinfo{person}{Subhasish Mitra}.} \bibinfo{year}{2012}\natexlab{}.
\newblock \showarticletitle{{ERSA}: Error Resilient System Architecture for Probabilistic Applications}.
\newblock \bibinfo{journal}{\emph{IEEE Trans. on Computer-Aided Design of Integrated Circuits and Systems}} \bibinfo{volume}{31}, \bibinfo{number}{4} (\bibinfo{year}{2012}), \bibinfo{pages}{546--558}.
\newblock


\bibitem[Cococcioni et~al\mbox{.}(2022)]%
        {riscv:lightweight}
\bibfield{author}{\bibinfo{person}{Marco Cococcioni}, \bibinfo{person}{Federico Rossi}, \bibinfo{person}{Emanuele Ruffaldi}, {and} \bibinfo{person}{Sergio Saponara}.} \bibinfo{year}{2022}\natexlab{}.
\newblock \showarticletitle{A Lightweight Posit Processing Unit for RISC-V Processors in Deep Neural Network Applications}.
\newblock \bibinfo{journal}{\emph{IEEE Trans. on Emerging Topics in Computing}} \bibinfo{volume}{10}, \bibinfo{number}{4} (\bibinfo{year}{2022}), \bibinfo{pages}{1898--1908}.
\newblock


\bibitem[Damsgaard et~al\mbox{.}(2023)]%
        {edge:survey}
\bibfield{author}{\bibinfo{person}{Hans~Jakob Damsgaard}, \bibinfo{person}{Aleksandr Ometov}, {and} \bibinfo{person}{Jari Nurmi}.} \bibinfo{year}{2023}\natexlab{}.
\newblock \showarticletitle{Approximation Opportunities in Edge Computing Hardware: A Systematic Literature Review}.
\newblock \bibinfo{journal}{\emph{Comput. Surveys}} \bibinfo{volume}{55}, \bibinfo{number}{12} (\bibinfo{year}{2023}), \bibinfo{pages}{1--49}.
\newblock


\bibitem[Darulova and Kuncak(2017)]%
        {2017_Darulova_ACMtoplas}
\bibfield{author}{\bibinfo{person}{Eva Darulova} {and} \bibinfo{person}{Viktor Kuncak}.} \bibinfo{year}{2017}\natexlab{}.
\newblock \showarticletitle{Towards a Compiler for Reals}.
\newblock \bibinfo{journal}{\emph{ACM Trans. on Programming Languages and Systems}} \bibinfo{volume}{39}, \bibinfo{number}{2} (\bibinfo{year}{2017}), \bibinfo{pages}{1--28}.
\newblock


\bibitem[Das et~al\mbox{.}(2024)]%
        {raha_2024}
\bibfield{author}{\bibinfo{person}{Arghadip Das}, \bibinfo{person}{Soumendu~Kumar Ghosh}, \bibinfo{person}{Arnab Raha}, {and} \bibinfo{person}{Vijay Raghunathan}.} \bibinfo{year}{2024}\natexlab{}.
\newblock \showarticletitle{Toward Energy-Efficient Collaborative Inference Using Multisystem Approximations}.
\newblock \bibinfo{journal}{\emph{IEEE Internet of Things Journal}} \bibinfo{volume}{11}, \bibinfo{number}{10} (\bibinfo{year}{2024}), \bibinfo{pages}{17989--18004}.
\newblock


\bibitem[de~Kruijf et~al\mbox{.}(2010)]%
        {relax}
\bibfield{author}{\bibinfo{person}{Marc de Kruijf}, \bibinfo{person}{Shuou Nomura}, {and} \bibinfo{person}{Karthikeyan Sankaralingam}.} \bibinfo{year}{2010}\natexlab{}.
\newblock \showarticletitle{Relax: An Architectural Framework for Software Recovery of Hardware Faults}. In \bibinfo{booktitle}{\emph{ACM/IEEE Int'l. Symposium on Computer Architecture (ISCA)}}. \bibinfo{pages}{497–508}.
\newblock


\bibitem[De~la Parra et~al\mbox{.}(2020)]%
        {de2020proxsim}
\bibfield{author}{\bibinfo{person}{Cecilia De~la Parra}, \bibinfo{person}{Andre Guntoro}, {and} \bibinfo{person}{Akash Kumar}.} \bibinfo{year}{2020}\natexlab{}.
\newblock \showarticletitle{ProxSim: GPU-based Simulation Framework for Cross-Layer Approximate DNN Optimization}. In \bibinfo{booktitle}{\emph{Design, Automation \& Test in Europe (DATE)}}. \bibinfo{pages}{1193--1198}.
\newblock


\bibitem[Ebrahimi-Azandaryani et~al\mbox{.}(2020)]%
        {2020_Ebrahimi_IEEEtcasii}
\bibfield{author}{\bibinfo{person}{Farhad Ebrahimi-Azandaryani}, \bibinfo{person}{Omid Akbari}, \bibinfo{person}{Mehdi Kamal}, \bibinfo{person}{Ali Afzali-Kusha}, {and} \bibinfo{person}{Massoud Pedram}.} \bibinfo{year}{2020}\natexlab{}.
\newblock \showarticletitle{Block-Based Carry Speculative Approximate Adder for Energy-Efficient Applications}.
\newblock \bibinfo{journal}{\emph{IEEE Trans. on Circuits and Systems II: Express Briefs}} \bibinfo{volume}{67}, \bibinfo{number}{1} (\bibinfo{year}{2020}), \bibinfo{pages}{137--141}.
\newblock


\bibitem[Esmaeilzadeh et~al\mbox{.}(2012a)]%
        {axprogramm:asplos}
\bibfield{author}{\bibinfo{person}{Hadi Esmaeilzadeh}, \bibinfo{person}{Adrian Sampson}, \bibinfo{person}{Luis Ceze}, {and} \bibinfo{person}{Doug Burger}.} \bibinfo{year}{2012}\natexlab{a}.
\newblock \showarticletitle{Architecture Support for Disciplined Approximate Programming}. In \bibinfo{booktitle}{\emph{ACM Int'l. Conference on Architectural Support for Programming Languages and Operating Systems (ASPLOS)}}. \bibinfo{pages}{301–312}.
\newblock


\bibitem[Esmaeilzadeh et~al\mbox{.}(2012b)]%
        {naccel:micro}
\bibfield{author}{\bibinfo{person}{Hadi Esmaeilzadeh}, \bibinfo{person}{Adrian Sampson}, \bibinfo{person}{Luis Ceze}, {and} \bibinfo{person}{Doug Burger}.} \bibinfo{year}{2012}\natexlab{b}.
\newblock \showarticletitle{Neural Acceleration for General-Purpose Approximate Programs}. In \bibinfo{booktitle}{\emph{IEEE/ACM Int'l. Symposium on Microarchitecture (MICRO)}}. \bibinfo{pages}{449--460}.
\newblock


\bibitem[Esposito et~al\mbox{.}(2018)]%
        {2018_Esposito_IEEEtcasi}
\bibfield{author}{\bibinfo{person}{Darjn Esposito}, \bibinfo{person}{Antonio Giuseppe~Maria Strollo}, \bibinfo{person}{Ettore Napoli}, \bibinfo{person}{Davide De~Caro}, {and} \bibinfo{person}{Nicola Petra}.} \bibinfo{year}{2018}\natexlab{}.
\newblock \showarticletitle{Approximate Multipliers Based on New Approximate Compressors}.
\newblock \bibinfo{journal}{\emph{IEEE Trans. on Circuits and Systems I: Regular Papers}} \bibinfo{volume}{65}, \bibinfo{number}{12} (\bibinfo{year}{2018}), \bibinfo{pages}{4169--4182}.
\newblock


\bibitem[Fan et~al\mbox{.}(2019)]%
        {fan2019axdnn}
\bibfield{author}{\bibinfo{person}{Yinghui Fan}, \bibinfo{person}{Xiaoxi Wu}, \bibinfo{person}{Jiying Dong}, {and} \bibinfo{person}{Zhi Qi}.} \bibinfo{year}{2019}\natexlab{}.
\newblock \showarticletitle{AxDNN: Towards the Cross-Layer Design of Approximate DNNs}. In \bibinfo{booktitle}{\emph{Asia and South Pacific Design Automation Conference (ASP-DAC)}}. \bibinfo{pages}{317--322}.
\newblock


\bibitem[Fang et~al\mbox{.}(2012)]%
        {2012_Fang_ATS}
\bibfield{author}{\bibinfo{person}{Yuntan Fang}, \bibinfo{person}{Huawei Li}, {and} \bibinfo{person}{Xiaowei Li}.} \bibinfo{year}{2012}\natexlab{}.
\newblock \showarticletitle{Soft{PCM}: Enhancing Energy Efficiency and Lifetime of Phase Change Memory in Video Applications via Approximate Write}. In \bibinfo{booktitle}{\emph{IEEE Asian Test Symposium (ATS)}}. \bibinfo{pages}{131--136}.
\newblock


\bibitem[Felzmann et~al\mbox{.}(2020)]%
        {riscv:control}
\bibfield{author}{\bibinfo{person}{Isaías Felzmann}, \bibinfo{person}{João~Fabrício Filho}, {and} \bibinfo{person}{Lucas Wanner}.} \bibinfo{year}{2020}\natexlab{}.
\newblock \showarticletitle{Risk-5: Controlled Approximations for RISC-V}.
\newblock \bibinfo{journal}{\emph{IEEE Trans. on Computer-Aided Design of Integrated Circuits and Systems}} \bibinfo{volume}{39}, \bibinfo{number}{11} (\bibinfo{year}{2020}), \bibinfo{pages}{4052--4063}.
\newblock


\bibitem[Fernando et~al\mbox{.}(2019)]%
        {2019_Fernando_ACMpapl}
\bibfield{author}{\bibinfo{person}{Vimuth Fernando}, \bibinfo{person}{Keyur Joshi}, {and} \bibinfo{person}{Sasa Misailovic}.} \bibinfo{year}{2019}\natexlab{}.
\newblock \showarticletitle{Verifying Safety and Accuracy of Approximate Parallel Programs via Canonical Sequentialization}.
\newblock \bibinfo{journal}{\emph{Proceedings of the ACM on Programming Languages}} \bibinfo{number}{119} (\bibinfo{year}{2019}), \bibinfo{pages}{1--29}.
\newblock


\bibitem[Frustaci et~al\mbox{.}(2016)]%
        {2016_Frustaci_TVLSI}
\bibfield{author}{\bibinfo{person}{Fabio Frustaci}, \bibinfo{person}{David Blaauw}, \bibinfo{person}{Dennis Sylvester}, {and} \bibinfo{person}{Massimo Alioto}.} \bibinfo{year}{2016}\natexlab{}.
\newblock \showarticletitle{Approximate {SRAM}s With Dynamic Energy-Quality Management}.
\newblock \bibinfo{journal}{\emph{IEEE Trans. on Very Large Scale Integration (VLSI) Systems}} \bibinfo{volume}{24}, \bibinfo{number}{6} (\bibinfo{year}{2016}), \bibinfo{pages}{2128--2141}.
\newblock


\bibitem[Frustaci et~al\mbox{.}(2020)]%
        {2020_Frustaci_IEEEtcasii}
\bibfield{author}{\bibinfo{person}{Fabio Frustaci}, \bibinfo{person}{Stefania Perri}, \bibinfo{person}{Pasquale Corsonello}, {and} \bibinfo{person}{Massimo Alioto}.} \bibinfo{year}{2020}\natexlab{}.
\newblock \showarticletitle{Approximate Multipliers With Dynamic Truncation for Energy Reduction via Graceful Quality Degradation}.
\newblock \bibinfo{journal}{\emph{IEEE Trans. on Circuits and Systems II: Express Briefs}} \bibinfo{volume}{67}, \bibinfo{number}{12} (\bibinfo{year}{2020}), \bibinfo{pages}{3427--3431}.
\newblock


\bibitem[Ganapathy et~al\mbox{.}(2020)]%
        {ganapathy2020dyvedeep}
\bibfield{author}{\bibinfo{person}{Sanjay Ganapathy}, \bibinfo{person}{Swagath Venkataramani}, \bibinfo{person}{Giridhur Sriraman}, \bibinfo{person}{Balaraman Ravindran}, {and} \bibinfo{person}{Anand Raghunathan}.} \bibinfo{year}{2020}\natexlab{}.
\newblock \showarticletitle{DyVEDeep: Dynamic Variable Effort Deep Neural Networks}.
\newblock \bibinfo{journal}{\emph{ACM Trans. on Embedded Computing Systems}} \bibinfo{volume}{19}, \bibinfo{number}{3} (\bibinfo{year}{2020}), \bibinfo{pages}{1--24}.
\newblock


\bibitem[Georgis et~al\mbox{.}(2019)]%
        {georgis}
\bibfield{author}{\bibinfo{person}{Georgios Georgis}, \bibinfo{person}{George Lentaris}, {and} \bibinfo{person}{Dionysios Reisis}.} \bibinfo{year}{2019}\natexlab{}.
\newblock \showarticletitle{{Acceleration Techniques and Evaluation on Multi-Core CPU, GPU and FPGA for Image Processing and Super-Resolution}}.
\newblock \bibinfo{journal}{\emph{Springer Journal of Real-Time Image Processing}} \bibinfo{volume}{16}, \bibinfo{number}{4} (\bibinfo{year}{2019}), \bibinfo{pages}{1207--1234}.
\newblock


\bibitem[Ghosh et~al\mbox{.}(2020)]%
        {ghosh2020approximate}
\bibfield{author}{\bibinfo{person}{Soumendu~Kumar Ghosh}, \bibinfo{person}{Arnab Raha}, {and} \bibinfo{person}{Vijay Raghunathan}.} \bibinfo{year}{2020}\natexlab{}.
\newblock \showarticletitle{Approximate Inference Systems (AxIS) End-to-End Approximations for Energy-Efficient Inference at the Edge}. In \bibinfo{booktitle}{\emph{ACM/IEEE Int'l. Symposium on Low Power Electronics and Design (ISLPED)}}. \bibinfo{pages}{7--12}.
\newblock


\bibitem[Ghosh et~al\mbox{.}(2023)]%
        {raha_axis_ext}
\bibfield{author}{\bibinfo{person}{Soumendu~Kumar Ghosh}, \bibinfo{person}{Arnab Raha}, {and} \bibinfo{person}{Vijay Raghunathan}.} \bibinfo{year}{2023}\natexlab{}.
\newblock \showarticletitle{Energy-Efficient Approximate Edge Inference Systems}.
\newblock \bibinfo{journal}{\emph{ACM Trans. on Embedded Computing Systems}} \bibinfo{volume}{22}, \bibinfo{number}{4} (\bibinfo{year}{2023}), \bibinfo{pages}{1--50}.
\newblock


\bibitem[Goiri et~al\mbox{.}(2015)]%
        {2015_Goiri_ASPLOS}
\bibfield{author}{\bibinfo{person}{Inigo Goiri}, \bibinfo{person}{Ricardo Bianchini}, \bibinfo{person}{Santosh Nagarakatte}, {and} \bibinfo{person}{Thu~D. Nguyen}.} \bibinfo{year}{2015}\natexlab{}.
\newblock \showarticletitle{Approx{H}adoop: Bringing Approximations to {M}ap{R}educe Frameworks}. In \bibinfo{booktitle}{\emph{ACM Int'l. Conference on Architectural Support for Programming Languages and Operating Systems (ASPLOS)}}. \bibinfo{pages}{383–397}.
\newblock


\bibitem[Gong et~al\mbox{.}(2023)]%
        {gong2023approxtrain}
\bibfield{author}{\bibinfo{person}{Jing Gong}, \bibinfo{person}{Hassaan Saadat}, \bibinfo{person}{Hasindu Gamaarachchi}, \bibinfo{person}{Haris Javaid}, \bibinfo{person}{Xiaobo~Sharon Hu}, {and} \bibinfo{person}{Sri Parameswaran}.} \bibinfo{year}{2023}\natexlab{}.
\newblock \showarticletitle{ApproxTrain: Fast Simulation of Approximate Multipliers for DNN Training and Inference}.
\newblock \bibinfo{journal}{\emph{IEEE Trans. on Computer-Aided Design of Integrated Circuits and Systems}} \bibinfo{volume}{42}, \bibinfo{number}{11} (\bibinfo{year}{2023}), \bibinfo{pages}{3505--3518}.
\newblock


\bibitem[Gong et~al\mbox{.}(2019)]%
        {gong2019ara}
\bibfield{author}{\bibinfo{person}{Yu Gong}, \bibinfo{person}{Bo Liu}, \bibinfo{person}{Wei Ge}, {and} \bibinfo{person}{Longxing Shi}.} \bibinfo{year}{2019}\natexlab{}.
\newblock \showarticletitle{ARA: Cross-Layer Approximate Computing Framework based Reconfigurable Architecture for CNNs}.
\newblock \bibinfo{journal}{\emph{Elsevier Microelectronics Journal}}  \bibinfo{volume}{87} (\bibinfo{year}{2019}), \bibinfo{pages}{33--44}.
\newblock


\bibitem[Guo et~al\mbox{.}(2022)]%
        {guo2022ant}
\bibfield{author}{\bibinfo{person}{Cong Guo} {et~al\mbox{.}}} \bibinfo{year}{2022}\natexlab{}.
\newblock \showarticletitle{ANT: Exploiting Adaptive Numerical Data Type for Low-Bit Deep Neural Network Quantization}. In \bibinfo{booktitle}{\emph{IEEE/ACM Int'l. Symposium on Microarchitecture (MICRO)}}. \bibinfo{pages}{1414--1433}.
\newblock


\bibitem[Guo and Rubio-Gonz\'{a}lez(2018)]%
        {2018_Guo_ISSTA}
\bibfield{author}{\bibinfo{person}{Hui Guo} {and} \bibinfo{person}{Cindy Rubio-Gonz\'{a}lez}.} \bibinfo{year}{2018}\natexlab{}.
\newblock \showarticletitle{Exploiting Community Structure for Floating-Point Precision Tuning}. In \bibinfo{booktitle}{\emph{ACM SIGSOFT Int'l. Symposium on Software Testing and Analysis (ISSTA)}}. \bibinfo{pages}{333–343}.
\newblock


\bibitem[Gupta et~al\mbox{.}(2013)]%
        {2013_Gupta_IEEEtcad}
\bibfield{author}{\bibinfo{person}{Vaibhav Gupta}, \bibinfo{person}{Debabrata Mohapatra}, \bibinfo{person}{Anand Raghunathan}, {and} \bibinfo{person}{Kaushik Roy}.} \bibinfo{year}{2013}\natexlab{}.
\newblock \showarticletitle{Low-Power Digital Signal Processing Using Approximate Adders}.
\newblock \bibinfo{journal}{\emph{IEEE Trans. on Computer-Aided Design of Integrated Circuits and Systems}} \bibinfo{volume}{32}, \bibinfo{number}{1} (\bibinfo{year}{2013}), \bibinfo{pages}{124--137}.
\newblock


\bibitem[Guthaus et~al\mbox{.}(2001)]%
        {2001_Guthaus_WWC}
\bibfield{author}{\bibinfo{person}{Matthew~R. Guthaus}, \bibinfo{person}{Jeffrey~S. Ringenberg}, \bibinfo{person}{Dan Ernst}, \bibinfo{person}{Todd~M. Austin}, \bibinfo{person}{Trevor Mudge}, {and} \bibinfo{person}{Richard~B. Brown}.} \bibinfo{year}{2001}\natexlab{}.
\newblock \showarticletitle{MiBench: A Free, Commercially Representative Embedded Benchmark Suite}. In \bibinfo{booktitle}{\emph{IEEE Int'l. Workshop on Workload Characterization (WWC)}}. \bibinfo{pages}{3--14}.
\newblock


\bibitem[Han and Orshansky(2013)]%
        {2013_Han_ETS}
\bibfield{author}{\bibinfo{person}{Jie Han} {and} \bibinfo{person}{Michael Orshansky}.} \bibinfo{year}{2013}\natexlab{}.
\newblock \showarticletitle{Approximate Computing: An Emerging Paradigm for Energy-Efficient Design}. In \bibinfo{booktitle}{\emph{IEEE European Test Symposium (ETS)}}. \bibinfo{pages}{1--6}.
\newblock


\bibitem[Han et~al\mbox{.}(2016a)]%
        {han2016eie}
\bibfield{author}{\bibinfo{person}{Song Han}, \bibinfo{person}{Xingyu Liu}, \bibinfo{person}{Huizi Mao}, \bibinfo{person}{Jing Pu}, \bibinfo{person}{Ardavan Pedram}, \bibinfo{person}{Mark~A Horowitz}, {and} \bibinfo{person}{William~J Dally}.} \bibinfo{year}{2016}\natexlab{a}.
\newblock \showarticletitle{{EIE}: Efficient Inference Engine on Compressed Deep Neural Network}.
\newblock \bibinfo{journal}{\emph{ACM SIGARCH Computer Architecture News}} \bibinfo{volume}{44}, \bibinfo{number}{3} (\bibinfo{year}{2016}), \bibinfo{pages}{243--254}.
\newblock


\bibitem[Han et~al\mbox{.}(2016b)]%
        {han2015deep}
\bibfield{author}{\bibinfo{person}{Song Han}, \bibinfo{person}{Huizi Mao}, {and} \bibinfo{person}{William~J Dally}.} \bibinfo{year}{2016}\natexlab{b}.
\newblock \showarticletitle{Deep Compression: Compressing Deep Neural Networks with Pruning, Trained Quantization and Huffman Coding}. In \bibinfo{booktitle}{\emph{Int'l. Conference on Learning Representations (ICLR)}}. \bibinfo{pages}{1--14}.
\newblock


\bibitem[Han et~al\mbox{.}(2015)]%
        {han2015learning}
\bibfield{author}{\bibinfo{person}{Song Han}, \bibinfo{person}{Jeff Pool}, \bibinfo{person}{John Tran}, {and} \bibinfo{person}{William Dally}.} \bibinfo{year}{2015}\natexlab{}.
\newblock \showarticletitle{Learning both Weights and Connections for Efficient Neural Networks}. In \bibinfo{booktitle}{\emph{Advances in Neural Information Processing Systems (NIPS)}}. \bibinfo{pages}{1135–1143}.
\newblock


\bibitem[Hanif et~al\mbox{.}(2019)]%
        {hanif2019cann}
\bibfield{author}{\bibinfo{person}{Muhammad~Abdullah Hanif}, \bibinfo{person}{Faiq Khalid}, {and} \bibinfo{person}{Muhammad Shafique}.} \bibinfo{year}{2019}\natexlab{}.
\newblock \showarticletitle{CANN: Curable Approximations for High-Performance Deep Neural Network Accelerators}. In \bibinfo{booktitle}{\emph{Design Automation Conference (DAC)}}. \bibinfo{pages}{1--6}.
\newblock


\bibitem[Hanif et~al\mbox{.}(2022)]%
        {hanif2022conlocnn}
\bibfield{author}{\bibinfo{person}{Muhammad~Abdullah Hanif}, \bibinfo{person}{Giuseppe~Maria Sarda}, \bibinfo{person}{Alberto Marchisio}, \bibinfo{person}{Guido Masera}, \bibinfo{person}{Maurizio Martina}, {and} \bibinfo{person}{Muhammad Shafique}.} \bibinfo{year}{2022}\natexlab{}.
\newblock \showarticletitle{CoNLoCNN: Exploiting Correlation and Non-Uniform Quantization for Energy-Efficient Low-precision Deep Convolutional Neural Networks}. In \bibinfo{booktitle}{\emph{Int'l. Joint Conference on Neural Networks (IJCNN)}}. \bibinfo{pages}{1--8}.
\newblock


\bibitem[Hanif and Shafique(2022)]%
        {hanif2022cross}
\bibfield{author}{\bibinfo{person}{Muhammad~Abdullah Hanif} {and} \bibinfo{person}{Muhammad Shafique}.} \bibinfo{year}{2022}\natexlab{}.
\newblock \showarticletitle{A Cross-Layer Approach Towards Developing Efficient Embedded Deep Learning Systems}.
\newblock \bibinfo{journal}{\emph{Elsevier Microprocessors and Microsystems}}  \bibinfo{volume}{88} (\bibinfo{year}{2022}), \bibinfo{pages}{103609}.
\newblock


\bibitem[Hashemi et~al\mbox{.}(2015)]%
        {2015_Hashemi_ICCAD}
\bibfield{author}{\bibinfo{person}{Soheil Hashemi}, \bibinfo{person}{R.~Iris Bahar}, {and} \bibinfo{person}{Sherief Reda}.} \bibinfo{year}{2015}\natexlab{}.
\newblock \showarticletitle{{DRUM}: A Dynamic Range Unbiased Multiplier for Approximate Applications}. In \bibinfo{booktitle}{\emph{Int'l. Conference on Computer-Aided Design (ICCAD)}}. \bibinfo{pages}{418--425}.
\newblock


\bibitem[Hashemi et~al\mbox{.}(2016)]%
        {2016_Hashemi_DAC}
\bibfield{author}{\bibinfo{person}{Soheil Hashemi}, \bibinfo{person}{R.~Iris Bahar}, {and} \bibinfo{person}{Sherief Reda}.} \bibinfo{year}{2016}\natexlab{}.
\newblock \showarticletitle{A Low-Power Dynamic Divider for Approximate Applications}. In \bibinfo{booktitle}{\emph{Design Automation Conference (DAC)}}. \bibinfo{pages}{1--6}.
\newblock


\bibitem[Hashemi et~al\mbox{.}(2018b)]%
        {hashemi2018approximate}
\bibfield{author}{\bibinfo{person}{Soheil Hashemi}, \bibinfo{person}{Hokchhay Tann}, \bibinfo{person}{Francesco Buttafuoco}, {and} \bibinfo{person}{Sherief Reda}.} \bibinfo{year}{2018}\natexlab{b}.
\newblock \showarticletitle{Approximate Computing for Biometric Security Systems: A Case Study on Iris Scanning}. In \bibinfo{booktitle}{\emph{Design, Automation \& Test in Europe (DATE)}}. \bibinfo{pages}{319--324}.
\newblock


\bibitem[Hashemi et~al\mbox{.}(2018a)]%
        {2018_Hashemi_DAC}
\bibfield{author}{\bibinfo{person}{Soheil Hashemi}, \bibinfo{person}{Hokchhay Tann}, {and} \bibinfo{person}{Sherief Reda}.} \bibinfo{year}{2018}\natexlab{a}.
\newblock \showarticletitle{{BLASYS}: Approximate Logic Synthesis Using Boolean Matrix Factorization}. In \bibinfo{booktitle}{\emph{Design Automation Conference (DAC)}}. \bibinfo{pages}{1--6}.
\newblock


\bibitem[He et~al\mbox{.}(2017)]%
        {he2017channel}
\bibfield{author}{\bibinfo{person}{Yihui He}, \bibinfo{person}{Xiangyu Zhang}, {and} \bibinfo{person}{Jian Sun}.} \bibinfo{year}{2017}\natexlab{}.
\newblock \showarticletitle{Channel Pruning for Accelerating Very Deep Neural Networks}. In \bibinfo{booktitle}{\emph{IEEE Int'l. Conference on Computer Vision (ICCV)}}. \bibinfo{pages}{1389--1397}.
\newblock


\bibitem[He et~al\mbox{.}(2022)]%
        {he2022filter}
\bibfield{author}{\bibinfo{person}{Zhiqiang He}, \bibinfo{person}{Yaguan Qian}, \bibinfo{person}{Yuqi Wang}, \bibinfo{person}{Bin Wang}, \bibinfo{person}{Xiaohui Guan}, \bibinfo{person}{Zhaoquan Gu}, \bibinfo{person}{Xiang Ling}, \bibinfo{person}{Shaoning Zeng}, \bibinfo{person}{Haijiang Wang}, {and} \bibinfo{person}{Wujie Zhou}.} \bibinfo{year}{2022}\natexlab{}.
\newblock \showarticletitle{Filter Pruning via Feature Discrimination in Deep Neural Networks}. In \bibinfo{booktitle}{\emph{European Conference on Computer Vision (ECCV)}}. \bibinfo{pages}{245--261}.
\newblock


\bibitem[Hoffmann et~al\mbox{.}(2009)]%
        {2009_Hoffmann_MIT}
\bibfield{author}{\bibinfo{person}{Henry Hoffmann}, \bibinfo{person}{Sasa Misailovic}, \bibinfo{person}{Stelios Sidiroglou}, \bibinfo{person}{Anant Agarwal}, {and} \bibinfo{person}{Martin~C. Rinard}.} \bibinfo{year}{2009}\natexlab{}.
\newblock \showarticletitle{Using Code Perforation to Improve Performance, Reduce Energy Consumption, and Respond to Failures}.
\newblock \bibinfo{journal}{\emph{Massachusetts Institute of Technology Technical Report}} (\bibinfo{year}{2009}), \bibinfo{pages}{1--21}.
\newblock


\bibitem[Hu et~al\mbox{.}(2019)]%
        {2019_Hu_MASCOTS}
\bibfield{author}{\bibinfo{person}{Guangyan Hu}, \bibinfo{person}{Sandro Rigo}, \bibinfo{person}{Desheng Zhang}, {and} \bibinfo{person}{Thu Nguyen}.} \bibinfo{year}{2019}\natexlab{}.
\newblock \showarticletitle{Approximation with Error Bounds in Spark}. In \bibinfo{booktitle}{\emph{IEEE Int'l. Symposium on Modeling, Analysis, and Simulation of Computer and Telecommunication Systems}}. \bibinfo{pages}{61--73}.
\newblock


\bibitem[Hu et~al\mbox{.}(2016)]%
        {hu2016network}
\bibfield{author}{\bibinfo{person}{Hengyuan Hu}, \bibinfo{person}{Rui Peng}, \bibinfo{person}{Yu-Wing Tai}, {and} \bibinfo{person}{Chi-Keung Tang}.} \bibinfo{year}{2016}\natexlab{}.
\newblock \showarticletitle{Network Trimming: A Data-Driven Neuron Pruning Approach towards Efficient Deep Architectures}.
\newblock \bibinfo{journal}{\emph{arXiv preprint arXiv:1607.03250}} (\bibinfo{year}{2016}), \bibinfo{pages}{1--9}.
\newblock


\bibitem[Hubara et~al\mbox{.}(2016)]%
        {hubara2016binarized}
\bibfield{author}{\bibinfo{person}{Itay Hubara}, \bibinfo{person}{Matthieu Courbariaux}, \bibinfo{person}{Daniel Soudry}, \bibinfo{person}{Ran El-Yaniv}, {and} \bibinfo{person}{Yoshua Bengio}.} \bibinfo{year}{2016}\natexlab{}.
\newblock \showarticletitle{Binarized Neural Networks}.
\newblock \bibinfo{journal}{\emph{Advances in Neural Information Processing Systems}}  \bibinfo{volume}{29} (\bibinfo{year}{2016}), \bibinfo{pages}{1--9}.
\newblock


\bibitem[Imani et~al\mbox{.}(2019)]%
        {2019_Imani_Date}
\bibfield{author}{\bibinfo{person}{Mohsen Imani}, \bibinfo{person}{Ricardo Garcia}, \bibinfo{person}{Andrew Huang}, {and} \bibinfo{person}{Tajana Rosing}.} \bibinfo{year}{2019}\natexlab{}.
\newblock \showarticletitle{{CADE}: Configurable Approximate Divider for Energy Efficiency}. In \bibinfo{booktitle}{\emph{Design, Automation \& Test in Europe (DATE)}}. \bibinfo{pages}{586--589}.
\newblock


\bibitem[{IoT Analytics}(2023)]%
        {iotanal}
\bibfield{author}{\bibinfo{person}{{IoT Analytics}}.} \bibinfo{year}{2023}\natexlab{}.
\newblock \bibinfo{booktitle}{\emph{State of IoT 2023}}.
\newblock
\urldef\tempurl%
\url{https://iot-analytics.com/number-connected-iot-devices/}
\showURL{%
\tempurl}


\bibitem[Jacob et~al\mbox{.}(2018)]%
        {jacob2018quantization}
\bibfield{author}{\bibinfo{person}{Benoit Jacob} {et~al\mbox{.}}} \bibinfo{year}{2018}\natexlab{}.
\newblock \showarticletitle{Quantization and Training of Neural Networks for Efficient Integer-Arithmetic-Only Inference}. In \bibinfo{booktitle}{\emph{IEEE Conference on Computer Vision and Pattern Recognition (CVPR)}}. \bibinfo{pages}{2704--2713}.
\newblock


\bibitem[Jain et~al\mbox{.}(2018)]%
        {jain2018compensated}
\bibfield{author}{\bibinfo{person}{Shubham Jain}, \bibinfo{person}{Swagath Venkataramani}, \bibinfo{person}{Vijayalakshmi Srinivasan}, \bibinfo{person}{Jungwook Choi}, \bibinfo{person}{Pierce Chuang}, {and} \bibinfo{person}{Leland Chang}.} \bibinfo{year}{2018}\natexlab{}.
\newblock \showarticletitle{Compensated-DNN: Energy Efficient Low-Precision Deep Neural Networks by Compensating Quantization Errors}. In \bibinfo{booktitle}{\emph{Design Automation Conference (DAC)}}. \bibinfo{pages}{1--6}.
\newblock


\bibitem[Jain et~al\mbox{.}(2019)]%
        {jain2019biscaled}
\bibfield{author}{\bibinfo{person}{Shubham Jain}, \bibinfo{person}{Swagath Venkataramani}, \bibinfo{person}{Vijayalakshmi Srinivasan}, \bibinfo{person}{Jungwook Choi}, \bibinfo{person}{Kailash Gopalakrishnan}, {and} \bibinfo{person}{Leland Chang}.} \bibinfo{year}{2019}\natexlab{}.
\newblock \showarticletitle{BiScaled-DNN: Quantizing Long-tailed Datastructures with Two Scale Factors for Deep Neural Networks}. In \bibinfo{booktitle}{\emph{Design Automation Conference (DAC)}}. \bibinfo{pages}{1--6}.
\newblock


\bibitem[Ji et~al\mbox{.}(2020)]%
        {ji2020error}
\bibfield{author}{\bibinfo{person}{Daehan Ji}, \bibinfo{person}{Dongyeob Shin}, {and} \bibinfo{person}{Jongsun Park}.} \bibinfo{year}{2020}\natexlab{}.
\newblock \showarticletitle{An Error Compensation Technique for Low-Voltage DNN Accelerators}.
\newblock \bibinfo{journal}{\emph{IEEE Trans. on Very Large Scale Integration (VLSI) Systems}} \bibinfo{volume}{29}, \bibinfo{number}{2} (\bibinfo{year}{2020}), \bibinfo{pages}{397--408}.
\newblock


\bibitem[Jiang et~al\mbox{.}(2016)]%
        {2016_Jiang_IEEEtc}
\bibfield{author}{\bibinfo{person}{Honglan Jiang}, \bibinfo{person}{Jie Han}, \bibinfo{person}{Fei Qiao}, {and} \bibinfo{person}{Fabrizio Lombardi}.} \bibinfo{year}{2016}\natexlab{}.
\newblock \showarticletitle{Approximate Radix-8 Booth Multipliers for Low-Power and High-Performance Operation}.
\newblock \bibinfo{journal}{\emph{IEEE Trans. on Computers}} \bibinfo{volume}{65}, \bibinfo{number}{8} (\bibinfo{year}{2016}), \bibinfo{pages}{2638--2644}.
\newblock


\bibitem[Jiang et~al\mbox{.}(2019)]%
        {2019_Jiang_IEEEtc}
\bibfield{author}{\bibinfo{person}{Honglan Jiang}, \bibinfo{person}{Leibo Liu}, \bibinfo{person}{Fabrizio Lombardi}, {and} \bibinfo{person}{Jie Han}.} \bibinfo{year}{2019}\natexlab{}.
\newblock \showarticletitle{Low-Power Unsigned Divider and Square Root Circuit Designs Using Adaptive Approximation}.
\newblock \bibinfo{journal}{\emph{IEEE Trans. on Computers}} \bibinfo{volume}{68}, \bibinfo{number}{11} (\bibinfo{year}{2019}), \bibinfo{pages}{1635--1646}.
\newblock


\bibitem[Jiao et~al\mbox{.}(2020a)]%
        {2020_Jiao_IEEEtcad}
\bibfield{author}{\bibinfo{person}{Xun Jiao}, \bibinfo{person}{Dongning Ma}, \bibinfo{person}{Wanli Chang}, {and} \bibinfo{person}{Yu Jiang}.} \bibinfo{year}{2020}\natexlab{a}.
\newblock \showarticletitle{{LEVAX}: An Input-Aware Learning-Based Error Model of Voltage-Scaled Functional Units}.
\newblock \bibinfo{journal}{\emph{IEEE Trans. on Computer-Aided Design of Integrated Circuits and Systems}} \bibinfo{volume}{39}, \bibinfo{number}{12} (\bibinfo{year}{2020}), \bibinfo{pages}{5032--5041}.
\newblock


\bibitem[Jiao et~al\mbox{.}(2020b)]%
        {tevot}
\bibfield{author}{\bibinfo{person}{Xun Jiao}, \bibinfo{person}{Dongning Ma}, \bibinfo{person}{Wanli Chang}, {and} \bibinfo{person}{Yu Jiang}.} \bibinfo{year}{2020}\natexlab{b}.
\newblock \showarticletitle{TEVoT: Timing Error Modeling of Functional Units under Dynamic Voltage and Temperature Variations}. In \bibinfo{booktitle}{\emph{Design Automation Conference (DAC)}}. \bibinfo{pages}{1--6}.
\newblock


\bibitem[Joshi et~al\mbox{.}(2019)]%
        {2019_Joshi_ICSE}
\bibfield{author}{\bibinfo{person}{Keyur Joshi}, \bibinfo{person}{Vimuth Fernando}, {and} \bibinfo{person}{Sasa Misailovic}.} \bibinfo{year}{2019}\natexlab{}.
\newblock \showarticletitle{Statistical Algorithmic Profiling for Randomized Approximate Programs}. In \bibinfo{booktitle}{\emph{ACM/IEEE Int'l. Conference on Software Engineering (ICSE)}}. \bibinfo{pages}{608--618}.
\newblock


\bibitem[Jouppi et~al\mbox{.}(2017)]%
        {google}
\bibfield{author}{\bibinfo{person}{Norman~P. Jouppi} {et~al\mbox{.}}} \bibinfo{year}{2017}\natexlab{}.
\newblock \showarticletitle{In-Datacenter Performance Analysis of a {T}ensor {P}rocessing {U}nit}. In \bibinfo{booktitle}{\emph{ACM/IEEE Int'l. Symposium on Computer Architecture (ISCA)}}. \bibinfo{pages}{1–12}.
\newblock


\bibitem[Jung et~al\mbox{.}(2015)]%
        {2015_Jung_MEMSYS}
\bibfield{author}{\bibinfo{person}{Matthias Jung}, \bibinfo{person}{\'{E}der Zulian}, \bibinfo{person}{Deepak~M. Mathew}, \bibinfo{person}{Matthias Herrmann}, \bibinfo{person}{Christian Brugger}, \bibinfo{person}{Christian Weis}, {and} \bibinfo{person}{Norbert Wehn}.} \bibinfo{year}{2015}\natexlab{}.
\newblock \showarticletitle{Omitting Refresh: A Case Study for Commodity and Wide {I/O DRAM}s}. In \bibinfo{booktitle}{\emph{Int'l. Symposium on Memory Systems (MEMSYS)}}. \bibinfo{pages}{85–91}.
\newblock


\bibitem[Kahng and Kang(2012)]%
        {2012_Kahng_DAC}
\bibfield{author}{\bibinfo{person}{Andrew~B. Kahng} {and} \bibinfo{person}{Seokhyeong Kang}.} \bibinfo{year}{2012}\natexlab{}.
\newblock \showarticletitle{Accuracy-Configurable Adder for Approximate Arithmetic Designs}. In \bibinfo{booktitle}{\emph{Design Automation Conference (DAC)}}. \bibinfo{pages}{820–825}.
\newblock


\bibitem[Kahng et~al\mbox{.}(2010)]%
        {vos:proc2}
\bibfield{author}{\bibinfo{person}{Andrew~B. Kahng}, \bibinfo{person}{Seokhyeong Kang}, \bibinfo{person}{Rakesh Kumar}, {and} \bibinfo{person}{John Sartori}.} \bibinfo{year}{2010}\natexlab{}.
\newblock \showarticletitle{Designing a Processor from the Ground up to Allow Voltage/Reliability Tradeoffs}. In \bibinfo{booktitle}{\emph{IEEE Int'l. Symposium on High Performance Computer Architecture}}. \bibinfo{pages}{1--11}.
\newblock


\bibitem[Kandula et~al\mbox{.}(2016)]%
        {2016_Kandula_MOD}
\bibfield{author}{\bibinfo{person}{Srikanth Kandula}, \bibinfo{person}{Anil Shanbhag}, \bibinfo{person}{Aleksandar Vitorovic}, \bibinfo{person}{Matthaios Olma}, \bibinfo{person}{Robert Grandl}, \bibinfo{person}{Surajit Chaudhuri}, {and} \bibinfo{person}{Bolin Ding}.} \bibinfo{year}{2016}\natexlab{}.
\newblock \showarticletitle{Quickr: Lazily Approximating Complex Ad-Hoc Queries in {B}ig {D}ata Clusters}. In \bibinfo{booktitle}{\emph{ACM SIGMOD Int'l. Conference on Management of Data (MOD)}}. \bibinfo{pages}{631–646}.
\newblock


\bibitem[Kanduri et~al\mbox{.}(2018)]%
        {2018_Kanduri_DAC}
\bibfield{author}{\bibinfo{person}{Anil Kanduri}, \bibinfo{person}{Antonio Miele}, \bibinfo{person}{Amir~M. Rahmani}, \bibinfo{person}{Pasi Liljeberg}, \bibinfo{person}{Cristiana Bolchini}, {and} \bibinfo{person}{Nikil Dutt}.} \bibinfo{year}{2018}\natexlab{}.
\newblock \showarticletitle{Approximation-Aware Coordinated Power/Performance Management for Heterogeneous Multi-cores}. In \bibinfo{booktitle}{\emph{Design Automation Conference (DAC)}}. \bibinfo{pages}{1--6}.
\newblock


\bibitem[Kang et~al\mbox{.}(2020)]%
        {2020_Kang_CGO}
\bibfield{author}{\bibinfo{person}{Seokwon Kang}, \bibinfo{person}{Kyunghwan Choi}, {and} \bibinfo{person}{Yongjun Park}.} \bibinfo{year}{2020}\natexlab{}.
\newblock \showarticletitle{Pre{S}caler: An Efficient System-Aware Precision Scaling Framework on Heterogeneous Systems}. In \bibinfo{booktitle}{\emph{IEEE/ACM Int'l. Symposium on Code Generation and Optimization}}. \bibinfo{pages}{280–292}.
\newblock


\bibitem[Karakoy et~al\mbox{.}(2019)]%
        {2019_Karakoy_ACMmacs}
\bibfield{author}{\bibinfo{person}{Mustafa Karakoy}, \bibinfo{person}{Orhan Kislal}, \bibinfo{person}{Xulong Tang}, \bibinfo{person}{Mahmut~Taylan Kandemir}, {and} \bibinfo{person}{Meenakshi Arunachalam}.} \bibinfo{year}{2019}\natexlab{}.
\newblock \showarticletitle{Architecture-Aware Approximate Computing}.
\newblock \bibinfo{journal}{\emph{Proceedings of the ACM on Measurement and Analysis of Computing Systems}} \bibinfo{volume}{3}, \bibinfo{number}{2} (\bibinfo{year}{2019}), \bibinfo{pages}{1--24}.
\newblock


\bibitem[Keramidas et~al\mbox{.}(2015)]%
        {2015_Keramidas_WAPCO}
\bibfield{author}{\bibinfo{person}{Georgios Keramidas}, \bibinfo{person}{Chrysa Kokkala}, {and} \bibinfo{person}{Iakovos Stamoulis}.} \bibinfo{year}{2015}\natexlab{}.
\newblock \showarticletitle{Clumsy Value Cache: An Approximate Memoization Technique for Mobile {GPU} Fragment Shaders}. In \bibinfo{booktitle}{\emph{Workshop on Approximate Computing (WAPCO)}}. \bibinfo{pages}{1--6}.
\newblock


\bibitem[Kim et~al\mbox{.}(2018a)]%
        {kim2018efficient}
\bibfield{author}{\bibinfo{person}{Min~Soo Kim}, \bibinfo{person}{Alberto~A Del~Barrio}, \bibinfo{person}{Leonardo~Tavares Oliveira}, \bibinfo{person}{Roman Hermida}, {and} \bibinfo{person}{Nader Bagherzadeh}.} \bibinfo{year}{2018}\natexlab{a}.
\newblock \showarticletitle{Efficient Mitchell’s Approximate Log Multipliers for Convolutional Neural Networks}.
\newblock \bibinfo{journal}{\emph{IEEE Trans. on Computers}} \bibinfo{volume}{68}, \bibinfo{number}{5} (\bibinfo{year}{2018}), \bibinfo{pages}{660--675}.
\newblock


\bibitem[Kim et~al\mbox{.}(2018b)]%
        {kim2018matic}
\bibfield{author}{\bibinfo{person}{Sung Kim}, \bibinfo{person}{Patrick Howe}, \bibinfo{person}{Thierry Moreau}, \bibinfo{person}{Armin Alaghi}, \bibinfo{person}{Luis Ceze}, {and} \bibinfo{person}{Visvesh Sathe}.} \bibinfo{year}{2018}\natexlab{b}.
\newblock \showarticletitle{MATIC: Learning Around Errors for Efficient Low-Voltage Neural Network Accelerators}. In \bibinfo{booktitle}{\emph{Design, Automation \& Test in Europe (DATE)}}. \bibinfo{pages}{1--6}.
\newblock


\bibitem[Kim et~al\mbox{.}(2021)]%
        {riscv:simil}
\bibfield{author}{\bibinfo{person}{Younghoon Kim}, \bibinfo{person}{Swagath Venkataramani}, \bibinfo{person}{Sanchari Sen}, {and} \bibinfo{person}{Anand Raghunathan}.} \bibinfo{year}{2021}\natexlab{}.
\newblock \showarticletitle{Value Similarity Extensions for Approximate Computing in General-Purpose Processors}. In \bibinfo{booktitle}{\emph{Design, Automation \& Test in Europe (DATE)}}. \bibinfo{pages}{481--486}.
\newblock


\bibitem[Kim et~al\mbox{.}(2013)]%
        {2013_Kim_ICCAD}
\bibfield{author}{\bibinfo{person}{Yongtae Kim}, \bibinfo{person}{Yong Zhang}, {and} \bibinfo{person}{Peng Li}.} \bibinfo{year}{2013}\natexlab{}.
\newblock \showarticletitle{An Energy Efficient Approximate Adder with Carry Skip for Error Resilient Neuromorphic {VLSI} Systems}. In \bibinfo{booktitle}{\emph{Int'l. Conference on Computer-Aided Design (ICCAD)}}. \bibinfo{pages}{130--137}.
\newblock


\bibitem[Kislal and Kandemir(2018)]%
        {2018_Kislal_ELSEclss}
\bibfield{author}{\bibinfo{person}{Orhan Kislal} {and} \bibinfo{person}{Mahmut~T. Kandemir}.} \bibinfo{year}{2018}\natexlab{}.
\newblock \showarticletitle{Data Access Skipping for Recursive Partitioning Methods}.
\newblock \bibinfo{journal}{\emph{Elsevier Computer Languages, Systems \& Structures}}  \bibinfo{volume}{53} (\bibinfo{year}{2018}), \bibinfo{pages}{143--162}.
\newblock


\bibitem[Koppula et~al\mbox{.}(2019)]%
        {koppula2019eden}
\bibfield{author}{\bibinfo{person}{Skanda Koppula}, \bibinfo{person}{Lois Orosa}, \bibinfo{person}{A~Giray Ya{\u{g}}l{\i}k{\c{c}}{\i}}, \bibinfo{person}{Roknoddin Azizi}, \bibinfo{person}{Taha Shahroodi}, \bibinfo{person}{Konstantinos Kanellopoulos}, {and} \bibinfo{person}{Onur Mutlu}.} \bibinfo{year}{2019}\natexlab{}.
\newblock \showarticletitle{EDEN: Enabling Energy-Efficient, High-Performance Deep Neural Network Inference Using Approximate DRAM}. In \bibinfo{booktitle}{\emph{IEEE/ACM Int'l. Symposium on Microarchitecture (MICRO)}}. \bibinfo{pages}{166--181}.
\newblock


\bibitem[Krishnamoorthi(2018)]%
        {krishnamoorthi2018quantizing}
\bibfield{author}{\bibinfo{person}{Raghuraman Krishnamoorthi}.} \bibinfo{year}{2018}\natexlab{}.
\newblock \showarticletitle{Quantizing Deep Convolutional Networks for Efficient Inference: A Whitepaper}.
\newblock \bibinfo{journal}{\emph{arXiv preprint arXiv:1806.08342}} (\bibinfo{year}{2018}), \bibinfo{pages}{1--36}.
\newblock


\bibitem[Krishnan et~al\mbox{.}(2016)]%
        {2016_Krishnan_WWW}
\bibfield{author}{\bibinfo{person}{Dhanya~R. Krishnan}, \bibinfo{person}{Do~Le Quoc}, \bibinfo{person}{Pramod Bhatotia}, \bibinfo{person}{Christof Fetzer}, {and} \bibinfo{person}{Rodrigo Rodrigues}.} \bibinfo{year}{2016}\natexlab{}.
\newblock \showarticletitle{Inc{A}pprox: A Data Analytics System for Incremental Approximate Computing}. In \bibinfo{booktitle}{\emph{Int'l. Conf.on World Wide Web}}. \bibinfo{pages}{1133–1144}.
\newblock


\bibitem[Kumar et~al\mbox{.}(2009)]%
        {2009_kumar_SQED}
\bibfield{author}{\bibinfo{person}{Animesh Kumar}, \bibinfo{person}{Jan Rabaey}, {and} \bibinfo{person}{Kannan Ramchandran}.} \bibinfo{year}{2009}\natexlab{}.
\newblock \showarticletitle{{SRAM} supply voltage scaling: A reliability perspective}. In \bibinfo{booktitle}{\emph{Int'l. Symposium on Quality Electronic Design (ISQED)}}. \bibinfo{pages}{782--787}.
\newblock


\bibitem[Kurdahi et~al\mbox{.}(2010)]%
        {2010_Kurdahi_IEEEtvlsi}
\bibfield{author}{\bibinfo{person}{Fadi~J. Kurdahi}, \bibinfo{person}{Ahmed Eltawil}, \bibinfo{person}{Kang Yi}, \bibinfo{person}{Stanley Cheng}, {and} \bibinfo{person}{Amin Khajeh}.} \bibinfo{year}{2010}\natexlab{}.
\newblock \showarticletitle{Low-Power Multimedia System Design by Aggressive Voltage Scaling}.
\newblock \bibinfo{journal}{\emph{IEEE Trans. on Very Large Scale Integration (VLSI) Systems}} \bibinfo{volume}{18}, \bibinfo{number}{5} (\bibinfo{year}{2010}), \bibinfo{pages}{852--856}.
\newblock


\bibitem[Kyrkou et~al\mbox{.}(2018)]%
        {kyrkou2018dronet}
\bibfield{author}{\bibinfo{person}{Christos Kyrkou}, \bibinfo{person}{George Plastiras}, \bibinfo{person}{Theocharis Theocharides}, \bibinfo{person}{Stylianos~I Venieris}, {and} \bibinfo{person}{Christos-Savvas Bouganis}.} \bibinfo{year}{2018}\natexlab{}.
\newblock \showarticletitle{DroNet: Efficient Convolutional Neural Network Detector for Real-Time UAV Applications}. In \bibinfo{booktitle}{\emph{Design, Automation \& Test in Europe (DATE)}}. \bibinfo{pages}{967--972}.
\newblock


\bibitem[Laguna et~al\mbox{.}(2019)]%
        {2019_Laguna_HPC}
\bibfield{author}{\bibinfo{person}{Ignacio Laguna}, \bibinfo{person}{Paul~C. Wood}, \bibinfo{person}{Ranvijay Singh}, {and} \bibinfo{person}{Saurabh Bagchi}.} \bibinfo{year}{2019}\natexlab{}.
\newblock \showarticletitle{{GPUM}ixer: Performance-Driven Floating-Point Tuning for {GPU} Scientific Applications}. In \bibinfo{booktitle}{\emph{ISC Int'l. Conference on High Performance Computing (HPC)}}. \bibinfo{pages}{227--246}.
\newblock


\bibitem[Lam and Hollingsworth(2018)]%
        {2018_Lam_SAGE}
\bibfield{author}{\bibinfo{person}{Michael~O. Lam} {and} \bibinfo{person}{Jeffrey~K. Hollingsworth}.} \bibinfo{year}{2018}\natexlab{}.
\newblock \showarticletitle{Fine-Grained Floating-Point Precision Analysis}.
\newblock \bibinfo{journal}{\emph{SAGE Int'l. Journal of High Performance Computing Applications}} \bibinfo{volume}{32}, \bibinfo{number}{2} (\bibinfo{year}{2018}), \bibinfo{pages}{231–245}.
\newblock


\bibitem[Lam et~al\mbox{.}(2013)]%
        {2013_Lam_ICS}
\bibfield{author}{\bibinfo{person}{Michael~O. Lam}, \bibinfo{person}{Jeffrey~K. Hollingsworth}, \bibinfo{person}{Bronis~R. de Supinski}, {and} \bibinfo{person}{Matthew~P. Legendre}.} \bibinfo{year}{2013}\natexlab{}.
\newblock \showarticletitle{Automatically Adapting Programs for Mixed-Precision Floating-Point Computation}. In \bibinfo{booktitle}{\emph{ACM Int'l. Conference on Supercomputing (ICS)}}. \bibinfo{pages}{369–378}.
\newblock


\bibitem[Laptev et~al\mbox{.}(2012)]%
        {2012_Laptev_VLDB}
\bibfield{author}{\bibinfo{person}{Nikolay Laptev}, \bibinfo{person}{Kai Zeng}, {and} \bibinfo{person}{Carlo Zaniolo}.} \bibinfo{year}{2012}\natexlab{}.
\newblock \showarticletitle{Early Accurate Results for Advanced Analytics on {M}ap{R}educe}.
\newblock \bibinfo{journal}{\emph{Proceedings of the VLDB Endowment}} \bibinfo{volume}{5}, \bibinfo{number}{10} (\bibinfo{year}{2012}), \bibinfo{pages}{1028–1039}.
\newblock


\bibitem[Lebedev and Lempitsky(2016)]%
        {lebedev2016fast}
\bibfield{author}{\bibinfo{person}{Vadim Lebedev} {and} \bibinfo{person}{Victor Lempitsky}.} \bibinfo{year}{2016}\natexlab{}.
\newblock \showarticletitle{Fast ConvNets Using Group-wise Brain Damage}. In \bibinfo{booktitle}{\emph{IEEE Conference on Computer Vision and Pattern Recognition (CVPR)}}. \bibinfo{pages}{2554--2564}.
\newblock


\bibitem[Lee et~al\mbox{.}(1997)]%
        {1997_Chunho_MICRO}
\bibfield{author}{\bibinfo{person}{Chunho Lee}, \bibinfo{person}{Miodrag Potkonjak}, {and} \bibinfo{person}{William~Henry Mangione-Smith}.} \bibinfo{year}{1997}\natexlab{}.
\newblock \showarticletitle{Media{B}ench: A Tool for Evaluating and Synthesizing Multimedia and Communications Systems}. In \bibinfo{booktitle}{\emph{IEEE/ACM Int'l. Symposium on Microarchitecture (MICRO)}}. \bibinfo{pages}{330--335}.
\newblock


\bibitem[Lee et~al\mbox{.}(2017)]%
        {2017_Lee_DATE}
\bibfield{author}{\bibinfo{person}{Seogoo Lee}, \bibinfo{person}{Lizy~K. John}, {and} \bibinfo{person}{Andreas Gerstlauer}.} \bibinfo{year}{2017}\natexlab{}.
\newblock \showarticletitle{High-Level Synthesis of Approximate Hardware under Joint Precision and Voltage Scaling}. In \bibinfo{booktitle}{\emph{Design, Automation \& Test in Europe (DATE)}}. \bibinfo{pages}{187--192}.
\newblock


\bibitem[Lentaris et~al\mbox{.}(2020)]%
        {lentaris_icecs}
\bibfield{author}{\bibinfo{person}{George Lentaris}, \bibinfo{person}{George Chatzitsompanis}, \bibinfo{person}{Vasileios Leon}, \bibinfo{person}{Kiamal Pekmestzi}, {and} \bibinfo{person}{Dimitrios Soudris}.} \bibinfo{year}{2020}\natexlab{}.
\newblock \showarticletitle{Combining Arithmetic Approximation Techniques for Improved CNN Circuit Design}. In \bibinfo{booktitle}{\emph{IEEE Int'l. Conference on Electronics, Circuits and Systems (ICECS)}}. \bibinfo{pages}{1--4}.
\newblock


\bibitem[Leon et~al\mbox{.}(2025)]%
        {mysurvey_pt1}
\bibfield{author}{\bibinfo{person}{Vasileios Leon}, \bibinfo{person}{Muhammad~Abdullah Hanif}, \bibinfo{person}{Giorgos Armeniakos}, \bibinfo{person}{Xun Jiao}, \bibinfo{person}{Muhammad Shafique}, \bibinfo{person}{Kiamal Pekmestzi}, {and} \bibinfo{person}{Dimitrios Soudris}.} \bibinfo{year}{2025}\natexlab{}.
\newblock \showarticletitle{{Approximate Computing Survey, Part I: Terminology and Software \& Hardware Approximation Techniques}}.
\newblock \bibinfo{journal}{\emph{Comput. Surveys}} \bibinfo{volume}{57}, \bibinfo{number}{7} (\bibinfo{year}{2025}), \bibinfo{pages}{1--36}.
\newblock


\bibitem[Leon et~al\mbox{.}(2022)]%
        {leon_lascas}
\bibfield{author}{\bibinfo{person}{Vasileios Leon}, \bibinfo{person}{Georgios Makris}, \bibinfo{person}{Sotirios Xydis}, \bibinfo{person}{Kiamal Pekmestzi}, {and} \bibinfo{person}{Dimitrios Soudris}.} \bibinfo{year}{2022}\natexlab{}.
\newblock \showarticletitle{MAx-DNN: Multi-Level Arithmetic Approximation for Energy-Efficient DNN Hardware Accelerators}. In \bibinfo{booktitle}{\emph{IEEE Latin America Symposium on Circuits and System (LASCAS)}}. \bibinfo{pages}{1--4}.
\newblock


\bibitem[Leon et~al\mbox{.}(2021a)]%
        {2021_Leon_ACMtecs}
\bibfield{author}{\bibinfo{person}{Vasileios Leon}, \bibinfo{person}{Theodora Paparouni}, \bibinfo{person}{Evangelos Petrongonas}, \bibinfo{person}{Dimitrios Soudris}, {and} \bibinfo{person}{Kiamal Pekmestzi}.} \bibinfo{year}{2021}\natexlab{a}.
\newblock \showarticletitle{Improving Power of {DSP} and {CNN} Hardware Accelerators Using Approximate Floating-Point Multipliers}.
\newblock \bibinfo{journal}{\emph{ACM Trans. on Embedded Computing Systems}} \bibinfo{volume}{20}, \bibinfo{number}{5} (\bibinfo{year}{2021}), \bibinfo{pages}{1--21}.
\newblock


\bibitem[Leon et~al\mbox{.}(2021b)]%
        {leon_qam}
\bibfield{author}{\bibinfo{person}{Vasileios Leon}, \bibinfo{person}{Ioannis Stratakos}, \bibinfo{person}{Giorgos Armeniakos}, \bibinfo{person}{George Lentaris}, {and} \bibinfo{person}{Dimitrios Soudris}.} \bibinfo{year}{2021}\natexlab{b}.
\newblock \showarticletitle{ApproxQAM: High-Order QAM Demodulation Circuits with Approximate Arithmetic}. In \bibinfo{booktitle}{\emph{Int'l. Conference on Modern Circuits and Systems Technologies (MOCAST)}}. \bibinfo{pages}{1--5}.
\newblock


\bibitem[Leon et~al\mbox{.}(2018a)]%
        {2018_Leon_IEEEtvlsi}
\bibfield{author}{\bibinfo{person}{Vasileios Leon}, \bibinfo{person}{Georgios Zervakis}, \bibinfo{person}{Dimitrios Soudris}, {and} \bibinfo{person}{Kiamal Pekmestzi}.} \bibinfo{year}{2018}\natexlab{a}.
\newblock \showarticletitle{Approximate Hybrid High Radix Encoding for Energy-Efficient Inexact Multipliers}.
\newblock \bibinfo{journal}{\emph{IEEE Trans. on Very Large Scale Integration (VLSI) Systems}} \bibinfo{volume}{26}, \bibinfo{number}{3} (\bibinfo{year}{2018}), \bibinfo{pages}{421--430}.
\newblock


\bibitem[Leon et~al\mbox{.}(2018b)]%
        {2018_Leon_IEEEmicro}
\bibfield{author}{\bibinfo{person}{Vasileios Leon}, \bibinfo{person}{Georgios Zervakis}, \bibinfo{person}{Sotirios Xydis}, \bibinfo{person}{Dimitrios Soudris}, {and} \bibinfo{person}{Kiamal Pekmestzi}.} \bibinfo{year}{2018}\natexlab{b}.
\newblock \showarticletitle{Walking through the Energy-Error Pareto Frontier of Approximate Multipliers}.
\newblock \bibinfo{journal}{\emph{IEEE Micro}} \bibinfo{volume}{38}, \bibinfo{number}{4} (\bibinfo{year}{2018}), \bibinfo{pages}{40--49}.
\newblock


\bibitem[Li et~al\mbox{.}(2019)]%
        {Li_DAC_2019}
\bibfield{author}{\bibinfo{person}{Fei Li}, \bibinfo{person}{Youyou Lu}, \bibinfo{person}{Zhongjie Wu}, {and} \bibinfo{person}{Jiwu Shu}.} \bibinfo{year}{2019}\natexlab{}.
\newblock \showarticletitle{ASCache: An Approximate SSD Cache for Error-Tolerant Applications}. In \bibinfo{booktitle}{\emph{Design Automation Conference (DAC)}}.
\newblock


\bibitem[Li et~al\mbox{.}(2017)]%
        {li2016pruning}
\bibfield{author}{\bibinfo{person}{Hao Li}, \bibinfo{person}{Asim Kadav}, \bibinfo{person}{Igor Durdanovic}, \bibinfo{person}{Hanan Samet}, {and} \bibinfo{person}{Hans~Peter Graf}.} \bibinfo{year}{2017}\natexlab{}.
\newblock \showarticletitle{Pruning Filters for Efficient ConvNets}. In \bibinfo{booktitle}{\emph{Int'l. Conference on Learning Representations (ICLR)}}. \bibinfo{pages}{1--13}.
\newblock


\bibitem[Li et~al\mbox{.}(2018)]%
        {2018_Li_ICS}
\bibfield{author}{\bibinfo{person}{Shikai Li}, \bibinfo{person}{Sunghyun Park}, {and} \bibinfo{person}{Scott Mahlke}.} \bibinfo{year}{2018}\natexlab{}.
\newblock \showarticletitle{Sculptor: Flexible Approximation with Selective Dynamic Loop Perforation}. In \bibinfo{booktitle}{\emph{ACM Int'l. Conference on Supercomputing (ICS)}}. \bibinfo{pages}{341–351}.
\newblock


\bibitem[Lin et~al\mbox{.}(2017)]%
        {2017_Lin_ISCAS}
\bibfield{author}{\bibinfo{person}{Yingyan Lin}, \bibinfo{person}{Charbel Sakr}, \bibinfo{person}{Yongjune Kim}, {and} \bibinfo{person}{Naresh Shanbhag}.} \bibinfo{year}{2017}\natexlab{}.
\newblock \showarticletitle{Predictive{N}et: An Energy-Efficient Convolutional Neural Network via Zero Prediction}. In \bibinfo{booktitle}{\emph{IEEE Int'l. Symposium on Circuits and Systems (ISCAS)}}. \bibinfo{pages}{1--4}.
\newblock


\bibitem[Linderman et~al\mbox{.}(2010)]%
        {2010_Linderman_CGO}
\bibfield{author}{\bibinfo{person}{Michael~D. Linderman}, \bibinfo{person}{Matthew Ho}, \bibinfo{person}{David~L. Dill}, \bibinfo{person}{Teresa~H. Meng}, {and} \bibinfo{person}{Garry~P. Nolan}.} \bibinfo{year}{2010}\natexlab{}.
\newblock \showarticletitle{Towards Program Optimization through Automated Analysis of Numerical Precision}. In \bibinfo{booktitle}{\emph{IEEE/ACM Int'l. Symposium on Code Generation and Optimization (CGO)}}. \bibinfo{pages}{230–237}.
\newblock


\bibitem[Liu and Zhang(2017)]%
        {2017_Liu_ICCAD}
\bibfield{author}{\bibinfo{person}{Gai Liu} {and} \bibinfo{person}{Zhiru Zhang}.} \bibinfo{year}{2017}\natexlab{}.
\newblock \showarticletitle{Statistically Certified Approximate Logic Synthesis}. In \bibinfo{booktitle}{\emph{Int'l. Conference on Computer-Aided Design (ICCAD)}}. \bibinfo{pages}{344--351}.
\newblock


\bibitem[Liu et~al\mbox{.}(2012)]%
        {2012_Liu_FAST}
\bibfield{author}{\bibinfo{person}{Ren-Shuo Liu}, \bibinfo{person}{Chia-Lin Yang}, {and} \bibinfo{person}{Wei Wu}.} \bibinfo{year}{2012}\natexlab{}.
\newblock \showarticletitle{Optimizing {NAND} Flash-Based {SSD}s via Retention Relaxation}. In \bibinfo{booktitle}{\emph{USENIX Conference on File and Storage Technologies (FAST)}}. \bibinfo{pages}{1--14}.
\newblock


\bibitem[Liu et~al\mbox{.}(2011)]%
        {2011_Liu_ASPLOS}
\bibfield{author}{\bibinfo{person}{Song Liu}, \bibinfo{person}{Karthik Pattabiraman}, \bibinfo{person}{Thomas Moscibroda}, {and} \bibinfo{person}{Benjamin~G. Zorn}.} \bibinfo{year}{2011}\natexlab{}.
\newblock \showarticletitle{Flikker: Saving {DRAM} Refresh-Power through Critical Data Partitioning}. In \bibinfo{booktitle}{\emph{ACM Int'l. Conference on Architectural Support for Programming Languages and Operating Systems (ASPLOS)}}. \bibinfo{pages}{213–224}.
\newblock


\bibitem[Liu et~al\mbox{.}(2018a)]%
        {2018_Liu_ARITH}
\bibfield{author}{\bibinfo{person}{Weiqiang Liu}, \bibinfo{person}{Jing Li}, \bibinfo{person}{Tao Xu}, \bibinfo{person}{Chenghua Wang}, \bibinfo{person}{Paolo Montuschi}, {and} \bibinfo{person}{Fabrizio Lombardi}.} \bibinfo{year}{2018}\natexlab{a}.
\newblock \showarticletitle{Combining Restoring Array and Logarithmic Dividers into an Approximate Hybrid Design}. In \bibinfo{booktitle}{\emph{IEEE Symposium on Computer Arithmetic (ARITH)}}. \bibinfo{pages}{92--98}.
\newblock


\bibitem[Liu et~al\mbox{.}(2017)]%
        {2017_Liu_IEEEtc}
\bibfield{author}{\bibinfo{person}{Weiqiang Liu}, \bibinfo{person}{Liangyu Qian}, \bibinfo{person}{Chenghua Wang}, \bibinfo{person}{Honglan Jiang}, \bibinfo{person}{Jie Han}, {and} \bibinfo{person}{Fabrizio Lombardi}.} \bibinfo{year}{2017}\natexlab{}.
\newblock \showarticletitle{Design of Approximate Radix-4 Booth Multipliers for Error-Tolerant Computing}.
\newblock \bibinfo{journal}{\emph{IEEE Trans. on Computers}} \bibinfo{volume}{66}, \bibinfo{number}{8} (\bibinfo{year}{2017}), \bibinfo{pages}{1435--1441}.
\newblock


\bibitem[Liu et~al\mbox{.}(2018b)]%
        {2018_Liu_IEEEtcasi}
\bibfield{author}{\bibinfo{person}{Weiqiang Liu}, \bibinfo{person}{Jiahua Xu}, \bibinfo{person}{Danye Wang}, \bibinfo{person}{Chenghua Wang}, \bibinfo{person}{Paolo Montuschi}, {and} \bibinfo{person}{Fabrizio Lombardi}.} \bibinfo{year}{2018}\natexlab{b}.
\newblock \showarticletitle{Design and Evaluation of Approximate Logarithmic Multipliers for Low Power Error-Tolerant Applications}.
\newblock \bibinfo{journal}{\emph{IEEE Trans. on Circuits and Systems I: Regular Papers}} \bibinfo{volume}{65}, \bibinfo{number}{9} (\bibinfo{year}{2018}), \bibinfo{pages}{2856--2868}.
\newblock


\bibitem[Liu et~al\mbox{.}(2010)]%
        {2010_Liu_IEEEtvlsi}
\bibfield{author}{\bibinfo{person}{Yang Liu}, \bibinfo{person}{Tong Zhang}, {and} \bibinfo{person}{Keshab~K. Parhi}.} \bibinfo{year}{2010}\natexlab{}.
\newblock \showarticletitle{Computation Error Analysis in Digital Signal Processing Systems With Overscaled Supply Voltage}.
\newblock \bibinfo{journal}{\emph{IEEE Trans. on Very Large Scale Integration (VLSI) Systems}} \bibinfo{volume}{18}, \bibinfo{number}{4} (\bibinfo{year}{2010}), \bibinfo{pages}{517--526}.
\newblock


\bibitem[Liu et~al\mbox{.}(2019)]%
        {2019_Liu_ISCA}
\bibfield{author}{\bibinfo{person}{Zhenhong Liu}, \bibinfo{person}{Amir Yazdanbakhsh}, \bibinfo{person}{Dong~Kai Wang}, \bibinfo{person}{Hadi Esmaeilzadeh}, {and} \bibinfo{person}{Nam~Sung Kim}.} \bibinfo{year}{2019}\natexlab{}.
\newblock \showarticletitle{Ax{M}emo: Hardware-Compiler Co-Design for Approximate Code Memoization}. In \bibinfo{booktitle}{\emph{ACM/IEEE Int'l. Symposium on Computer Architecture (ISCA)}}. \bibinfo{pages}{685–697}.
\newblock


\bibitem[Lou et~al\mbox{.}(2021)]%
        {lou2021dynamic}
\bibfield{author}{\bibinfo{person}{Wei Lou}, \bibinfo{person}{Lei Xun}, \bibinfo{person}{Amin Sabet}, \bibinfo{person}{Jia Bi}, \bibinfo{person}{Jonathon Hare}, {and} \bibinfo{person}{Geoff~V Merrett}.} \bibinfo{year}{2021}\natexlab{}.
\newblock \showarticletitle{Dynamic-OFA: Runtime DNN Architecture Switching for Performance Scaling on Heterogeneous Embedded Platforms}. In \bibinfo{booktitle}{\emph{IEEE/CVF Conference on Computer Vision and Pattern Recognition Workshops (CVPRW)}}. \bibinfo{pages}{3110--3118}.
\newblock


\bibitem[Ma et~al\mbox{.}(2021)]%
        {date21}
\bibfield{author}{\bibinfo{person}{Dongning Ma}, \bibinfo{person}{Rahul Thapa}, \bibinfo{person}{Xingjian Wang}, \bibinfo{person}{Xun Jiao}, {and} \bibinfo{person}{Cong Hao}.} \bibinfo{year}{2021}\natexlab{}.
\newblock \showarticletitle{Workload-Aware Approximate Computing Configuration}. In \bibinfo{booktitle}{\emph{Design, Automation \& Test in Europe (DATE)}}. \bibinfo{pages}{920--925}.
\newblock


\bibitem[Ma et~al\mbox{.}(2022)]%
        {devot}
\bibfield{author}{\bibinfo{person}{Dongning Ma}, \bibinfo{person}{Xinqiao Zhang}, \bibinfo{person}{Ke Huang}, \bibinfo{person}{Yu Jiang}, \bibinfo{person}{Wanli Chang}, {and} \bibinfo{person}{Xun Jiao}.} \bibinfo{year}{2022}\natexlab{}.
\newblock \showarticletitle{DEVoT: Dynamic Delay Modeling of Functional Units Under Voltage and Temperature Variations}.
\newblock \bibinfo{journal}{\emph{IEEE Trans. on Computer-Aided Design of Integrated Circuits and Systems}} \bibinfo{volume}{41}, \bibinfo{number}{4} (\bibinfo{year}{2022}), \bibinfo{pages}{827--839}.
\newblock


\bibitem[Mahajan et~al\mbox{.}(2016)]%
        {axaccel:isca}
\bibfield{author}{\bibinfo{person}{Divya Mahajan}, \bibinfo{person}{Amir Yazdanbaksh}, \bibinfo{person}{Jongse Park}, \bibinfo{person}{Bradley Thwaites}, {and} \bibinfo{person}{Hadi Esmaeilzadeh}.} \bibinfo{year}{2016}\natexlab{}.
\newblock \showarticletitle{Towards Statistical Guarantees in Controlling Quality Tradeoffs for Approximate Acceleration}. In \bibinfo{booktitle}{\emph{ACM/IEEE Annual Int'l. Symposium on Computer Architecture (ISCA)}}. \bibinfo{pages}{66--77}.
\newblock


\bibitem[Manikantta~Reddy et~al\mbox{.}(2020)]%
        {2020_Manikanta_IEEEtvlsi}
\bibfield{author}{\bibinfo{person}{K. Manikantta~Reddy}, \bibinfo{person}{M.~H. Vasantha}, \bibinfo{person}{Y.~B. Nithin~Kumar}, {and} \bibinfo{person}{Devesh Dwivedi}.} \bibinfo{year}{2020}\natexlab{}.
\newblock \showarticletitle{Design of Approximate Booth Squarer for Error-Tolerant Computing}.
\newblock \bibinfo{journal}{\emph{IEEE Trans. on Very Large Scale Integration Systems}} \bibinfo{volume}{28}, \bibinfo{number}{5} (\bibinfo{year}{2020}), \bibinfo{pages}{1230--1241}.
\newblock


\bibitem[Meng et~al\mbox{.}(2009)]%
        {2009_Meng_IPDPS}
\bibfield{author}{\bibinfo{person}{Jiayuan Meng}, \bibinfo{person}{Srimat Chakradhar}, {and} \bibinfo{person}{Anand Raghunathan}.} \bibinfo{year}{2009}\natexlab{}.
\newblock \showarticletitle{Best-Effort Parallel Execution Framework for Recognition and Mining Applications}. In \bibinfo{booktitle}{\emph{IEEE Int'l. Symposium on Parallel Distributed Processing (IPDPS)}}. \bibinfo{pages}{1--12}.
\newblock


\bibitem[Mengt et~al\mbox{.}(2010)]%
        {2010_Mengt_IPDPS}
\bibfield{author}{\bibinfo{person}{Jiayuan Mengt}, \bibinfo{person}{Anand Raghunathan}, \bibinfo{person}{Srimat Chakradhar}, {and} \bibinfo{person}{Surendra Byna}.} \bibinfo{year}{2010}\natexlab{}.
\newblock \showarticletitle{Exploiting the Forgiving Nature of Applications for Scalable Parallel Execution}. In \bibinfo{booktitle}{\emph{IEEE Int'l. Symposium on Parallel Distributed Processing}}. \bibinfo{pages}{1--12}.
\newblock


\bibitem[Menon et~al\mbox{.}(2018)]%
        {2018_Menon_SC}
\bibfield{author}{\bibinfo{person}{Harshitha Menon}, \bibinfo{person}{Michael~O. Lam}, \bibinfo{person}{Daniel Osei-Kuffuor}, \bibinfo{person}{Markus Schordan}, \bibinfo{person}{Scott Lloyd}, \bibinfo{person}{Kathryn Mohror}, {and} \bibinfo{person}{Jeffrey Hittinger}.} \bibinfo{year}{2018}\natexlab{}.
\newblock \showarticletitle{{ADAPT}: Algorithmic Differentiation Applied to Floating-Point Precision Tuning}. In \bibinfo{booktitle}{\emph{ACM/IEEE SC, Int'l. Conference for High Performance Computing, Networking, Storage and Analysis}}. \bibinfo{pages}{614--626}.
\newblock


\bibitem[Miguel et~al\mbox{.}(2016)]%
        {2016_Joshua_MICRO}
\bibfield{author}{\bibinfo{person}{Joshua~San Miguel}, \bibinfo{person}{Jorge Albericio}, \bibinfo{person}{Natalie~Enright Jerger}, {and} \bibinfo{person}{Aamer Jaleel}.} \bibinfo{year}{2016}\natexlab{}.
\newblock \showarticletitle{The Bunker Cache for Spatio-Value Approximation}. In \bibinfo{booktitle}{\emph{IEEE/ACM Int'l. Symposium on Microarchitecture (MICRO)}}. \bibinfo{pages}{1--12}.
\newblock


\bibitem[Miguel et~al\mbox{.}(2015)]%
        {2015_Joshua_MICRO}
\bibfield{author}{\bibinfo{person}{Joshua~San Miguel}, \bibinfo{person}{Jorge Albericio}, \bibinfo{person}{Andreas Moshovos}, {and} \bibinfo{person}{Natalie~Enright Jerger}.} \bibinfo{year}{2015}\natexlab{}.
\newblock \showarticletitle{Doppelgänger: A Cache for Approximate Computing}. In \bibinfo{booktitle}{\emph{IEEE/ACM Int'l. Symposium on Microarchitecture (MICRO)}}. \bibinfo{pages}{50--61}.
\newblock


\bibitem[Miguel et~al\mbox{.}(2014)]%
        {2014_Miguel_MICRO}
\bibfield{author}{\bibinfo{person}{Joshua~San Miguel}, \bibinfo{person}{Mario Badr}, {and} \bibinfo{person}{Natalie~Enright Jerger}.} \bibinfo{year}{2014}\natexlab{}.
\newblock \showarticletitle{Load Value Approximation}. In \bibinfo{booktitle}{\emph{IEEE/ACM Int'l. Symposium on Microarchitecture (MICRO)}}. \bibinfo{pages}{127--139}.
\newblock


\bibitem[Misailovic et~al\mbox{.}(2014)]%
        {2014_Misailovic_OOPSLA}
\bibfield{author}{\bibinfo{person}{Sasa Misailovic}, \bibinfo{person}{Michael Carbin}, \bibinfo{person}{Sara Achour}, \bibinfo{person}{Zichao Qi}, {and} \bibinfo{person}{Martin~C. Rinard}.} \bibinfo{year}{2014}\natexlab{}.
\newblock \showarticletitle{Chisel: Reliability- and Accuracy-Aware Optimization of Approximate Computational Kernels}. In \bibinfo{booktitle}{\emph{ACM SIGPLAN Int'l. Conference on Object-Oriented Programming, Systems, Languages, and Applications (OOPSLA)}}. \bibinfo{pages}{309–328}.
\newblock


\bibitem[Misailovic et~al\mbox{.}(2013)]%
        {2013_Misailovic_ACMtecs}
\bibfield{author}{\bibinfo{person}{Sasa Misailovic}, \bibinfo{person}{Deokhwan Kim}, {and} \bibinfo{person}{Martin~C. Rinard}.} \bibinfo{year}{2013}\natexlab{}.
\newblock \showarticletitle{Parallelizing Sequential Programs with Statistical Accuracy Tests}.
\newblock \bibinfo{journal}{\emph{ACM Trans. on Embedded Computing Systems}} \bibinfo{volume}{12}, \bibinfo{number}{2s} (\bibinfo{year}{2013}), \bibinfo{pages}{1--26}.
\newblock


\bibitem[Misailovic et~al\mbox{.}(2010)]%
        {2010_Misailovic_ICSE}
\bibfield{author}{\bibinfo{person}{Sasa Misailovic}, \bibinfo{person}{Stelios Sidiroglou}, \bibinfo{person}{Henry Hoffmann}, {and} \bibinfo{person}{Martin~C. Rinard}.} \bibinfo{year}{2010}\natexlab{}.
\newblock \showarticletitle{Quality of Service Profiling}. In \bibinfo{booktitle}{\emph{ACM/IEEE Int'l. Conference on Software Engineering (ICSE)}}, Vol.~\bibinfo{volume}{1}. \bibinfo{pages}{25--34}.
\newblock


\bibitem[Misailovic et~al\mbox{.}(2012)]%
        {2012_Misailovic_RACES}
\bibfield{author}{\bibinfo{person}{Sasa Misailovic}, \bibinfo{person}{Stelios Sidiroglou}, {and} \bibinfo{person}{Martin~C. Rinard}.} \bibinfo{year}{2012}\natexlab{}.
\newblock \showarticletitle{Dancing with Uncertainty}. In \bibinfo{booktitle}{\emph{ACM Workshop on Relaxing Synchronization for Multicore and Manycore Scalability (RACES)}}. \bibinfo{pages}{51–60}.
\newblock


\bibitem[Mishra et~al\mbox{.}(2014)]%
        {2014_Mishra_WACAS}
\bibfield{author}{\bibinfo{person}{Asit~K. Mishra}, \bibinfo{person}{Rajkishore Barik}, {and} \bibinfo{person}{Somnath Paul}.} \bibinfo{year}{2014}\natexlab{}.
\newblock \showarticletitle{i{ACT}: A Software-Hardware Framework for Understanding the Scope of Approximate Computing}. In \bibinfo{booktitle}{\emph{Workshop on Approximate Computing Across the System Stack}}. \bibinfo{pages}{1--6}.
\newblock


\bibitem[Mitra et~al\mbox{.}(2017)]%
        {2017_Mitra_CGO}
\bibfield{author}{\bibinfo{person}{Subrata Mitra}, \bibinfo{person}{Manish~K. Gupta}, \bibinfo{person}{Sasa Misailovic}, {and} \bibinfo{person}{Saurabh Bagchi}.} \bibinfo{year}{2017}\natexlab{}.
\newblock \showarticletitle{Phase-Aware Optimization in Approximate Computing}. In \bibinfo{booktitle}{\emph{IEEE/ACM Int'l. Symposium on Code Generation and Optimization (CGO)}}. \bibinfo{pages}{185--196}.
\newblock


\bibitem[Mittal(2016)]%
        {2016_Mittal_ACMsrv}
\bibfield{author}{\bibinfo{person}{Sparsh Mittal}.} \bibinfo{year}{2016}\natexlab{}.
\newblock \showarticletitle{A Survey of Techniques for Approximate Computing}.
\newblock \bibinfo{journal}{\emph{Comput. Surveys}} \bibinfo{volume}{48}, \bibinfo{number}{4} (\bibinfo{year}{2016}), \bibinfo{pages}{1--33}.
\newblock


\bibitem[Mittal and Vetter(2015)]%
        {heterog}
\bibfield{author}{\bibinfo{person}{Sparsh Mittal} {and} \bibinfo{person}{Jeffrey~S. Vetter}.} \bibinfo{year}{2015}\natexlab{}.
\newblock \showarticletitle{{A Survey of CPU-GPU Heterogeneous Computing Techniques}}.
\newblock \bibinfo{journal}{\emph{Comput. Surveys}} \bibinfo{volume}{47}, \bibinfo{number}{4} (\bibinfo{year}{2015}), \bibinfo{pages}{1--35}.
\newblock


\bibitem[Mohapatra et~al\mbox{.}(2011)]%
        {2011_Mohapatra_DATE}
\bibfield{author}{\bibinfo{person}{Debabrata Mohapatra}, \bibinfo{person}{Vinay~K. Chippa}, \bibinfo{person}{Anand Raghunathan}, {and} \bibinfo{person}{Kaushik Roy}.} \bibinfo{year}{2011}\natexlab{}.
\newblock \showarticletitle{Design of Voltage-Scalable Meta-Functions for Approximate Computing}. In \bibinfo{booktitle}{\emph{Design, Automation \& Test in Europe (DATE)}}. \bibinfo{pages}{1--6}.
\newblock


\bibitem[Momeni et~al\mbox{.}(2015)]%
        {2015_Momeni_IEEEtc}
\bibfield{author}{\bibinfo{person}{Amir Momeni}, \bibinfo{person}{Jie Han}, \bibinfo{person}{Paolo Montuschi}, {and} \bibinfo{person}{Fabrizio Lombardi}.} \bibinfo{year}{2015}\natexlab{}.
\newblock \showarticletitle{Design and Analysis of Approximate Compressors for Multiplication}.
\newblock \bibinfo{journal}{\emph{IEEE Trans. on Computers}} \bibinfo{volume}{64}, \bibinfo{number}{4} (\bibinfo{year}{2015}), \bibinfo{pages}{984--994}.
\newblock


\bibitem[Moreau et~al\mbox{.}(2018)]%
        {2018_Moreau_IEEEesl}
\bibfield{author}{\bibinfo{person}{Thierry Moreau}, \bibinfo{person}{Joshua San~Miguel}, \bibinfo{person}{Mark Wyse}, \bibinfo{person}{James Bornholt}, \bibinfo{person}{Armin Alaghi}, \bibinfo{person}{Luis Ceze}, \bibinfo{person}{Natalie Enright~Jerger}, {and} \bibinfo{person}{Adrian Sampson}.} \bibinfo{year}{2018}\natexlab{}.
\newblock \showarticletitle{A Taxonomy of General Purpose Approximate Computing Techniques}.
\newblock \bibinfo{journal}{\emph{IEEE Embedded Systems Letters}} \bibinfo{volume}{10}, \bibinfo{number}{1} (\bibinfo{year}{2018}), \bibinfo{pages}{2--5}.
\newblock


\bibitem[Moreau et~al\mbox{.}(2015)]%
        {snnap}
\bibfield{author}{\bibinfo{person}{Thierry Moreau}, \bibinfo{person}{Mark Wyse}, \bibinfo{person}{Jacob Nelson}, \bibinfo{person}{Adrian Sampson}, \bibinfo{person}{Hadi Esmaeilzadeh}, \bibinfo{person}{Luis Ceze}, {and} \bibinfo{person}{Mark Oskin}.} \bibinfo{year}{2015}\natexlab{}.
\newblock \showarticletitle{SNNAP: Approximate Computing on Programmable SoCs via Neural Acceleration}. In \bibinfo{booktitle}{\emph{IEEE Int'l. Symposium on High Performance Computer Architecture (HPCA)}}. \bibinfo{pages}{603--614}.
\newblock


\bibitem[Mrazek et~al\mbox{.}(2017)]%
        {2017_Mrazek_DATE}
\bibfield{author}{\bibinfo{person}{Vojtech Mrazek}, \bibinfo{person}{Radek Hrbacek}, \bibinfo{person}{Zdenek Vasicek}, {and} \bibinfo{person}{Lukas Sekanina}.} \bibinfo{year}{2017}\natexlab{}.
\newblock \showarticletitle{Evo{A}pprox8b: Library of Approximate Adders and Multipliers for Circuit Design and Benchmarking of Approximation Methods}. In \bibinfo{booktitle}{\emph{Design, Automation \& Test in Europe (DATE)}}. \bibinfo{pages}{258--261}.
\newblock


\bibitem[Mrazek et~al\mbox{.}(2016)]%
        {mrazek2016design}
\bibfield{author}{\bibinfo{person}{Vojtech Mrazek}, \bibinfo{person}{Syed~Shakib Sarwar}, \bibinfo{person}{Lukas Sekanina}, \bibinfo{person}{Zdenek Vasicek}, {and} \bibinfo{person}{Kaushik Roy}.} \bibinfo{year}{2016}\natexlab{}.
\newblock \showarticletitle{Design of Power-Efficient Approximate Multipliers for Approximate Artificial Neural Networks}. In \bibinfo{booktitle}{\emph{IEEE/ACM Int'l. Conference on Computer-Aided Design (ICCAD)}}. \bibinfo{pages}{1--7}.
\newblock


\bibitem[Mrazek et~al\mbox{.}(2020)]%
        {mrazek2020libraries}
\bibfield{author}{\bibinfo{person}{Vojtech Mrazek}, \bibinfo{person}{Lukas Sekanina}, {and} \bibinfo{person}{Zdenek Vasicek}.} \bibinfo{year}{2020}\natexlab{}.
\newblock \showarticletitle{Libraries of Approximate Circuits: Automated Design and Application in CNN Accelerators}.
\newblock \bibinfo{journal}{\emph{IEEE Journal on Emerging and Selected Topics in Circuits and Systems}} \bibinfo{volume}{10}, \bibinfo{number}{4} (\bibinfo{year}{2020}), \bibinfo{pages}{406--418}.
\newblock


\bibitem[Mrazek et~al\mbox{.}(2019)]%
        {mrazek2019alwann}
\bibfield{author}{\bibinfo{person}{Vojtech Mrazek}, \bibinfo{person}{Zdenek Vas{\'\i}cek}, \bibinfo{person}{Luk{\'a}s Sekanina}, \bibinfo{person}{Muhammad~Abdullah Hanif}, {and} \bibinfo{person}{Muhammad Shafique}.} \bibinfo{year}{2019}\natexlab{}.
\newblock \showarticletitle{ALWANN: Automatic Layer-Wise Approximation of Deep Neural Network Accelerators without Retraining}. In \bibinfo{booktitle}{\emph{IEEE/ACM Int'l. Conference on Computer-Aided Design (ICCAD)}}. \bibinfo{pages}{1--8}.
\newblock


\bibitem[Mubarik et~al\mbox{.}(2020)]%
        {Mubarik:MICRO:2020:printedml}
\bibfield{author}{\bibinfo{person}{Muhammad~Husnain Mubarik}, \bibinfo{person}{Dennis~D. Weller}, \bibinfo{person}{Nathaniel Bleier}, \bibinfo{person}{Matthew Tomei}, \bibinfo{person}{Jasmin Aghassi-Hagmann}, \bibinfo{person}{Mehdi~B. Tahoori}, {and} \bibinfo{person}{Rakesh Kumar}.} \bibinfo{year}{2020}\natexlab{}.
\newblock \showarticletitle{Printed Machine Learning Classifiers}. In \bibinfo{booktitle}{\emph{IEEE/ACM Int'l. Symposium on Microarchitecture (MICRO)}}. \bibinfo{pages}{73--87}.
\newblock


\bibitem[Narayanan et~al\mbox{.}(2006)]%
        {2006_Narayan_IISWC}
\bibfield{author}{\bibinfo{person}{Ramanathan Narayanan}, \bibinfo{person}{Berkin Ozisikyilmaz}, \bibinfo{person}{Joseph Zambreno}, \bibinfo{person}{Gokhan Memik}, {and} \bibinfo{person}{Alok Choudhary}.} \bibinfo{year}{2006}\natexlab{}.
\newblock \showarticletitle{Mine{B}ench: A Benchmark Suite for Data Mining Workloads}. In \bibinfo{booktitle}{\emph{IEEE Int'l. Symposium on Workload Characterization (IISWC)}}. \bibinfo{pages}{182--188}.
\newblock


\bibitem[Ndour et~al\mbox{.}(2019)]%
        {riscv:bw}
\bibfield{author}{\bibinfo{person}{Genevi\`{e}ve Ndour}, \bibinfo{person}{Tiago~Trevisan Jost}, \bibinfo{person}{Anca Molnos}, \bibinfo{person}{Yves Durand}, {and} \bibinfo{person}{Arnaud Tisserand}.} \bibinfo{year}{2019}\natexlab{}.
\newblock \showarticletitle{Evaluation of Variable Bit-Width Units in a RISC-V Processor for Approximate Computing}. In \bibinfo{booktitle}{\emph{ACM Int'l. Conference on Computing Frontiers}}. \bibinfo{pages}{344–349}.
\newblock


\bibitem[Nepal et~al\mbox{.}(2019)]%
        {2019_Nepal_IEEEtetc}
\bibfield{author}{\bibinfo{person}{Kumud Nepal}, \bibinfo{person}{Soheil Hashemi}, \bibinfo{person}{Hokchhay Tann}, \bibinfo{person}{R.~Iris Bahar}, {and} \bibinfo{person}{Sherief Reda}.} \bibinfo{year}{2019}\natexlab{}.
\newblock \showarticletitle{Automated High-Level Generation of Low-Power Approximate Computing Circuits}.
\newblock \bibinfo{journal}{\emph{IEEE Trans. on Emerging Topics in Computing}} \bibinfo{volume}{7}, \bibinfo{number}{1} (\bibinfo{year}{2019}), \bibinfo{pages}{18--30}.
\newblock


\bibitem[Nepal et~al\mbox{.}(2014)]%
        {2014_Nepal_DATE}
\bibfield{author}{\bibinfo{person}{Kumud Nepal}, \bibinfo{person}{Yueting Li}, \bibinfo{person}{R.~Iris Bahar}, {and} \bibinfo{person}{Sherief Reda}.} \bibinfo{year}{2014}\natexlab{}.
\newblock \showarticletitle{{ABACUS}: A Technique for Automated Behavioral Synthesis of Approximate Computing Circuits}. In \bibinfo{booktitle}{\emph{Design, Automation \& Test in Europe (DATE)}}. \bibinfo{pages}{1--6}.
\newblock


\bibitem[Ottavi et~al\mbox{.}(2020)]%
        {riscv:mixed}
\bibfield{author}{\bibinfo{person}{Gianmarco Ottavi}, \bibinfo{person}{Angelo Garofalo}, \bibinfo{person}{Giuseppe Tagliavini}, \bibinfo{person}{Francesco Conti}, \bibinfo{person}{Luca Benini}, {and} \bibinfo{person}{Davide Rossi}.} \bibinfo{year}{2020}\natexlab{}.
\newblock \showarticletitle{A Mixed-Precision RISC-V Processor for Extreme-Edge DNN Inference}. In \bibinfo{booktitle}{\emph{IEEE Computer Society Annual Symposium on VLSI (ISVLSI)}}. \bibinfo{pages}{512--517}.
\newblock


\bibitem[Panda et~al\mbox{.}(2017a)]%
        {panda2017energy1}
\bibfield{author}{\bibinfo{person}{Priyadarshini Panda}, \bibinfo{person}{Abhronil Sengupta}, {and} \bibinfo{person}{Kaushik Roy}.} \bibinfo{year}{2017}\natexlab{a}.
\newblock \showarticletitle{Energy-Efficient and Improved Image Recognition with Conditional Deep Learning}.
\newblock \bibinfo{journal}{\emph{ACM Journal on Emerging Technologies in Computing Systems}} \bibinfo{volume}{13}, \bibinfo{number}{3} (\bibinfo{year}{2017}), \bibinfo{pages}{1--21}.
\newblock


\bibitem[Panda et~al\mbox{.}(2016)]%
        {panda2016cross}
\bibfield{author}{\bibinfo{person}{Priyadarshini Panda}, \bibinfo{person}{Abhronil Sengupta}, \bibinfo{person}{Syed~Shakib Sarwar}, \bibinfo{person}{Gopalakrishnan Srinivasan}, \bibinfo{person}{Swagath Venkataramani}, \bibinfo{person}{Anand Raghunathan}, {and} \bibinfo{person}{Kaushik Roy}.} \bibinfo{year}{2016}\natexlab{}.
\newblock \showarticletitle{Cross-Layer Approximations for Neuromorphic Computing: From Devices to Circuits and Systems}. In \bibinfo{booktitle}{\emph{Design Automation Conference (DAC)}}. \bibinfo{pages}{1--6}.
\newblock


\bibitem[Panda et~al\mbox{.}(2017b)]%
        {panda2017energy2}
\bibfield{author}{\bibinfo{person}{Priyadarshini Panda}, \bibinfo{person}{Swagath Venkataramani}, \bibinfo{person}{Abhronil Sengupta}, \bibinfo{person}{Anand Raghunathan}, {and} \bibinfo{person}{Kaushik Roy}.} \bibinfo{year}{2017}\natexlab{b}.
\newblock \showarticletitle{Energy-Efficient Object Detection Using Semantic Decomposition}.
\newblock \bibinfo{journal}{\emph{IEEE Trans. on Very Large Scale Integration (VLSI) Systems}} \bibinfo{volume}{25}, \bibinfo{number}{9} (\bibinfo{year}{2017}), \bibinfo{pages}{2673--2677}.
\newblock


\bibitem[Pandey et~al\mbox{.}(2019)]%
        {2019_Pandey_DAC}
\bibfield{author}{\bibinfo{person}{Pramesh Pandey}, \bibinfo{person}{Prabal Basu}, \bibinfo{person}{Koushik Chakraborty}, {and} \bibinfo{person}{Sanghamitra Roy}.} \bibinfo{year}{2019}\natexlab{}.
\newblock \showarticletitle{Green{TPU}: Improving Timing Error Resilience of a Near-Threshold Tensor Processing Unit}. In \bibinfo{booktitle}{\emph{Design Automation Conference (DAC)}}. \bibinfo{pages}{1--6}.
\newblock


\bibitem[Parashar et~al\mbox{.}(2017)]%
        {parashar2017scnn}
\bibfield{author}{\bibinfo{person}{Angshuman Parashar}, \bibinfo{person}{Minsoo Rhu}, \bibinfo{person}{Anurag Mukkara}, \bibinfo{person}{Antonio Puglielli}, \bibinfo{person}{Rangharajan Venkatesan}, \bibinfo{person}{Brucek Khailany}, \bibinfo{person}{Joel Emer}, \bibinfo{person}{Stephen~W Keckler}, {and} \bibinfo{person}{William~J Dally}.} \bibinfo{year}{2017}\natexlab{}.
\newblock \showarticletitle{SCNN: An Accelerator for Compressed-sparse Convolutional Neural Networks}.
\newblock \bibinfo{journal}{\emph{ACM SIGARCH Computer Architecture News}} \bibinfo{volume}{45}, \bibinfo{number}{2} (\bibinfo{year}{2017}), \bibinfo{pages}{27--40}.
\newblock


\bibitem[Park et~al\mbox{.}(2015)]%
        {2015_Park_FSE}
\bibfield{author}{\bibinfo{person}{Jongse Park}, \bibinfo{person}{Hadi Esmaeilzadeh}, \bibinfo{person}{Xin Zhang}, \bibinfo{person}{Mayur Naik}, {and} \bibinfo{person}{William Harris}.} \bibinfo{year}{2015}\natexlab{}.
\newblock \showarticletitle{Flex{J}ava: Language Support for Safe and Modular Approximate Programming}. In \bibinfo{booktitle}{\emph{ACM SIGSOFT Symposium and European Conference on Foundations of Software Engineering (FSE)}}. \bibinfo{pages}{745–757}.
\newblock


\bibitem[Park et~al\mbox{.}(2014)]%
        {2014_Park_GIT}
\bibfield{author}{\bibinfo{person}{Jongse Park}, \bibinfo{person}{Xin Zhang}, \bibinfo{person}{Kangqi Ni}, \bibinfo{person}{Hadi Esmaeilzadeh}, {and} \bibinfo{person}{Mayur Naik}.} \bibinfo{year}{2014}\natexlab{}.
\newblock \showarticletitle{Exp{AX}: A Framework for Automating Approximate Programming}.
\newblock \bibinfo{journal}{\emph{Georgia Institute of Technology Technical Report}} (\bibinfo{year}{2014}), \bibinfo{pages}{1--17}.
\newblock


\bibitem[Park et~al\mbox{.}(2021)]%
        {samsung}
\bibfield{author}{\bibinfo{person}{Jun-Seok Park} {et~al\mbox{.}}} \bibinfo{year}{2021}\natexlab{}.
\newblock \showarticletitle{A 6K-MAC Feature-Map-Sparsity-Aware Neural Processing Unit in 5nm Flagship Mobile SoC}. In \bibinfo{booktitle}{\emph{IEEE Int'l. Solid- State Circuits Conference (ISSCC)}}, Vol.~\bibinfo{volume}{64}. \bibinfo{pages}{152--154}.
\newblock


\bibitem[Park et~al\mbox{.}(2019)]%
        {2019_Park_MOD}
\bibfield{author}{\bibinfo{person}{Yongjoo Park}, \bibinfo{person}{Jingyi Qing}, \bibinfo{person}{Xiaoyang Shen}, {and} \bibinfo{person}{Barzan Mozafari}.} \bibinfo{year}{2019}\natexlab{}.
\newblock \showarticletitle{Blink{ML}: Efficient Maximum Likelihood Estimation with Probabilistic Guarantees}. In \bibinfo{booktitle}{\emph{ACM SIGMOD Int'l. Conference on Management of Data (MOD)}}. \bibinfo{pages}{1135–1152}.
\newblock


\bibitem[Pashaeifar et~al\mbox{.}(2018)]%
        {2018_Pashaeifar_IEEEtvlsi}
\bibfield{author}{\bibinfo{person}{Masoud Pashaeifar}, \bibinfo{person}{Mehdi Kamal}, \bibinfo{person}{Ali Afzali-Kusha}, {and} \bibinfo{person}{Massoud Pedram}.} \bibinfo{year}{2018}\natexlab{}.
\newblock \showarticletitle{Approximate Reverse Carry Propagate Adder for Energy-Efficient {DSP} Applications}.
\newblock \bibinfo{journal}{\emph{IEEE Trans. on Very Large Scale Integration (VLSI) Systems}} \bibinfo{volume}{26}, \bibinfo{number}{11} (\bibinfo{year}{2018}), \bibinfo{pages}{2530--2541}.
\newblock


\bibitem[Pilipović et~al\mbox{.}(2021)]%
        {2021_Pilipovic_IEEEtcasi}
\bibfield{author}{\bibinfo{person}{Ratko Pilipović}, \bibinfo{person}{Patricio Bulić}, {and} \bibinfo{person}{Uroš Lotrič}.} \bibinfo{year}{2021}\natexlab{}.
\newblock \showarticletitle{A Two-Stage Operand Trimming Approximate Logarithmic Multiplier}.
\newblock \bibinfo{journal}{\emph{IEEE Trans. on Circuits and Systems I: Regular Papers}} \bibinfo{volume}{68}, \bibinfo{number}{6} (\bibinfo{year}{2021}), \bibinfo{pages}{2535--2545}.
\newblock


\bibitem[Plastiras et~al\mbox{.}(2018)]%
        {plastiras2018efficient}
\bibfield{author}{\bibinfo{person}{George Plastiras}, \bibinfo{person}{Christos Kyrkou}, {and} \bibinfo{person}{Theocharis Theocharides}.} \bibinfo{year}{2018}\natexlab{}.
\newblock \showarticletitle{Efficient ConvNet-based Object Detection for Unmanned Aerial Vehicles by Selective Tile Processing}. In \bibinfo{booktitle}{\emph{Int'l. Conf. on Distributed Smart Cameras (ICDSC)}}. \bibinfo{pages}{1--6}.
\newblock


\bibitem[Pouchet(2015)]%
        {2015_Pouchet_POLY}
\bibfield{author}{\bibinfo{person}{Louis-Noel Pouchet}.} \bibinfo{year}{2015}\natexlab{}.
\newblock \bibinfo{booktitle}{\emph{Poly{B}ench}}.
\newblock
\urldef\tempurl%
\url{http://sourceforge.net/projects/polybench/}
\showURL{%
\tempurl}


\bibitem[Pozo and Miller(2004)]%
        {2004_Pozo_SCI}
\bibfield{author}{\bibinfo{person}{Roldan Pozo} {and} \bibinfo{person}{Bruce Miller}.} \bibinfo{year}{2004}\natexlab{}.
\newblock \bibinfo{booktitle}{\emph{Sci{M}ark 2.0}}.
\newblock
\urldef\tempurl%
\url{http://math.nist.gov/scimark2/}
\showURL{%
\tempurl}


\bibitem[Quoc et~al\mbox{.}(2017a)]%
        {2017_Quoc_ATC}
\bibfield{author}{\bibinfo{person}{Do~Le Quoc}, \bibinfo{person}{Martin Beck}, \bibinfo{person}{Pramod Bhatotia}, \bibinfo{person}{Ruichuan Chen}, \bibinfo{person}{Christof Fetzer}, {and} \bibinfo{person}{Thorsten Strufe}.} \bibinfo{year}{2017}\natexlab{a}.
\newblock \showarticletitle{Priv{A}pprox: Privacy-Preserving Stream Analytics}. In \bibinfo{booktitle}{\emph{USENIX Annual Technical Conference (ATC)}}. \bibinfo{pages}{659–672}.
\newblock


\bibitem[Quoc et~al\mbox{.}(2017b)]%
        {2017_Quoc_MIDDL}
\bibfield{author}{\bibinfo{person}{Do~Le Quoc}, \bibinfo{person}{Ruichuan Chen}, \bibinfo{person}{Pramod Bhatotia}, \bibinfo{person}{Christof Fetzer}, \bibinfo{person}{Volker Hilt}, {and} \bibinfo{person}{Thorsten Strufe}.} \bibinfo{year}{2017}\natexlab{b}.
\newblock \showarticletitle{Stream{A}pprox: Approximate Computing for Stream Analytics}. In \bibinfo{booktitle}{\emph{ACM/IFIP/USENIX Int'l. Middleware Conference}}. \bibinfo{pages}{185–197}.
\newblock


\bibitem[Ragavan et~al\mbox{.}(2016)]%
        {2016_Ragavan_ISVLSI}
\bibfield{author}{\bibinfo{person}{Rengarajan Ragavan}, \bibinfo{person}{Cedric Killian}, {and} \bibinfo{person}{Olivier Sentieys}.} \bibinfo{year}{2016}\natexlab{}.
\newblock \showarticletitle{Adaptive Overclocking and Error Correction Based on Dynamic Speculation Window}. In \bibinfo{booktitle}{\emph{IEEE Computer Society Annual Symposium on VLSI (ISVLSI)}}. \bibinfo{pages}{325--330}.
\newblock


\bibitem[Raha and Raghunathan(2017a)]%
        {qlut_2017}
\bibfield{author}{\bibinfo{person}{Arnab Raha} {and} \bibinfo{person}{Vijay Raghunathan}.} \bibinfo{year}{2017}\natexlab{a}.
\newblock \showarticletitle{{qLUT: Input-Aware Quantized Table Lookup for Energy-Efficient Approximate Accelerators}}.
\newblock \bibinfo{journal}{\emph{ACM Trans. on Embedded Computing Systems}} \bibinfo{volume}{16}, \bibinfo{number}{5s} (\bibinfo{year}{2017}), \bibinfo{pages}{130:1--130:23}.
\newblock


\bibitem[Raha and Raghunathan(2017b)]%
        {raha2017towards}
\bibfield{author}{\bibinfo{person}{Arnab Raha} {and} \bibinfo{person}{Vijay Raghunathan}.} \bibinfo{year}{2017}\natexlab{b}.
\newblock \showarticletitle{Towards Full-System Energy-Accuracy Tradeoffs: A Case Study of An Approximate Smart Camera System}. In \bibinfo{booktitle}{\emph{Design Automation Conference (DAC)}}. \bibinfo{pages}{1--6}.
\newblock


\bibitem[Raha and Raghunathan(2018)]%
        {raha2018approximating}
\bibfield{author}{\bibinfo{person}{Arnab Raha} {and} \bibinfo{person}{Vijay Raghunathan}.} \bibinfo{year}{2018}\natexlab{}.
\newblock \showarticletitle{Approximating Beyond the Processor: Exploring Full-System Energy-Accuracy Tradeoffs in a Smart Camera System}.
\newblock \bibinfo{journal}{\emph{IEEE Trans. on Very Large Scale Integration (VLSI) Systems}} \bibinfo{volume}{26}, \bibinfo{number}{12} (\bibinfo{year}{2018}), \bibinfo{pages}{2884--2897}.
\newblock


\bibitem[Raha et~al\mbox{.}(2017)]%
        {raha_dram}
\bibfield{author}{\bibinfo{person}{Arnab Raha}, \bibinfo{person}{Soubhagya Sutar}, \bibinfo{person}{Hrishikesh Jayakumar}, {and} \bibinfo{person}{Vijay Raghunathan}.} \bibinfo{year}{2017}\natexlab{}.
\newblock \showarticletitle{Quality Configurable Approximate DRAM}.
\newblock \bibinfo{journal}{\emph{IEEE Trans. on Computers}} \bibinfo{volume}{66}, \bibinfo{number}{7} (\bibinfo{year}{2017}), \bibinfo{pages}{1172--1187}.
\newblock


\bibitem[Raha et~al\mbox{.}(2015)]%
        {2015_Raha_DATE}
\bibfield{author}{\bibinfo{person}{Arnab Raha}, \bibinfo{person}{Swagath Venkataramani}, \bibinfo{person}{Vijay Raghunathan}, {and} \bibinfo{person}{Anand Raghunathan}.} \bibinfo{year}{2015}\natexlab{}.
\newblock \showarticletitle{Quality Configurable Reduce-and-Rank for Energy Efficient Approximate Computing}. In \bibinfo{booktitle}{\emph{Design, Automation \& Test in Europe (DATE)}}. \bibinfo{pages}{665--670}.
\newblock


\bibitem[Rahimi et~al\mbox{.}(2013)]%
        {2013_Rahimi_IEEEtcasii}
\bibfield{author}{\bibinfo{person}{Abbas Rahimi}, \bibinfo{person}{Luca Benini}, {and} \bibinfo{person}{Rajesh~K. Gupta}.} \bibinfo{year}{2013}\natexlab{}.
\newblock \showarticletitle{Spatial Memoization: Concurrent Instruction Reuse to Correct Timing Errors in {SIMD} Architectures}.
\newblock \bibinfo{journal}{\emph{IEEE Trans. on Circuits and Systems II: Express Briefs}} \bibinfo{volume}{60}, \bibinfo{number}{12} (\bibinfo{year}{2013}), \bibinfo{pages}{847--851}.
\newblock


\bibitem[Ramasubramanian et~al\mbox{.}(2013)]%
        {2013_Ramasubramanian_DAC}
\bibfield{author}{\bibinfo{person}{Shankar~Ganesh Ramasubramanian}, \bibinfo{person}{Swagath Venkataramani}, \bibinfo{person}{Adithya Parandhaman}, {and} \bibinfo{person}{Anand Raghunathan}.} \bibinfo{year}{2013}\natexlab{}.
\newblock \showarticletitle{Relax-and-Retime: A Methodology for Energy-Efficient Recovery Based Design}. In \bibinfo{booktitle}{\emph{Design Automation Conference (DAC)}}. \bibinfo{pages}{1--6}.
\newblock


\bibitem[Ranjan et~al\mbox{.}(2020)]%
        {2020_Ranjan_TVLSI}
\bibfield{author}{\bibinfo{person}{Ashish Ranjan}, \bibinfo{person}{Arnab Raha}, \bibinfo{person}{Vijay Raghunathan}, {and} \bibinfo{person}{Anand Raghunathan}.} \bibinfo{year}{2020}\natexlab{}.
\newblock \showarticletitle{Approximate Memory Compression}.
\newblock \bibinfo{journal}{\emph{IEEE Trans. on Very Large Scale Integration (VLSI) Systems}} \bibinfo{volume}{28}, \bibinfo{number}{4} (\bibinfo{year}{2020}), \bibinfo{pages}{980--991}.
\newblock


\bibitem[Ranjan et~al\mbox{.}(2014)]%
        {2014_Ranjan_DATE}
\bibfield{author}{\bibinfo{person}{Ashish Ranjan}, \bibinfo{person}{Arnab Raha}, \bibinfo{person}{Swagath Venkataramani}, \bibinfo{person}{Kaushik Roy}, {and} \bibinfo{person}{Anand Raghunathan}.} \bibinfo{year}{2014}\natexlab{}.
\newblock \showarticletitle{{ASLAN}: Synthesis of Approximate Sequential Circuits}. In \bibinfo{booktitle}{\emph{Design, Automation \& Test in Europe (DATE)}}. \bibinfo{pages}{1--6}.
\newblock


\bibitem[Ranjan et~al\mbox{.}(2015)]%
        {2015_Ranjan_DAC}
\bibfield{author}{\bibinfo{person}{Ashish Ranjan}, \bibinfo{person}{Swagath Venkataramani}, \bibinfo{person}{Xuanyao Fong}, \bibinfo{person}{Kaushik Roy}, {and} \bibinfo{person}{Anand Raghunathan}.} \bibinfo{year}{2015}\natexlab{}.
\newblock \showarticletitle{Approximate Storage for Energy Efficient Spintronic Memories}. In \bibinfo{booktitle}{\emph{Design Automation Conference (DAC)}}. \bibinfo{pages}{1--6}.
\newblock


\bibitem[Rastegari et~al\mbox{.}(2016)]%
        {rastegari2016xnor}
\bibfield{author}{\bibinfo{person}{Mohammad Rastegari}, \bibinfo{person}{Vicente Ordonez}, \bibinfo{person}{Joseph Redmon}, {and} \bibinfo{person}{Ali Farhadi}.} \bibinfo{year}{2016}\natexlab{}.
\newblock \showarticletitle{XNOR-Net: ImageNet Classification Using Binary Convolutional Neural Networks}. In \bibinfo{booktitle}{\emph{European Conference on Computer Vision (ECCV)}}. \bibinfo{pages}{525--542}.
\newblock


\bibitem[Renganarayana et~al\mbox{.}(2012)]%
        {2012_Renganarayana_RACES}
\bibfield{author}{\bibinfo{person}{Lakshminarayanan Renganarayana}, \bibinfo{person}{Vijayalakshmi Srinivasan}, \bibinfo{person}{Ravi Nair}, {and} \bibinfo{person}{Daniel Prener}.} \bibinfo{year}{2012}\natexlab{}.
\newblock \showarticletitle{Programming with Relaxed Synchronization}. In \bibinfo{booktitle}{\emph{ACM Workshop on Relaxing Synchronization for Multicore and Manycore Scalability (RACES)}}. \bibinfo{pages}{41–50}.
\newblock


\bibitem[Reviriego et~al\mbox{.}(2022)]%
        {2022_Reviriego_TEPC}
\bibfield{author}{\bibinfo{person}{Pedro Reviriego}, \bibinfo{person}{Shanshan Liu}, \bibinfo{person}{Otmar Ertl}, \bibinfo{person}{Farzad Niknia}, {and} \bibinfo{person}{Fabrizio Lombardi}.} \bibinfo{year}{2022}\natexlab{}.
\newblock \showarticletitle{Computing the Similarity Estimate Using Approximate Memory}.
\newblock \bibinfo{journal}{\emph{IEEE Trans. on Emerging Topics in Computing}} \bibinfo{volume}{10}, \bibinfo{number}{3} (\bibinfo{year}{2022}), \bibinfo{pages}{1593--1604}.
\newblock


\bibitem[Rinard(2006)]%
        {2006_Rinard_ICS}
\bibfield{author}{\bibinfo{person}{Martin~C. Rinard}.} \bibinfo{year}{2006}\natexlab{}.
\newblock \showarticletitle{Probabilistic Accuracy Bounds for Fault-Tolerant Computations That Discard Tasks}. In \bibinfo{booktitle}{\emph{ACM Int'l. Conference on Supercomputing (ICS)}}. \bibinfo{pages}{324–334}.
\newblock


\bibitem[Rinard(2007)]%
        {2007_Rinard_OOPSLA}
\bibfield{author}{\bibinfo{person}{Martin~C. Rinard}.} \bibinfo{year}{2007}\natexlab{}.
\newblock \showarticletitle{Using Early Phase Termination to Eliminate Load Imbalances at Barrier Synchronization Points}. In \bibinfo{booktitle}{\emph{ACM SIGPLAN Int'l. Conference on Object-Oriented Programming, Systems, Languages, and Applications (OOPSLA)}}. \bibinfo{pages}{369–386}.
\newblock


\bibitem[Rubio-Gonz\'{a}lez et~al\mbox{.}(2016)]%
        {2016_Rubio_ICSE}
\bibfield{author}{\bibinfo{person}{Cindy Rubio-Gonz\'{a}lez}, \bibinfo{person}{Cuong Nguyen}, \bibinfo{person}{Benjamin Mehne}, \bibinfo{person}{Koushik Sen}, \bibinfo{person}{James Demmel}, \bibinfo{person}{William Kahan}, \bibinfo{person}{Costin Iancu}, \bibinfo{person}{Wim Lavrijsen}, \bibinfo{person}{David~H. Bailey}, {and} \bibinfo{person}{David Hough}.} \bibinfo{year}{2016}\natexlab{}.
\newblock \showarticletitle{Floating-Point Precision Tuning Using {B}lame {A}nalysis}. In \bibinfo{booktitle}{\emph{IEEE/ACM Int'l. Conference on Software Engineering (ICSE)}}. \bibinfo{pages}{1074–1085}.
\newblock


\bibitem[Rubio-González et~al\mbox{.}(2013)]%
        {2013_Rubio_SC}
\bibfield{author}{\bibinfo{person}{Cindy Rubio-González}, \bibinfo{person}{Cuong Nguyen}, \bibinfo{person}{Hong~Diep Nguyen}, \bibinfo{person}{James Demmel}, \bibinfo{person}{William Kahan}, \bibinfo{person}{Koushik Sen}, \bibinfo{person}{David~H. Bailey}, \bibinfo{person}{Costin Iancu}, {and} \bibinfo{person}{David Hough}.} \bibinfo{year}{2013}\natexlab{}.
\newblock \showarticletitle{Precimonious: Tuning Assistant for Floating-Point Precision}. In \bibinfo{booktitle}{\emph{SC13: Int'l. Conference on High Performance Computing, Networking, Storage and Analysis}}. \bibinfo{pages}{1--12}.
\newblock


\bibitem[Saadat et~al\mbox{.}(2018)]%
        {saadat2018minimally}
\bibfield{author}{\bibinfo{person}{Hassaan Saadat}, \bibinfo{person}{Haseeb Bokhari}, {and} \bibinfo{person}{Sri Parameswaran}.} \bibinfo{year}{2018}\natexlab{}.
\newblock \showarticletitle{Minimally Biased Multipliers for Approximate Integer and Floating-Point Multiplication}.
\newblock \bibinfo{journal}{\emph{IEEE Trans. on Computer-Aided Design of Integrated Circuits and Systems}} \bibinfo{volume}{37}, \bibinfo{number}{11} (\bibinfo{year}{2018}), \bibinfo{pages}{2623--2635}.
\newblock


\bibitem[Saadat et~al\mbox{.}(2020)]%
        {2020_Saadat_DATE}
\bibfield{author}{\bibinfo{person}{Hassaan Saadat}, \bibinfo{person}{Haris Javaid}, \bibinfo{person}{Aleksandar Ignjatovic}, {and} \bibinfo{person}{Sri Parameswaran}.} \bibinfo{year}{2020}\natexlab{}.
\newblock \showarticletitle{{REALM}: Reduced-Error Approximate Log-based Integer Multiplier}. In \bibinfo{booktitle}{\emph{Design, Automation \& Test in Europe (DATE)}}. \bibinfo{pages}{1366--1371}.
\newblock


\bibitem[Saadat et~al\mbox{.}(2019)]%
        {2019_Saadat_DAC}
\bibfield{author}{\bibinfo{person}{Hassaan Saadat}, \bibinfo{person}{Haris Javaid}, {and} \bibinfo{person}{Sri Parameswaran}.} \bibinfo{year}{2019}\natexlab{}.
\newblock \showarticletitle{Approximate Integer and Floating-Point Dividers with Near-Zero Error Bias}. In \bibinfo{booktitle}{\emph{Design Automation Conference (DAC)}}. \bibinfo{pages}{1--6}.
\newblock


\bibitem[Sabetzadeh et~al\mbox{.}(2019)]%
        {2019_Sabetzadeh_IEEEtcasi}
\bibfield{author}{\bibinfo{person}{Farnaz Sabetzadeh}, \bibinfo{person}{Mohammad~Hossein Moaiyeri}, {and} \bibinfo{person}{Mohammad Ahmadinejad}.} \bibinfo{year}{2019}\natexlab{}.
\newblock \showarticletitle{A Majority-Based Imprecise Multiplier for Ultra-Efficient Approximate Image Multiplication}.
\newblock \bibinfo{journal}{\emph{IEEE Trans. on Circuits and Systems I: Regular Papers}} \bibinfo{volume}{66}, \bibinfo{number}{11} (\bibinfo{year}{2019}), \bibinfo{pages}{4200--4208}.
\newblock


\bibitem[Salajegheh et~al\mbox{.}(2011)]%
        {2011_Salajegheh_FAST}
\bibfield{author}{\bibinfo{person}{Mastooreh Salajegheh}, \bibinfo{person}{Yue Wang}, \bibinfo{person}{Kevin Fu}, \bibinfo{person}{Anxiao Jiang}, {and} \bibinfo{person}{Erik Learned-Miller}.} \bibinfo{year}{2011}\natexlab{}.
\newblock \showarticletitle{Exploiting Half-Wits: Smarter Storage for Low-Power Devices}. In \bibinfo{booktitle}{\emph{USENIX Conference on File and Storage Technologies (FAST)}}. \bibinfo{pages}{1--14}.
\newblock


\bibitem[Samadi et~al\mbox{.}(2014)]%
        {2014_Samadi_ASPLOS}
\bibfield{author}{\bibinfo{person}{Mehrzad Samadi}, \bibinfo{person}{Davoud~Anoushe Jamshidi}, \bibinfo{person}{Janghaeng Lee}, {and} \bibinfo{person}{Scott Mahlke}.} \bibinfo{year}{2014}\natexlab{}.
\newblock \showarticletitle{Paraprox: Pattern-Based Approximation for Data Parallel Applications}. In \bibinfo{booktitle}{\emph{ACM Int'l. Conference on Architectural Support for Programming Languages and Operating Systems (ASPLOS)}}. \bibinfo{pages}{35–50}.
\newblock


\bibitem[Samadi et~al\mbox{.}(2013)]%
        {2013_Samadi_MICRO}
\bibfield{author}{\bibinfo{person}{Mehrzad Samadi}, \bibinfo{person}{Janghaeng Lee}, \bibinfo{person}{D.~Anoushe Jamshidi}, \bibinfo{person}{Amir Hormati}, {and} \bibinfo{person}{Scott Mahlke}.} \bibinfo{year}{2013}\natexlab{}.
\newblock \showarticletitle{{SAGE}: Self-Tuning Approximation for Graphics Engines}. In \bibinfo{booktitle}{\emph{IEEE/ACM Int'l. Symposium on Microarchitecture (MICRO)}}. \bibinfo{pages}{13--24}.
\newblock


\bibitem[Sampaio et~al\mbox{.}(2015)]%
        {2015_Sampaio_CASES}
\bibfield{author}{\bibinfo{person}{Felipe Sampaio}, \bibinfo{person}{Muhammad Shafique}, \bibinfo{person}{Bruno Zatt}, \bibinfo{person}{Sergio Bampi}, {and} \bibinfo{person}{Jörg Henkel}.} \bibinfo{year}{2015}\natexlab{}.
\newblock \showarticletitle{Approximation-Aware Multi-Level Cells {STT-RAM} Cache Architecture}. In \bibinfo{booktitle}{\emph{Int'l. Conference on Compilers, Architecture and Synthesis for Embedded Systems (CASES)}}. \bibinfo{pages}{79--88}.
\newblock


\bibitem[Sampson et~al\mbox{.}(2015)]%
        {2015_Sampson_UOW}
\bibfield{author}{\bibinfo{person}{Adrian Sampson}, \bibinfo{person}{Andr{\'e} Baixo}, \bibinfo{person}{Benjamin Ransford}, \bibinfo{person}{Thierry Moreau}, \bibinfo{person}{Joshua Yip}, \bibinfo{person}{Luis Ceze}, {and} \bibinfo{person}{Mark Oskin}.} \bibinfo{year}{2015}\natexlab{}.
\newblock \showarticletitle{{ACCEPT}: A Programmer-Guided Compiler Framework for Practical Approximate Computing}.
\newblock \bibinfo{journal}{\emph{University of Washington Technical Report}} (\bibinfo{year}{2015}), \bibinfo{pages}{1--14}.
\newblock


\bibitem[Sampson et~al\mbox{.}(2011)]%
        {2011_Sampson_PLDI}
\bibfield{author}{\bibinfo{person}{Adrian Sampson}, \bibinfo{person}{Werner Dietl}, \bibinfo{person}{Emily Fortuna}, \bibinfo{person}{Danushen Gnanapragasam}, \bibinfo{person}{Luis Ceze}, {and} \bibinfo{person}{Dan Grossman}.} \bibinfo{year}{2011}\natexlab{}.
\newblock \showarticletitle{Ener{J}: Approximate Data Types for Safe and General Low-Power Computation}. In \bibinfo{booktitle}{\emph{ACM SIGPLAN Conference on Programming Language Design and Implementation (PLDI)}}. \bibinfo{pages}{164–174}.
\newblock


\bibitem[Sampson et~al\mbox{.}(2014)]%
        {2014_Sampson_ACMtc}
\bibfield{author}{\bibinfo{person}{Adrian Sampson}, \bibinfo{person}{Jacob Nelson}, \bibinfo{person}{Karin Strauss}, {and} \bibinfo{person}{Luis Ceze}.} \bibinfo{year}{2014}\natexlab{}.
\newblock \showarticletitle{Approximate Storage in Solid-State Memories}.
\newblock \bibinfo{journal}{\emph{ACM Trans. on Computer Systems}} \bibinfo{volume}{32}, \bibinfo{number}{3} (\bibinfo{year}{2014}), \bibinfo{pages}{1--23}.
\newblock


\bibitem[Sartori and Kumar(2011)]%
        {vos:proc1}
\bibfield{author}{\bibinfo{person}{John Sartori} {and} \bibinfo{person}{Rakesh Kumar}.} \bibinfo{year}{2011}\natexlab{}.
\newblock \showarticletitle{Architecting Processors to Allow Voltage/Reliability Tradeoffs}. In \bibinfo{booktitle}{\emph{Int'l. Conference on Compilers, Architectures and Synthesis for Embedded Systems (CASES)}}. \bibinfo{pages}{115--124}.
\newblock


\bibitem[Schkufza et~al\mbox{.}(2014)]%
        {2014_Schkufza_PLDI}
\bibfield{author}{\bibinfo{person}{Eric Schkufza}, \bibinfo{person}{Rahul Sharma}, {and} \bibinfo{person}{Alex Aiken}.} \bibinfo{year}{2014}\natexlab{}.
\newblock \showarticletitle{Stochastic Optimization of Floating-Point Programs with Tunable Precision}. In \bibinfo{booktitle}{\emph{ACM SIGPLAN Conference on Programming Language Design and Implementation (PLDI)}}. \bibinfo{pages}{53–64}.
\newblock


\bibitem[Schlachter et~al\mbox{.}(2017)]%
        {2017_Schlachter_IEEEtvlsi}
\bibfield{author}{\bibinfo{person}{Jeremy Schlachter}, \bibinfo{person}{Vincent Camus}, \bibinfo{person}{Krishna~V. Palem}, {and} \bibinfo{person}{Christian Enz}.} \bibinfo{year}{2017}\natexlab{}.
\newblock \showarticletitle{Design and Applications of Approximate Circuits by Gate-Level Pruning}.
\newblock \bibinfo{journal}{\emph{IEEE Trans. on Very Large Scale Integration (VLSI) Systems}} \bibinfo{volume}{25}, \bibinfo{number}{5} (\bibinfo{year}{2017}), \bibinfo{pages}{1694--1702}.
\newblock


\bibitem[Shafique et~al\mbox{.}(2016)]%
        {2016_Shafique_DAC}
\bibfield{author}{\bibinfo{person}{Muhammad Shafique}, \bibinfo{person}{Rehan Hafiz}, \bibinfo{person}{Semeen Rehman}, \bibinfo{person}{Walaa El-Harouni}, {and} \bibinfo{person}{Jörg Henkel}.} \bibinfo{year}{2016}\natexlab{}.
\newblock \showarticletitle{Cross-Layer Approximate Computing: From Logic to Architectures}. In \bibinfo{booktitle}{\emph{Design Automation Conference (DAC)}}. \bibinfo{pages}{1--6}.
\newblock


\bibitem[Shi et~al\mbox{.}(2013)]%
        {2013_Shi_FCCM}
\bibfield{author}{\bibinfo{person}{Kan Shi}, \bibinfo{person}{David Boland}, {and} \bibinfo{person}{George~A. Constantinides}.} \bibinfo{year}{2013}\natexlab{}.
\newblock \showarticletitle{Accuracy-Performance Tradeoffs on an {FPGA} through Overclocking}. In \bibinfo{booktitle}{\emph{IEEE Int'l. Symposium on Field-Programmable Custom Computing Machines (FCCM)}}. \bibinfo{pages}{29--36}.
\newblock


\bibitem[Shi et~al\mbox{.}(2014)]%
        {2014_Shi_DAC}
\bibfield{author}{\bibinfo{person}{Kan Shi}, \bibinfo{person}{David Boland}, \bibinfo{person}{Edward Stott}, \bibinfo{person}{Samuel Bayliss}, {and} \bibinfo{person}{George~A. Constantinides}.} \bibinfo{year}{2014}\natexlab{}.
\newblock \showarticletitle{Datapath Synthesis for Overclocking: Online Arithmetic for Latency-Accuracy Trade-offs}. In \bibinfo{booktitle}{\emph{Design Automation Conference (DAC)}}. \bibinfo{pages}{1--6}.
\newblock


\bibitem[Shi et~al\mbox{.}(2015)]%
        {2015_Shi_IEEEcal}
\bibfield{author}{\bibinfo{person}{Qingchuan Shi}, \bibinfo{person}{Henry Hoffmann}, {and} \bibinfo{person}{Omer Khan}.} \bibinfo{year}{2015}\natexlab{}.
\newblock \showarticletitle{A Cross-Layer Multicore Architecture to Tradeoff Program Accuracy and Resilience Overheads}.
\newblock \bibinfo{journal}{\emph{IEEE Computer Architecture Letters}} \bibinfo{volume}{14}, \bibinfo{number}{2} (\bibinfo{year}{2015}), \bibinfo{pages}{85--89}.
\newblock


\bibitem[Shin et~al\mbox{.}(2019)]%
        {shin2019sensitivity}
\bibfield{author}{\bibinfo{person}{Dongyeob Shin}, \bibinfo{person}{Wonseok Choi}, \bibinfo{person}{Jongsun Park}, {and} \bibinfo{person}{Swaroop Ghosh}.} \bibinfo{year}{2019}\natexlab{}.
\newblock \showarticletitle{Sensitivity-Based Error Resilient Techniques With Heterogeneous Multiply–Accumulate Unit for Voltage Scalable Deep Neural Network Accelerators}.
\newblock \bibinfo{journal}{\emph{IEEE Journal on Emerging and Selected Topics in Circuits and Systems}} \bibinfo{volume}{9}, \bibinfo{number}{3} (\bibinfo{year}{2019}), \bibinfo{pages}{520--531}.
\newblock


\bibitem[Sidiroglou-Douskos et~al\mbox{.}(2011)]%
        {2011_Sidiroglou_FCE}
\bibfield{author}{\bibinfo{person}{Stelios Sidiroglou-Douskos}, \bibinfo{person}{Sasa Misailovic}, \bibinfo{person}{Henry Hoffmann}, {and} \bibinfo{person}{Martin~C. Rinard}.} \bibinfo{year}{2011}\natexlab{}.
\newblock \showarticletitle{Managing Performance vs. Accuracy Trade-Offs with Loop Perforation}. In \bibinfo{booktitle}{\emph{ACM SIGSOFT Symposium and European Conference on Foundations of Software Engineering (FSE)}}. \bibinfo{pages}{124–134}.
\newblock


\bibitem[Sonnino et~al\mbox{.}(2024)]%
        {sonnino2023daism}
\bibfield{author}{\bibinfo{person}{Lorenzo Sonnino}, \bibinfo{person}{Shaswot Shresthamali}, \bibinfo{person}{Yuan He}, {and} \bibinfo{person}{Masaaki Kondo}.} \bibinfo{year}{2024}\natexlab{}.
\newblock \showarticletitle{DAISM: Digital Approximate In-SRAM Multiplier-Based Accelerator for DNN Training and Inference}. In \bibinfo{booktitle}{\emph{Design, Automation \& Test in Europe (DATE)}}. \bibinfo{pages}{1--6}.
\newblock


\bibitem[Sorber et~al\mbox{.}(2007)]%
        {2007_Sorber_SenSys}
\bibfield{author}{\bibinfo{person}{Jacob Sorber}, \bibinfo{person}{Alexander Kostadinov}, \bibinfo{person}{Matthew Garber}, \bibinfo{person}{Matthew Brennan}, \bibinfo{person}{Mark~D. Corner}, {and} \bibinfo{person}{Emery~D. Berger}.} \bibinfo{year}{2007}\natexlab{}.
\newblock \showarticletitle{Eon: A Language and Runtime System for Perpetual Systems}. In \bibinfo{booktitle}{\emph{ACM Int'l. Conference on Embedded Networked Sensor Systems (SenSys)}}. \bibinfo{pages}{161–174}.
\newblock


\bibitem[Source(2023)]%
        {riscv}
\bibfield{author}{\bibinfo{person}{The~Open Source}.} \bibinfo{year}{2023}\natexlab{}.
\newblock \showarticletitle{History – RISC-V International}.
\newblock  (\bibinfo{year}{2023}).
\newblock
\urldef\tempurl%
\url{https://riscv.org/about/history/}
\showURL{%
\tempurl}


\bibitem[Spantidi et~al\mbox{.}(2021)]%
        {spantidi2021positive}
\bibfield{author}{\bibinfo{person}{Ourania Spantidi}, \bibinfo{person}{Georgios Zervakis}, \bibinfo{person}{Iraklis Anagnostopoulos}, \bibinfo{person}{Hussam Amrouch}, {and} \bibinfo{person}{J{\"o}rg Henkel}.} \bibinfo{year}{2021}\natexlab{}.
\newblock \showarticletitle{Positive/Negative Approximate Multipliers for DNN Accelerators}. In \bibinfo{booktitle}{\emph{IEEE/ACM Int'l. Conference On Computer Aided Design (ICCAD)}}. \bibinfo{pages}{1--9}.
\newblock


\bibitem[Sreeram and Pande(2010)]%
        {2010_Sreeram_IISWC}
\bibfield{author}{\bibinfo{person}{Jaswanth Sreeram} {and} \bibinfo{person}{Santosh Pande}.} \bibinfo{year}{2010}\natexlab{}.
\newblock \showarticletitle{Exploiting Approximate Value Locality for Data Synchronization on Multi-Core Processors}. In \bibinfo{booktitle}{\emph{IEEE Int'l. Symposium on Workload Characterization (IISWC)}}. \bibinfo{pages}{1--10}.
\newblock


\bibitem[Stanley-Marbell et~al\mbox{.}(2020)]%
        {2020_Stanley_ACMsrv}
\bibfield{author}{\bibinfo{person}{Phillip Stanley-Marbell} {et~al\mbox{.}}} \bibinfo{year}{2020}\natexlab{}.
\newblock \showarticletitle{Exploiting Errors for Efficiency: A Survey from Circuits to Applications}.
\newblock \bibinfo{journal}{\emph{Comput. Surveys}} \bibinfo{volume}{53}, \bibinfo{number}{3} (\bibinfo{year}{2020}), \bibinfo{pages}{1--39}.
\newblock


\bibitem[Stitt and Campbell(2020)]%
        {2020_Stitt_ACMtecs}
\bibfield{author}{\bibinfo{person}{Greg Stitt} {and} \bibinfo{person}{David Campbell}.} \bibinfo{year}{2020}\natexlab{}.
\newblock \showarticletitle{{PANDORA}: An Architecture-Independent Parallelizing Approximation-Discovery Framework}.
\newblock \bibinfo{journal}{\emph{ACM Trans. on Embedded Computing Systems}} \bibinfo{volume}{19}, \bibinfo{number}{5} (\bibinfo{year}{2020}), \bibinfo{pages}{1--17}.
\newblock


\bibitem[Strollo et~al\mbox{.}(2020)]%
        {2020_Strollo_IEEEtcasi}
\bibfield{author}{\bibinfo{person}{Antonio Giuseppe~Maria Strollo}, \bibinfo{person}{Ettore Napoli}, \bibinfo{person}{Davide De~Caro}, \bibinfo{person}{Nicola Petra}, {and} \bibinfo{person}{Gennaro~Di Meo}.} \bibinfo{year}{2020}\natexlab{}.
\newblock \showarticletitle{Comparison and Extension of Approximate 4-2 Compressors for Low-Power Approximate Multipliers}.
\newblock \bibinfo{journal}{\emph{IEEE Trans. on Circuits and Systems I: Regular Papers}} \bibinfo{volume}{67}, \bibinfo{number}{9} (\bibinfo{year}{2020}), \bibinfo{pages}{3021--3034}.
\newblock


\bibitem[Sui et~al\mbox{.}(2021)]%
        {sui2021chip}
\bibfield{author}{\bibinfo{person}{Yang Sui}, \bibinfo{person}{Miao Yin}, \bibinfo{person}{Yi Xie}, \bibinfo{person}{Huy Phan}, \bibinfo{person}{Saman Aliari~Zonouz}, {and} \bibinfo{person}{Bo Yuan}.} \bibinfo{year}{2021}\natexlab{}.
\newblock \showarticletitle{CHIP: CHannel Independence-based Pruning for Compact Neural Networks}.
\newblock \bibinfo{journal}{\emph{Advances in Neural Information Processing Systems}}  \bibinfo{volume}{34} (\bibinfo{year}{2021}), \bibinfo{pages}{24604--24616}.
\newblock


\bibitem[Taheri et~al\mbox{.}(2023)]%
        {taheri2023deepaxe}
\bibfield{author}{\bibinfo{person}{Mahdi Taheri}, \bibinfo{person}{Mohammad Riazati}, \bibinfo{person}{Mohammad~Hasan Ahmadilivani}, \bibinfo{person}{Maksim Jenihhin}, \bibinfo{person}{Masoud Daneshtalab}, \bibinfo{person}{Jaan Raik}, \bibinfo{person}{Mikael Sjodin}, {and} \bibinfo{person}{Bjorn Lisper}.} \bibinfo{year}{2023}\natexlab{}.
\newblock \showarticletitle{DeepAxe: A Framework for Exploration of Approximation and Reliability Trade-offs in DNN Accelerators}. In \bibinfo{booktitle}{\emph{Int'l. Symposium on Quality Electronic Design (ISQED)}}. \bibinfo{pages}{1--8}.
\newblock


\bibitem[Tan et~al\mbox{.}(2015)]%
        {2015_Tan_ASP-DAC}
\bibfield{author}{\bibinfo{person}{Cheng Tan}, \bibinfo{person}{Thannirmalai~Somu Muthukaruppan}, \bibinfo{person}{Tulika Mitra}, {and} \bibinfo{person}{Ju Lei}.} \bibinfo{year}{2015}\natexlab{}.
\newblock \showarticletitle{Approximation-Aware Scheduling on Heterogeneous Multi-Core Architectures}. In \bibinfo{booktitle}{\emph{Asia and South Pacific Design Automation Conference (ASP-DAC)}}. \bibinfo{pages}{618--623}.
\newblock


\bibitem[Tan and Motani(2020)]%
        {tan2020dropnet}
\bibfield{author}{\bibinfo{person}{Chong Min~John Tan} {and} \bibinfo{person}{Mehul Motani}.} \bibinfo{year}{2020}\natexlab{}.
\newblock \showarticletitle{DropNet: Reducing Neural Network Complexity via Iterative Pruning}. In \bibinfo{booktitle}{\emph{Int'l. Conference on Machine Learning (ICML)}}. \bibinfo{pages}{9356--9366}.
\newblock


\bibitem[Tan and Le(2019)]%
        {tan2019efficientnet}
\bibfield{author}{\bibinfo{person}{Mingxing Tan} {and} \bibinfo{person}{Quoc Le}.} \bibinfo{year}{2019}\natexlab{}.
\newblock \showarticletitle{EfficientNet: Rethinking Model Scaling for Convolutional Neural Networks}. In \bibinfo{booktitle}{\emph{Int'l. Conference on Machine Learning (ICML)}}. \bibinfo{pages}{6105--6114}.
\newblock


\bibitem[Tann et~al\mbox{.}(2016)]%
        {tann2016runtime}
\bibfield{author}{\bibinfo{person}{Hokchhay Tann}, \bibinfo{person}{Soheil Hashemi}, \bibinfo{person}{R~Iris Bahar}, {and} \bibinfo{person}{Sherief Reda}.} \bibinfo{year}{2016}\natexlab{}.
\newblock \showarticletitle{Runtime Configurable Deep Neural Networks for Energy-Accuracy Trade-off}. In \bibinfo{booktitle}{\emph{Conference on Hardware/Software Codesign and System Synthesis}}. \bibinfo{pages}{1--10}.
\newblock


\bibitem[Teimoori et~al\mbox{.}(2018)]%
        {2018_Teimoori_DATE}
\bibfield{author}{\bibinfo{person}{Mohammad~Taghi Teimoori}, \bibinfo{person}{Muhammad~Abdullah Hanif}, \bibinfo{person}{Alireza Ejlali}, {and} \bibinfo{person}{Muhammad Shafique}.} \bibinfo{year}{2018}\natexlab{}.
\newblock \showarticletitle{Ad{AM}: Adaptive Approximation Management for the Non-Volatile Memory Hierarchies}. In \bibinfo{booktitle}{\emph{Design, Automation \& Test in Europe Conference (DATE)}}. \bibinfo{pages}{785--790}.
\newblock


\bibitem[Tian et~al\mbox{.}(2015)]%
        {2015_Tian_GLSVLSI}
\bibfield{author}{\bibinfo{person}{Ye Tian}, \bibinfo{person}{Qian Zhang}, \bibinfo{person}{Ting Wang}, \bibinfo{person}{Feng Yuan}, {and} \bibinfo{person}{Qiang Xu}.} \bibinfo{year}{2015}\natexlab{}.
\newblock \showarticletitle{Approx{MA}: Approximate Memory Access for Dynamic Precision Scaling}. In \bibinfo{booktitle}{\emph{Great Lakes Symposium on VLSI (GLSVLSI)}}. \bibinfo{pages}{337–342}.
\newblock


\bibitem[Tolpin et~al\mbox{.}(2016)]%
        {2016_Tolpin_IFL}
\bibfield{author}{\bibinfo{person}{David Tolpin}, \bibinfo{person}{Jan-Willem van~de Meent}, \bibinfo{person}{Hongseok Yang}, {and} \bibinfo{person}{Frank Wood}.} \bibinfo{year}{2016}\natexlab{}.
\newblock \showarticletitle{Design and Implementation of Probabilistic Programming Language Anglican}. In \bibinfo{booktitle}{\emph{Symposium on Implementation and Application of Functional Programming Languages (IFL)}}. \bibinfo{pages}{1--12}.
\newblock


\bibitem[Tseng et~al\mbox{.}(2013)]%
        {2013_Tseng_DAC}
\bibfield{author}{\bibinfo{person}{Hung-Wei Tseng}, \bibinfo{person}{Laura~M. Grupp}, {and} \bibinfo{person}{Steven Swanson}.} \bibinfo{year}{2013}\natexlab{}.
\newblock \showarticletitle{Underpowering {NAND} Flash: Profits and Perils}. In \bibinfo{booktitle}{\emph{Design Automation Conference (DAC)}}. \bibinfo{pages}{1--6}.
\newblock


\bibitem[Tziantzioulis et~al\mbox{.}(2018)]%
        {2018_Tziantzioulis_IEEEmicro}
\bibfield{author}{\bibinfo{person}{Georgios Tziantzioulis}, \bibinfo{person}{Nikos Hardavellas}, {and} \bibinfo{person}{Simone Campanoni}.} \bibinfo{year}{2018}\natexlab{}.
\newblock \showarticletitle{Temporal Approximate Function Memoization}.
\newblock \bibinfo{journal}{\emph{IEEE Micro}} \bibinfo{volume}{38}, \bibinfo{number}{4} (\bibinfo{year}{2018}), \bibinfo{pages}{60--70}.
\newblock


\bibitem[Ullah et~al\mbox{.}(2018)]%
        {2018_Ullah_DAC}
\bibfield{author}{\bibinfo{person}{Salim Ullah}, \bibinfo{person}{Sanjeev~Sripadraj Murthy}, {and} \bibinfo{person}{Akash Kumar}.} \bibinfo{year}{2018}\natexlab{}.
\newblock \showarticletitle{{SMA}pprox{L}ib: Library of {FPGA}-based Approximate Multipliers}. In \bibinfo{booktitle}{\emph{Design Automation Conference (DAC)}}. \bibinfo{pages}{1--6}.
\newblock


\bibitem[Vahdat et~al\mbox{.}(2019)]%
        {2019_Vahdat_IEEEtvlsi}
\bibfield{author}{\bibinfo{person}{Shaghayegh Vahdat}, \bibinfo{person}{Mehdi Kamal}, \bibinfo{person}{Ali Afzali-Kusha}, {and} \bibinfo{person}{Massoud Pedram}.} \bibinfo{year}{2019}\natexlab{}.
\newblock \showarticletitle{{TOSAM}: An Energy-Efficient Truncation- and Rounding-Based Scalable Approximate Multiplier}.
\newblock \bibinfo{journal}{\emph{IEEE Trans. on Very Large Scale Integration (VLSI) Systems}} \bibinfo{volume}{27}, \bibinfo{number}{5} (\bibinfo{year}{2019}), \bibinfo{pages}{1161--1173}.
\newblock


\bibitem[Vahdat et~al\mbox{.}(2017)]%
        {2017_Vahdat_DATE}
\bibfield{author}{\bibinfo{person}{Shaghayegh Vahdat}, \bibinfo{person}{Mehdi Kamal}, \bibinfo{person}{Ali Afzali-Kusha}, \bibinfo{person}{Massoud Pedram}, {and} \bibinfo{person}{Zainalabedin Navabi}.} \bibinfo{year}{2017}\natexlab{}.
\newblock \showarticletitle{Trunc{A}pp: A Truncation-Based Approximate Divider for Energy Efficient {DSP} Applications}. In \bibinfo{booktitle}{\emph{Design, Automation \& Test in Europe (DATE)}}. \bibinfo{pages}{1635--1638}.
\newblock


\bibitem[Vasicek et~al\mbox{.}(2019)]%
        {2019_Vasicek_DATE}
\bibfield{author}{\bibinfo{person}{Zdenek Vasicek}, \bibinfo{person}{Vojtech Mrazek}, {and} \bibinfo{person}{Lukas Sekanina}.} \bibinfo{year}{2019}\natexlab{}.
\newblock \showarticletitle{Automated Circuit Approximation Method Driven by Data Distribution}. In \bibinfo{booktitle}{\emph{Design, Automation \& Test in Europe (DATE)}}. \bibinfo{pages}{96--101}.
\newblock


\bibitem[Vasicek and Sekanina(2015)]%
        {2015_Vasicek_IEEEtec}
\bibfield{author}{\bibinfo{person}{Zdenek Vasicek} {and} \bibinfo{person}{Lukas Sekanina}.} \bibinfo{year}{2015}\natexlab{}.
\newblock \showarticletitle{Evolutionary Approach to Approximate Digital Circuits Design}.
\newblock \bibinfo{journal}{\emph{IEEE Trans. on Evolutionary Computation}} \bibinfo{volume}{19}, \bibinfo{number}{3} (\bibinfo{year}{2015}), \bibinfo{pages}{432--444}.
\newblock


\bibitem[Vassiliadis et~al\mbox{.}(2015)]%
        {2015_Vassiliadis_PPoPP}
\bibfield{author}{\bibinfo{person}{Vassilis Vassiliadis}, \bibinfo{person}{Konstantinos Parasyris}, \bibinfo{person}{Charalambos Chalios}, \bibinfo{person}{Christos~D. Antonopoulos}, \bibinfo{person}{Spyros Lalis}, \bibinfo{person}{Nikolaos Bellas}, \bibinfo{person}{Hans Vandierendonck}, {and} \bibinfo{person}{Dimitrios~S. Nikolopoulos}.} \bibinfo{year}{2015}\natexlab{}.
\newblock \showarticletitle{A Programming Model and Runtime System for Significance-Aware Energy-Efficient Computing}. In \bibinfo{booktitle}{\emph{ACM SIGPLAN Symposium on Principles and Practice of Parallel Programming (PPoPP)}}. \bibinfo{pages}{275–276}.
\newblock


\bibitem[Vassiliadis et~al\mbox{.}(2016)]%
        {2016_Vassiliadis_CGO}
\bibfield{author}{\bibinfo{person}{Vassilis Vassiliadis}, \bibinfo{person}{Jan Riehme}, \bibinfo{person}{Jens Deussen}, \bibinfo{person}{Konstantinos Parasyris}, \bibinfo{person}{Christos~D. Antonopoulos}, \bibinfo{person}{Nikolaos Bellas}, \bibinfo{person}{Spyros Lalis}, {and} \bibinfo{person}{Uwe Naumann}.} \bibinfo{year}{2016}\natexlab{}.
\newblock \showarticletitle{Towards Automatic Significance Analysis for Approximate Computing}. In \bibinfo{booktitle}{\emph{IEEE/ACM Int'l. Symposium on Code Generation and Optimization (CGO)}}. \bibinfo{pages}{182–193}.
\newblock


\bibitem[Venkatachalam et~al\mbox{.}(2019)]%
        {2019_Venkatachalam_IEEEtc}
\bibfield{author}{\bibinfo{person}{Suganthi Venkatachalam}, \bibinfo{person}{Elizabeth Adams}, \bibinfo{person}{Hyuk~Jae Lee}, {and} \bibinfo{person}{Seok-Bum Ko}.} \bibinfo{year}{2019}\natexlab{}.
\newblock \showarticletitle{Design and Analysis of Area and Power Efficient Approximate Booth Multipliers}.
\newblock \bibinfo{journal}{\emph{IEEE Trans. on Computers}} \bibinfo{volume}{68}, \bibinfo{number}{11} (\bibinfo{year}{2019}), \bibinfo{pages}{1697--1703}.
\newblock


\bibitem[Venkataramani et~al\mbox{.}(2020)]%
        {swagath:2020}
\bibfield{author}{\bibinfo{person}{Swagath Venkataramani} {et~al\mbox{.}}} \bibinfo{year}{2020}\natexlab{}.
\newblock \showarticletitle{Efficient AI System Design With Cross-Layer Approximate Computing}.
\newblock \bibinfo{journal}{\emph{Proc. IEEE}} \bibinfo{volume}{108}, \bibinfo{number}{12} (\bibinfo{year}{2020}), \bibinfo{pages}{2232--2250}.
\newblock


\bibitem[Venkataramani et~al\mbox{.}(2015a)]%
        {venkataramani2015sapphire}
\bibfield{author}{\bibinfo{person}{Swagath Venkataramani}, \bibinfo{person}{Victor Bahl}, \bibinfo{person}{Xian-Sheng Hua}, \bibinfo{person}{Jie Liu}, \bibinfo{person}{Jin Li}, \bibinfo{person}{Matthai Phillipose}, \bibinfo{person}{Bodhi Priyantha}, {and} \bibinfo{person}{Mohammed Shoaib}.} \bibinfo{year}{2015}\natexlab{a}.
\newblock \showarticletitle{SAPPHIRE: An Always-on Context-Aware Computer Vision System for Portable Devices}. In \bibinfo{booktitle}{\emph{Design, Automation \& Test in Europe (DATE)}}. \bibinfo{pages}{1491--1496}.
\newblock


\bibitem[Venkataramani et~al\mbox{.}(2015b)]%
        {2015_Venkataramani_DAC}
\bibfield{author}{\bibinfo{person}{Swagath Venkataramani}, \bibinfo{person}{Srimat~T. Chakradhar}, \bibinfo{person}{Kaushik Roy}, {and} \bibinfo{person}{Anand Raghunathan}.} \bibinfo{year}{2015}\natexlab{b}.
\newblock \showarticletitle{Approximate Computing and the Quest for Computing Efficiency}. In \bibinfo{booktitle}{\emph{Design Automation Conference (DAC)}}. \bibinfo{pages}{1--6}.
\newblock


\bibitem[Venkataramani et~al\mbox{.}(2015c)]%
        {venkataramani2015scalable}
\bibfield{author}{\bibinfo{person}{Swagath Venkataramani}, \bibinfo{person}{Anand Raghunathan}, \bibinfo{person}{Jie Liu}, {and} \bibinfo{person}{Mohammed Shoaib}.} \bibinfo{year}{2015}\natexlab{c}.
\newblock \showarticletitle{Scalable-Effort Classifiers for Energy-Efficient Machine Learning}. In \bibinfo{booktitle}{\emph{Design Automation Conference (DAC)}}. \bibinfo{pages}{1--6}.
\newblock


\bibitem[Venkataramani et~al\mbox{.}(2014)]%
        {venkataramani2014axnn}
\bibfield{author}{\bibinfo{person}{Swagath Venkataramani}, \bibinfo{person}{Ashish Ranjan}, \bibinfo{person}{Kaushik Roy}, {and} \bibinfo{person}{Anand Raghunathan}.} \bibinfo{year}{2014}\natexlab{}.
\newblock \showarticletitle{AxNN: Energy-Efficient Neuromorphic Systems Using Approximate Computing}. In \bibinfo{booktitle}{\emph{ACM/IEEE Int'l. Symposium on Low Power Electronics and Design (ISLPED)}}. \bibinfo{pages}{27--32}.
\newblock


\bibitem[Venkataramani et~al\mbox{.}(2013)]%
        {2013_Venkataramani_DATE}
\bibfield{author}{\bibinfo{person}{Swagath Venkataramani}, \bibinfo{person}{Kaushik Roy}, {and} \bibinfo{person}{Anand Raghunathan}.} \bibinfo{year}{2013}\natexlab{}.
\newblock \showarticletitle{Substitute-and-Simplify: A Unified Design Paradigm for Approximate and Quality Configurable Circuits}. In \bibinfo{booktitle}{\emph{Design, Automation \& Test in Europe (DATE)}}. \bibinfo{pages}{1367--1372}.
\newblock


\bibitem[Venkataramani et~al\mbox{.}(2012)]%
        {2012_Venkataramani_DAC}
\bibfield{author}{\bibinfo{person}{Swagath Venkataramani}, \bibinfo{person}{Amit Sabne}, \bibinfo{person}{Vivek Kozhikkottu}, \bibinfo{person}{Kaushik Roy}, {and} \bibinfo{person}{Anand Raghunathan}.} \bibinfo{year}{2012}\natexlab{}.
\newblock \showarticletitle{{SALSA}: Systematic Logic Synthesis of Approximate Circuits}. In \bibinfo{booktitle}{\emph{Design Automation Conference (DAC)}}. \bibinfo{pages}{796--801}.
\newblock


\bibitem[Wang et~al\mbox{.}(2021)]%
        {wang2020neural}
\bibfield{author}{\bibinfo{person}{Huan Wang}, \bibinfo{person}{Can Qin}, \bibinfo{person}{Yulun Zhang}, {and} \bibinfo{person}{Yun Fu}.} \bibinfo{year}{2021}\natexlab{}.
\newblock \showarticletitle{Neural Pruning via Growing Regularization}. In \bibinfo{booktitle}{\emph{Int'l. Conference on Learning Representations (ICLR)}}. \bibinfo{pages}{1--16}.
\newblock


\bibitem[Wang et~al\mbox{.}(2020)]%
        {2020_Wang_IEEEtc}
\bibfield{author}{\bibinfo{person}{Jing Wang}, \bibinfo{person}{Xin Fu}, \bibinfo{person}{Xu Wang}, \bibinfo{person}{Shubo Liu}, \bibinfo{person}{Lan Gao}, {and} \bibinfo{person}{Weigong Zhang}.} \bibinfo{year}{2020}\natexlab{}.
\newblock \showarticletitle{Enabling Energy-Efficient and Reliable Neural Network via Neuron-Level Voltage Scaling}.
\newblock \bibinfo{journal}{\emph{IEEE Trans. on Computers}} \bibinfo{volume}{69}, \bibinfo{number}{10} (\bibinfo{year}{2020}), \bibinfo{pages}{1460--1473}.
\newblock


\bibitem[Wang et~al\mbox{.}(2019)]%
        {wang2019haq}
\bibfield{author}{\bibinfo{person}{Kuan Wang}, \bibinfo{person}{Zhijian Liu}, \bibinfo{person}{Yujun Lin}, \bibinfo{person}{Ji Lin}, {and} \bibinfo{person}{Song Han}.} \bibinfo{year}{2019}\natexlab{}.
\newblock \showarticletitle{HAQ: Hardware-Aware Automated Quantization with Mixed Precision}. In \bibinfo{booktitle}{\emph{IEEE/CVF Conference on Computer Vision and Pattern Recognition (CVPR)}}. \bibinfo{pages}{8612--8620}.
\newblock


\bibitem[Wang et~al\mbox{.}(2017)]%
        {2017_Wang_IEEEtvlsi}
\bibfield{author}{\bibinfo{person}{Ying Wang}, \bibinfo{person}{Jiachao Deng}, \bibinfo{person}{Yuntan Fang}, \bibinfo{person}{Huawei Li}, {and} \bibinfo{person}{Xiaowei Li}.} \bibinfo{year}{2017}\natexlab{}.
\newblock \showarticletitle{Resilience-Aware Frequency Tuning for Neural-Network-Based Approximate Computing Chips}.
\newblock \bibinfo{journal}{\emph{IEEE Trans. on Very Large Scale Integration (VLSI) Systems}} \bibinfo{volume}{25}, \bibinfo{number}{10} (\bibinfo{year}{2017}), \bibinfo{pages}{2736--2748}.
\newblock


\bibitem[Waris et~al\mbox{.}(2020)]%
        {2020_Waris_IEEEtcasii}
\bibfield{author}{\bibinfo{person}{Haroon Waris}, \bibinfo{person}{Chenghua Wang}, {and} \bibinfo{person}{Weiqiang Liu}.} \bibinfo{year}{2020}\natexlab{}.
\newblock \showarticletitle{Hybrid Low Radix Encoding-Based Approximate Booth Multipliers}.
\newblock \bibinfo{journal}{\emph{IEEE Trans. on Circuits and Systems II: Express Briefs}} \bibinfo{volume}{67}, \bibinfo{number}{12} (\bibinfo{year}{2020}), \bibinfo{pages}{3367--3371}.
\newblock


\bibitem[Waris et~al\mbox{.}(2021)]%
        {2021_Waris_IEEEtcasii}
\bibfield{author}{\bibinfo{person}{Haroon Waris}, \bibinfo{person}{Chenghua Wang}, \bibinfo{person}{Weiqiang Liu}, {and} \bibinfo{person}{Fabrizio Lombardi}.} \bibinfo{year}{2021}\natexlab{}.
\newblock \showarticletitle{AxBMs: Approximate Radix-8 Booth Multipliers for High-Performance FPGA-Based Accelerators}.
\newblock \bibinfo{journal}{\emph{IEEE Trans. on Circuits and Systems II: Express Briefs}} \bibinfo{volume}{68}, \bibinfo{number}{5} (\bibinfo{year}{2021}), \bibinfo{pages}{1566--1570}.
\newblock


\bibitem[Wen et~al\mbox{.}(2016)]%
        {wen2016learning}
\bibfield{author}{\bibinfo{person}{Wei Wen}, \bibinfo{person}{Chunpeng Wu}, \bibinfo{person}{Yandan Wang}, \bibinfo{person}{Yiran Chen}, {and} \bibinfo{person}{Hai Li}.} \bibinfo{year}{2016}\natexlab{}.
\newblock \showarticletitle{Learning Structured Sparsity in Deep Neural Networks}.
\newblock \bibinfo{journal}{\emph{Advances in Neural Information Processing Systems}}  \bibinfo{volume}{29} (\bibinfo{year}{2016}), \bibinfo{pages}{1--10}.
\newblock


\bibitem[Wen et~al\mbox{.}(2018)]%
        {2018_Wen_ICDCS}
\bibfield{author}{\bibinfo{person}{Zhenyu Wen}, \bibinfo{person}{Do~Le Quoc}, \bibinfo{person}{Pramod Bhatotia}, \bibinfo{person}{Ruichuan Chen}, {and} \bibinfo{person}{Myungjin Lee}.} \bibinfo{year}{2018}\natexlab{}.
\newblock \showarticletitle{Approx{I}o{T}: Approximate Analytics for Edge Computing}. In \bibinfo{booktitle}{\emph{IEEE Int'l. Conference on Distributed Computing Systems (ICDCS)}}. \bibinfo{pages}{411--421}.
\newblock


\bibitem[Woo et~al\mbox{.}(1995)]%
        {1995_Woo_ISCA}
\bibfield{author}{\bibinfo{person}{Steven~Cameron Woo}, \bibinfo{person}{Moriyoshi Ohara}, \bibinfo{person}{Evan Torrie}, \bibinfo{person}{Jaswinder~Pal Singh}, {and} \bibinfo{person}{Anoop Gupta}.} \bibinfo{year}{1995}\natexlab{}.
\newblock \showarticletitle{The {SPLASH-2} Programs: Characterization and Methodological Considerations}. In \bibinfo{booktitle}{\emph{ACM/IEEE Annual Int'l. Symposium on Computer Architecture (ISCA)}}. \bibinfo{pages}{24–36}.
\newblock


\bibitem[Xu et~al\mbox{.}(2016)]%
        {2016_Xu_IEEEdt}
\bibfield{author}{\bibinfo{person}{Qiang Xu}, \bibinfo{person}{Todd Mytkowicz}, {and} \bibinfo{person}{Nam~Sung Kim}.} \bibinfo{year}{2016}\natexlab{}.
\newblock \showarticletitle{Approximate Computing: A Survey}.
\newblock \bibinfo{journal}{\emph{IEEE Design \& Test}} \bibinfo{volume}{33}, \bibinfo{number}{1} (\bibinfo{year}{2016}), \bibinfo{pages}{8--22}.
\newblock


\bibitem[Yazdanbakhsh et~al\mbox{.}(2015a)]%
        {2015_Yazdanbakhsh_DATE}
\bibfield{author}{\bibinfo{person}{Amir Yazdanbakhsh} {et~al\mbox{.}}} \bibinfo{year}{2015}\natexlab{a}.
\newblock \showarticletitle{Axilog: Language Support for Approximate Hardware Design}. In \bibinfo{booktitle}{\emph{Design, Automation \& Test in Europe (DATE)}}. \bibinfo{pages}{812--817}.
\newblock


\bibitem[Yazdanbakhsh et~al\mbox{.}(2017)]%
        {2017_Yazdanbakhsh_IEEEdt}
\bibfield{author}{\bibinfo{person}{Amir Yazdanbakhsh}, \bibinfo{person}{Divya Mahajan}, \bibinfo{person}{Hadi Esmaeilzadeh}, {and} \bibinfo{person}{Pejman Lotfi-Kamran}.} \bibinfo{year}{2017}\natexlab{}.
\newblock \showarticletitle{Ax{B}ench: A Multiplatform Benchmark Suite for Approximate Computing}.
\newblock \bibinfo{journal}{\emph{IEEE Design \& Test}} \bibinfo{volume}{34}, \bibinfo{number}{2} (\bibinfo{year}{2017}), \bibinfo{pages}{60--68}.
\newblock


\bibitem[Yazdanbakhsh et~al\mbox{.}(2016)]%
        {yazdanbakhsh2016axbench}
\bibfield{author}{\bibinfo{person}{Amir Yazdanbakhsh}, \bibinfo{person}{Divya Mahajan}, \bibinfo{person}{Pejman Lotfi-Kamran}, {and} \bibinfo{person}{Hadi Esmaeilzadeh}.} \bibinfo{year}{2016}\natexlab{}.
\newblock \bibinfo{booktitle}{\emph{AxBench: A Benchmark Suite for Approximate Computing Across the System Stack}}.
\newblock \bibinfo{type}{{T}echnical {R}eport}. \bibinfo{institution}{Georgia Institute of Technology}.
\newblock


\bibitem[Yazdanbakhsh et~al\mbox{.}(2015b)]%
        {gpu:accel}
\bibfield{author}{\bibinfo{person}{Amir Yazdanbakhsh}, \bibinfo{person}{Jongse Park}, \bibinfo{person}{Hardik Sharma}, \bibinfo{person}{Pejman Lotfi-Kamran}, {and} \bibinfo{person}{Hadi Esmaeilzadeh}.} \bibinfo{year}{2015}\natexlab{b}.
\newblock \showarticletitle{Neural Acceleration for GPU Throughput Processors}. In \bibinfo{booktitle}{\emph{IEEE/ACM Int'l. Symposium on Microarchitecture (MICRO)}}. \bibinfo{pages}{482--493}.
\newblock


\bibitem[Ye et~al\mbox{.}(2013)]%
        {2013_Ye_ICCAD}
\bibfield{author}{\bibinfo{person}{Rong Ye}, \bibinfo{person}{Ting Wang}, \bibinfo{person}{Feng Yuan}, \bibinfo{person}{Rakesh Kumar}, {and} \bibinfo{person}{Qiang Xu}.} \bibinfo{year}{2013}\natexlab{}.
\newblock \showarticletitle{On Reconfiguration-Oriented Approximate Adder Design and Its Application}. In \bibinfo{booktitle}{\emph{Int'l. Conference on Computer-Aided Design (ICCAD)}}. \bibinfo{pages}{48--54}.
\newblock


\bibitem[Yesil et~al\mbox{.}(2018)]%
        {2018_Yesil_IEEEmicro}
\bibfield{author}{\bibinfo{person}{Serif Yesil}, \bibinfo{person}{Ismail Akturk}, {and} \bibinfo{person}{Ulya~R. Karpuzcu}.} \bibinfo{year}{2018}\natexlab{}.
\newblock \showarticletitle{Toward Dynamic Precision Scaling}.
\newblock \bibinfo{journal}{\emph{IEEE Micro}} \bibinfo{volume}{38}, \bibinfo{number}{4} (\bibinfo{year}{2018}), \bibinfo{pages}{30--39}.
\newblock


\bibitem[Yu et~al\mbox{.}(2017)]%
        {yu2017scalpel}
\bibfield{author}{\bibinfo{person}{Jiecao Yu}, \bibinfo{person}{Andrew Lukefahr}, \bibinfo{person}{David Palframan}, \bibinfo{person}{Ganesh Dasika}, \bibinfo{person}{Reetuparna Das}, {and} \bibinfo{person}{Scott Mahlke}.} \bibinfo{year}{2017}\natexlab{}.
\newblock \showarticletitle{Scalpel: Customizing DNN Pruning to the Underlying Hardware Parallelism}.
\newblock \bibinfo{journal}{\emph{ACM SIGARCH Computer Architecture News}} \bibinfo{volume}{45}, \bibinfo{number}{2} (\bibinfo{year}{2017}), \bibinfo{pages}{548--560}.
\newblock


\bibitem[Yvinec et~al\mbox{.}(2021)]%
        {yvinec2021red}
\bibfield{author}{\bibinfo{person}{Edouard Yvinec}, \bibinfo{person}{Arnaud Dapogny}, \bibinfo{person}{Matthieu Cord}, {and} \bibinfo{person}{Kevin Bailly}.} \bibinfo{year}{2021}\natexlab{}.
\newblock \showarticletitle{RED : Looking for Redundancies for Data-Free Structured Compression of Deep Neural Networks}.
\newblock \bibinfo{journal}{\emph{Advances in Neural Information Processing Systems}}  \bibinfo{volume}{34} (\bibinfo{year}{2021}), \bibinfo{pages}{20863--20873}.
\newblock


\bibitem[Yvinec et~al\mbox{.}(2022)]%
        {yvinec2022red++}
\bibfield{author}{\bibinfo{person}{Edouard Yvinec}, \bibinfo{person}{Arnaud Dapogny}, \bibinfo{person}{Matthieu Cord}, {and} \bibinfo{person}{Kevin Bailly}.} \bibinfo{year}{2022}\natexlab{}.
\newblock \showarticletitle{RED++ : Data-Free Pruning of Deep Neural Networks via Input Splitting and Output Merging}.
\newblock \bibinfo{journal}{\emph{IEEE Trans. on Pattern Analysis and Machine Intelligence}} \bibinfo{volume}{45}, \bibinfo{number}{3} (\bibinfo{year}{2022}), \bibinfo{pages}{3664--3676}.
\newblock


\bibitem[Zeinali et~al\mbox{.}(2018)]%
        {2018_Zeinali_IEEEtcasii}
\bibfield{author}{\bibinfo{person}{Behzad Zeinali}, \bibinfo{person}{Dimitrios Karsinos}, {and} \bibinfo{person}{Farshad Moradi}.} \bibinfo{year}{2018}\natexlab{}.
\newblock \showarticletitle{Progressive Scaled {STT-RAM} for Approximate Computing in Multimedia Applications}.
\newblock \bibinfo{journal}{\emph{IEEE Trans. on Circuits and Systems II: Express Briefs}} \bibinfo{volume}{65}, \bibinfo{number}{7} (\bibinfo{year}{2018}), \bibinfo{pages}{938--942}.
\newblock


\bibitem[Zendegani et~al\mbox{.}(2017)]%
        {2017_Zendegani_IEEEtvlsi}
\bibfield{author}{\bibinfo{person}{Reza Zendegani}, \bibinfo{person}{Mehdi Kamal}, \bibinfo{person}{Milad Bahadori}, \bibinfo{person}{Ali Afzali-Kusha}, {and} \bibinfo{person}{Massoud Pedram}.} \bibinfo{year}{2017}\natexlab{}.
\newblock \showarticletitle{Ro{B}A Multiplier: A Rounding-Based Approximate Multiplier for High-Speed yet Energy-Efficient Digital Signal Processing}.
\newblock \bibinfo{journal}{\emph{IEEE Trans. on Very Large Scale Integration (VLSI) Systems}} \bibinfo{volume}{25}, \bibinfo{number}{2} (\bibinfo{year}{2017}), \bibinfo{pages}{393--401}.
\newblock


\bibitem[Zendegani et~al\mbox{.}(2016)]%
        {2016_Zendegani_DATE}
\bibfield{author}{\bibinfo{person}{Reza Zendegani}, \bibinfo{person}{Mehdi Kamal}, \bibinfo{person}{Arash Fayyazi}, \bibinfo{person}{Ali Afzali-Kusha}, \bibinfo{person}{Saeed Safari}, {and} \bibinfo{person}{Massoud Pedram}.} \bibinfo{year}{2016}\natexlab{}.
\newblock \showarticletitle{{SEERAD}: A High Speed yet Energy-Efficient Rounding-Based Approximate Divider}. In \bibinfo{booktitle}{\emph{Design, Automation \& Test in Europe (DATE)}}. \bibinfo{pages}{1481--1484}.
\newblock


\bibitem[Zervakis et~al\mbox{.}(2018)]%
        {2018_Zervakis_IEEEtvlsi}
\bibfield{author}{\bibinfo{person}{Georgios Zervakis}, \bibinfo{person}{Fotios Ntouskas}, \bibinfo{person}{Sotirios Xydis}, \bibinfo{person}{Dimitrios Soudris}, {and} \bibinfo{person}{Kiamal Pekmestzi}.} \bibinfo{year}{2018}\natexlab{}.
\newblock \showarticletitle{{VOS}sim: A Framework for Enabling Fast Voltage Overscaling Simulation for Approximate Computing Circuits}.
\newblock \bibinfo{journal}{\emph{IEEE Trans. on Very Large Scale Integration (VLSI) Systems}} \bibinfo{volume}{26}, \bibinfo{number}{6} (\bibinfo{year}{2018}), \bibinfo{pages}{1204--1208}.
\newblock


\bibitem[Zervakis et~al\mbox{.}(2019)]%
        {2019_Zervakis_IEEEtcasii}
\bibfield{author}{\bibinfo{person}{Georgios Zervakis}, \bibinfo{person}{Sotirios Xydis}, \bibinfo{person}{Dimitrios Soudris}, {and} \bibinfo{person}{Kiamal Pekmestzi}.} \bibinfo{year}{2019}\natexlab{}.
\newblock \showarticletitle{Multi-Level Approximate Accelerator Synthesis Under Voltage Island Constraints}.
\newblock \bibinfo{journal}{\emph{IEEE Trans. on Circuits and Systems II: Express Briefs}} \bibinfo{volume}{66}, \bibinfo{number}{4} (\bibinfo{year}{2019}), \bibinfo{pages}{607--611}.
\newblock


\bibitem[Zhang and Sanchez(2018)]%
        {2018_Zhang_IEEEcal}
\bibfield{author}{\bibinfo{person}{Guowei Zhang} {and} \bibinfo{person}{Daniel Sanchez}.} \bibinfo{year}{2018}\natexlab{}.
\newblock \showarticletitle{Leveraging Hardware Caches for Memoization}.
\newblock \bibinfo{journal}{\emph{IEEE Computer Architecture Letters}} \bibinfo{volume}{17}, \bibinfo{number}{1} (\bibinfo{year}{2018}), \bibinfo{pages}{59--63}.
\newblock


\bibitem[Zhang et~al\mbox{.}(2018)]%
        {2018_Zhang_DAC}
\bibfield{author}{\bibinfo{person}{Jeff Zhang}, \bibinfo{person}{Kartheek Rangineni}, \bibinfo{person}{Zahra Ghodsi}, {and} \bibinfo{person}{Siddharth Garg}.} \bibinfo{year}{2018}\natexlab{}.
\newblock \showarticletitle{Th{U}nder{V}olt: Enabling Aggressive Voltage Underscaling and Timing Error Resilience for Energy Efficient Deep Learning Accelerators}. In \bibinfo{booktitle}{\emph{Design Automation Conference (DAC)}}. \bibinfo{pages}{1--6}.
\newblock


\bibitem[Zhang et~al\mbox{.}(2015)]%
        {2015_Zhang_DATE}
\bibfield{author}{\bibinfo{person}{Qian Zhang}, \bibinfo{person}{Ting Wang}, \bibinfo{person}{Ye Tian}, \bibinfo{person}{Feng Yuan}, {and} \bibinfo{person}{Qiang Xu}.} \bibinfo{year}{2015}\natexlab{}.
\newblock \showarticletitle{Approx{ANN}: An {A}pproximate {C}omputing {F}ramework for {A}rtificial {N}eural {N}etwork}. In \bibinfo{booktitle}{\emph{Design, Automation \& Test in Europe (DATE)}}. \bibinfo{pages}{701--706}.
\newblock


\bibitem[Zhang et~al\mbox{.}(2016)]%
        {2016_Zhang_VLDB}
\bibfield{author}{\bibinfo{person}{Xuhong Zhang}, \bibinfo{person}{Jun Wang}, {and} \bibinfo{person}{Jiangling Yin}.} \bibinfo{year}{2016}\natexlab{}.
\newblock \showarticletitle{Sapprox: Enabling Efficient and Accurate Approximations on Sub-Datasets with Distribution-Aware Online Sampling}.
\newblock \bibinfo{journal}{\emph{Proceedings of the VLDB Endowment}} \bibinfo{volume}{10}, \bibinfo{number}{3} (\bibinfo{year}{2016}), \bibinfo{pages}{109–120}.
\newblock


\bibitem[Zhou et~al\mbox{.}(2016)]%
        {zhou2016dorefa}
\bibfield{author}{\bibinfo{person}{Shuchang Zhou}, \bibinfo{person}{Yuxin Wu}, \bibinfo{person}{Zekun Ni}, \bibinfo{person}{Xinyu Zhou}, \bibinfo{person}{He Wen}, {and} \bibinfo{person}{Yuheng Zou}.} \bibinfo{year}{2016}\natexlab{}.
\newblock \showarticletitle{DoReFa-Net: Training Low Bitwidth Convolutional Neural Networks with Low Bitwidth Gradients}.
\newblock \bibinfo{journal}{\emph{arXiv preprint arXiv:1606.06160}} (\bibinfo{year}{2016}), \bibinfo{pages}{1--13}.
\newblock


\bibitem[Zhu et~al\mbox{.}(2022)]%
        {2022_Zhu_IEEEtcasii}
\bibfield{author}{\bibinfo{person}{Feiyu Zhu}, \bibinfo{person}{Shaowei Zhen}, \bibinfo{person}{Xilin Yi}, \bibinfo{person}{Haoran Pei}, \bibinfo{person}{Bowen Hou}, {and} \bibinfo{person}{Yajuan He}.} \bibinfo{year}{2022}\natexlab{}.
\newblock \showarticletitle{Design of Approximate Radix-256 Booth Encoding for Error-Tolerant Computing}.
\newblock \bibinfo{journal}{\emph{IEEE Trans. on Circuits and Systems II: Express Briefs}} \bibinfo{volume}{69}, \bibinfo{number}{4} (\bibinfo{year}{2022}), \bibinfo{pages}{2286--2290}.
\newblock


\end{thebibliography}

\end{document}